\documentclass[reprint, aps, pra, twoside, floatfix, nofootinbib, longbibliography, superscriptaddress]{revtex4-2}
\usepackage[english]{babel}
\usepackage{float}
\usepackage[letterpaper,top=2cm,bottom=2cm,left=3cm,right=3cm,marginparwidth=1.75cm]{geometry}

\usepackage{graphicx,adjustbox}
\usepackage[colorlinks=true, allcolors=blue]{hyperref}
\usepackage{amsmath,amssymb}
\usepackage{xcolor}

\usepackage{orcidlink}
\usepackage{siunitx}
\usepackage{tabularray}
\usepackage{subfigure}
\usepackage{enumitem}

\usepackage[final]{changes} 

\begin{document}
\title{Exact amplitudes of  parametric processes in driven Josephson circuits}

\author{Roman~Baskov\,\orcidlink{0000-0001-9285-9469}}
\email{roman.baskov@yale.edu}
\author{Daniel~K.~Weiss\,\orcidlink{0000-0002-1952-896X}}
\author{Steven~M.~Girvin\,\orcidlink{0000-0002-6470-5494}} \affiliation{Departments of Applied Physics and Physics, Yale University, New Haven, CT 06520, USA} 
\affiliation{Yale Quantum Institute, 17 Hillhouse Ave., New Haven, CT 06520, USA}

\begin{abstract}
We present a general approach for analyzing arbitrary parametric processes in Josephson circuits within a single degree of freedom approximation. Introducing a systematic normal-ordered expansion for the  Hamiltonian of parametrically driven superconducting circuits we present a flexible procedure to describe parametric processes and to compare \added{and optimize} different circuit designs for particular applications. We obtain formally exact amplitudes (`supercoefficients') of these parametric processes for driven SNAIL-based and SQUID-based circuits.  The corresponding amplitudes contain  complete information about the circuit topology, the form of the nonlinearity, and the parametric drive, making them, in particular, well-suited for the study of the strong drive regime. We present a closed-form expression for supercoefficients describing circuits without stray inductors and a tractable formulation for those with it. We demonstrate the versatility of the approach by applying it to the estimation of Kerr-cat qubit Hamiltonian parameters and by examining the criterion for the emergence of chaos in Kerr-cat qubits. Additionally, we extend the approach to multi-degree-of-freedom circuits comprising multiple linear modes weakly coupled to a single nonlinear mode. We apply this generalized framework to study the activation of a beam-splitter interaction between two cavities coupled via driven nonlinear elements. Finally, utilizing the flexibility of the proposed approach, we separately derive supercoefficients for the higher-harmonics model of Josephson junctions, circuits with multiple drives, and the expansion of the Hamiltonian in the exact eigenstate basis for Josephson circuits with specific symmetries.
\end{abstract}

\setcounter{page}{1}
\maketitle

\section{Introduction}
Since the inception of circuit QED \cite{Blais2004,Wallraff_cQED_2004,Blais2021}, which initially focused on demonstrating the feasibility of superconducting circuits as building blocks for quantum computing, there has been a tremendous expansion in both the versatility of circuit designs and the range of utilized properties. These include, for example, multi-loop circuits \cite{Aoki2024,hua2024,bhandari2024symmetricallythreadedsquidsgeneration}, multiple drives \cite{Royer2018,Lu2023}, multiphoton processes \cite{Leghtas2015,xiao2023diagrammatic,hajr2024},  strong parametric drives \cite{Shillito2022,googleRWA,google_beyondRWA,dumas24}, and higher-harmonic Josephson junctions \cite{Willsch2024,wang2024systematicstudyhighejec}.

However, along with the increased serviceability of more complex circuit schemes come more complicated system dynamics  and the need to control unwanted processes attributed to higher-order nonlinearities and strong drives\cite{Grimm2020,jaya2023interference,Cohen23,chávezcarlos2024}. For parametrically driven circuits, such an increase in complexity  
necessitates the development of a theoretical model which can flexibly account for circuit topology and extend beyond the weak-drive approximation, entering  the regime where the low-order nonlinearity expansion of the model Hamiltonian is no longer valid. For practical applications, it is often useful to make the parametric driving as strong as possible, but not so strong that undesirable effects (\textit{e.g}, chaos, multiphoton resonances) destroy the coherence.  The methods presented here can aid in circuit optimization to achieve the desired goals.

In this regard, several proposals have been made for modeling circuits with arbitrary topologies, both for single and multiple degrees of freedom (DOF) \cite{BlackboxQuantization,sandro2023,Minev2021}, as well as for the renormalization of parametric process amplitudes up to the required order of drive amplitude and nonlinearities through an iterative procedure \cite{xiao2023diagrammatic,jaya2022lind,frattini2022squeezed}. However, these approaches face tractability issues when the drive is strong or higher-order nonlinearities significantly affect process amplitudes, necessitating an increasing number of terms in the model Hamiltonian \cite{frattini2022squeezed,ChapmanStijn2023,GarciaMata2024effectiveversus}. Addressing this requires a compact (or closed-form) expression that captures the circuit topology and accounts for the strong drive. Previous works \cite{smith2016,Verney2019,petrescu2023,chapple2024} have accomplished this in specific cases.

In this paper, we present a general approach that allows us to overcome this problem for parametrically driven circuits within single degree of freedom (DOF) approximation and to obtain formally exact amplitudes of parametric processes for Josephson circuits. The corresponding amplitudes contain full information about the circuit configuration and seamlessly account for the increasing complexity of such circuits.

\added{The paper is organized as follows. In Section~\ref{sec:eff_ham_kpo}, we present the general normal-ordered Hamiltonian for an arbitrary single DOF potential and introduce formally exact amplitudes—referred to as ‘supercoefficients’—for parametric processes. Closed-form expressions for these supercoefficients are derived for circuits with specific symmetries in Section~\ref{sec:sc_for_sym_circuits}. In Section~\ref{sec:applications}, we demonstrate the versatility of the supercoefficient approach by applying it to Kerr-cat dynamics and tunable beam-splitter interactions. Detailed derivations are provided in the Appendices.}

\section{General normal-ordered Hamiltonian for single DOF circuits}
\label{sec:eff_ham_kpo}

We start with the circuit QED Hamiltonian for a capacitively driven circuit with an arbitrary single DOF potential  \cite{Blais2021,frattini2022squeezed}
\begin{equation}\label{eq:full_cqed_ham}
\hat H = 4E_C(\hat{n}-n_g)^2+U(\hat{\varphi})+\hat{n}\Omega_d\cos{(\omega_d t+\theta)},
\end{equation}
where $n_g$ is a charge offset, $E_C$ is effective capacitive energy of the superconducting dipole and $\Omega_d$, $\omega_d$, $\theta$ are the amplitude, the frequency and the phase of the parametric drive. At this point we deliberately consider the potential energy in its most general form, delaying circuit-specific information to as late a stage as possible.   

Hereafter, we adopt a standard approximation~\cite{Koch2007,frattini2022squeezed}, where $\hat\varphi$ is treated as non-compact and the charge offset is gauged away ($n_g = 0$), an approach that works well in many cases. This allows us to use the harmonic oscillator basis defined by the quadratic part of the Hamiltonian. \added{Later in the paper, we also consider the exact eigenstate basis to go beyond this approximation 
to treat circuits where offset charge is important or where there are multiple potential minima.}

Firstly, to obtain the normal-ordered expansion of Hamiltonian (\ref{eq:full_cqed_ham}) we extract the quadratic part of the Hamiltonian and perform a displacement transformation based on the response of the system to the drive (see Appendix~\ref{app:eff_ham_derivaiton})
\begin{align}\label{eq:ham_with_quad_term}
\hat{{H}}&=\omega_0\hat{a}^\dag\hat{a}\nonumber
{+}{{U}\left({{\varphi}_\mathrm{zpf}}\left(\hat{a}+\hat{a}^\dag\right){+}\Tilde{\Pi}\cos{(\omega_d{t}{+}\gamma)}{+}\varphi_0\right)}\nonumber\\&\quad-\frac{E_Jc_2}{2}{\left({{\varphi}_\mathrm{zpf}}\left(\hat{a}+\hat{a}^\dag\right){+}\Tilde{\Pi}\cos{(\omega_d{t}{+}\gamma)}\right)^2},
\end{align}
where the effective drive amplitude is given by the linear-response 
$\Tilde{\Pi} = \Omega_d{\omega}_d/(\omega_d^2{-}\omega_0^2)$ and the phase  by
$\gamma=\theta{-}\pi/2$. Also, we have defined ${\partial_\varphi^2{U}(\varphi)}|_{\varphi_0}=E_Jc_2$ to distinguish the energy properties from the features of the circuit topology with potential minimum $\varphi_0$ defined by the usual condition ${\partial_\varphi U}|_{\varphi_0}=0$, $\omega_0=\sqrt{8E_CE_Jc_2}$ (see Appendix~\ref{app:eff_ham_derivaiton}). The energy scale $E_J$ is the largest Josephson energy of the individual Josephson junction (JJ) in the circuit. In addition,  we employ the bosonic basis,  $\hat{{n}}=-i{n}_\mathrm{zpf}{(\hat{a}-\hat{a}^\dag)}$, $\hat{{\varphi}}={\varphi}_\mathrm{zpf}{(\hat{a}+\hat{a}^\dag)}+\varphi_0$,
where $\varphi_\mathrm{zpf} = \sqrt[\leftroot{-2}\uproot{2}4]{{2E_C}/{E_Jc_2}}$ and $n_\mathrm{zpf}=1/(2\varphi_\mathrm{zpf})$ are the zero-point fluctuations of the phase and the charge for the ground state of the quadratic part of the Hamiltonian that describes small oscillations of the phase near the minimum of potential. 

Secondly, we can reformulate the usual expansion of the potential near $\varphi_0$ 
\begin{align}
&{U}({{\varphi}_\mathrm{zpf}}\left(\hat{a}+\hat{a}^\dag\right)+{\Tilde{\Pi}}\cos{(\omega_d{t}
+\gamma)}+\varphi_0)=\nonumber\\
&e^{\left({{\varphi}_\mathrm{zpf}}\left(\hat{a}{+}\hat{a}^\dag\right){+}{\Tilde{\Pi}}\cos{(\omega_d{t}{+}\gamma)}\right){\partial_\varphi}}\left.{U}(\varphi)\right|_{\varphi_0}
\end{align}
by utilizing the definition of the translation operator $\hat{T}(x)f(r) = f(r+x)$, $\hat{T}(x) = \exp(x{\partial_r})$.
Then, using expressions for the normal-ordered expansion of the operator function and the Jacobi-Anger expansion for the drive term in the exponent (see Ref.~\cite{wilcox1967} and Appendix~\ref{app:eff_ham_derivaiton}),  the Hamiltonian can be written in the following concise form
\begin{align}
\label{eq:ham_supercoeff} \hat{{H}}=\omega_0\hat{a}^\dag\hat{a}+&\sum\limits_{ n,l,p=0}^{\{l,p\}^\prime}{C}_{nl,p}(\hat{a}^{\dag n}\hat{a}^{n+l}+\hat{a}^{\dag n+l}\hat{a}^{n})\nonumber\\&\times(e^{ip(\omega_d{t}{+}\gamma)}+e^{-ip(\omega_d{t}{+}\gamma)}),\hspace{5mm}
\end{align}
where we introduce the  `supercoefficients' $C_{nl,p}$ for the nonlinear part of the Hamiltonian, which encapsulate complete information about the circuit configuration. \added{The superscription $\{l,p\}^\prime$ denotes that for terms with $l=0$ and $p=0$, an additional $1/2$ multiplier should be applied.}

In the most general form, the supercoefficients (SCs) can be written as an infinite sum 
\begin{equation}\label{eq:supercoefficiennt_as_sum}
{C_{nl,p}}{=}{\sum\limits_{k,m}^{\mathrm{S}\geq3}}
\frac{c_{2n{+}l{+}2k{+}p{+}2m}E_J\varphi_\mathrm{zpf}^{2n{+}l{+}2m}{\Tilde{\Pi}}^{2k{+}p}}{m!k!(k{+}p)!n!(n{+}l)!2^{m+2k+p}},
\end{equation}
where $c_n{=}{\partial^n_\varphi{U}(\varphi)}|_{\varphi_0}/E_J$ is a dimensionless nonlinear coefficient. 
\added{Here, $\mathrm{S}=2n+l+p+2m+2k$ and the condition $S\geq 3$ on the indices in the sum originates from the subtraction of the quadratic term in (\ref{eq:ham_with_quad_term}). Overall, the combined index $\mathrm{S}$ reflects the number of harmonic oscillator quanta and drive photons contributing to each term in the sum. Here, the index $m$ reflects the contribution from the non-commutativity of bosonic operators. The indices $l$ and $p$ reflect the number of oscillator and drive photons responsible for the coupling of different oscillator states, respectively (see Eq.~\eqref{eq:ham_supercoeff} and, for example, Ref.~\cite{xiao2023diagrammatic}). The indices $n$ and $k$ describe the number of oscillator and drive photon pairs involved in the renormalization of the uncoupled energy spacing between Fock states ($l=0$, $p=0$) and the coupling strength, due to the population of the oscillator, $\langle\hat{a}^\dag\hat{a}\rangle$, and the drive, $\tilde{\Pi}^2$. In particular, this includes the Stark shift and Kerr nonlinearity.} 

The representation of the Hamiltonian in Eq.~(\ref{eq:ham_supercoeff}) is more convenient than the more common expansion \cite{Blais2021,frattini2022squeezed}
\begin{equation}\label{eq:ham_gn} \hat{{H}}=\omega_0\hat{a}^\dag\hat{a}{+}\sum\limits_{ n\geq3}{g_n}\left(\hat{a}{+}\hat{a}^\dag{+}\Pi e^{-i\omega_d{t}}{+}\Pi^{\ast} e^{i\omega_d{t}}\right)^n,
\end{equation}
with $g_n=c_n\varphi_\mathrm{zpf}^nE_J/n!$ and $\Pi=n_\mathrm{zpf}\Tilde{\Pi}e^{i\gamma}$ for a capacitive drive, as it separates out the individual terms for each parametric process with amplitudes $C_{nl,p}$.  These amplitudes contain comprehensive information about the circuit and make it easier to distinguish and track how different properties—such as drive power, circuit topology, and zero-point fluctuations—contribute to the various parametric processes. 
Additionally, the higher level of abstraction provided by normal ordering and the explicit expressions for the supercoefficients make this approach particularly well-suited for deriving an effective Hamiltonian of the required order to obtain the desired parametric processes. Although we have considered a single capacitive drive, the approach is easily generalized with minor modifications for the case of flux-driven circuits and multiple drives (see Appendix \ref{app:eff_ham_derivaiton}).

\section{Supercoefficients for symmetric Josephson circuits}
\label{sec:sc_for_sym_circuits}

\begin{table*}[t]
\begin{tblr}{
 row{odd} = {bg=azure9},
 row{1} = {bg=azure3, fg=white, font=\sffamily
 },
 colspec={X[l,0.7]X[c,1]X[c,2.2]},
}
Circuit Design & Potential $U(\varphi)$& Supercoefficients $C_{nl,p}$\\
\SetCell[r=2]{l,m}{
\includegraphics[width=0.2\textwidth]{ 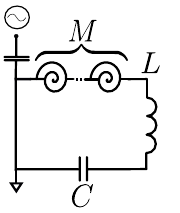}}
& {$-M\alpha_\mathrm{s} E_J\cos{\frac{\varphi}{M}}-$\\$MNE_J\cos\left(\frac{\varphi}{MN}-
\frac{\varphi_{e}}{N}\right),$\\for $L=0$} & {$\frac{({-}1)^{\left \lfloor\frac{2n{+}l{+}p}{2}\right \rfloor}E_J{\varphi}_\mathrm{zpf}^{2n{+}l}}{n!(n+l)!M^{2n+l-1}}\Bigg({\alpha_\mathrm{s}}\left\{\!\begin{aligned}
&{-}\cos{({\varphi_0}/{M})}\\[0.125ex]
&\sin{({\varphi_0}/{M})}\\[0.125ex]
\end{aligned}\right\}_{l{+}p}J_{p}(\frac{\Tilde{\Pi}}{M})e^{-\frac{{\varphi}_\mathrm{zpf}^2}{2M^2}}$ \\
${+}\left\{\!\begin{aligned}
&{-}\cos{\left(\frac{\varphi_0}{MN}{-}\frac{\varphi_{e}}{N}\right)}  \\[0.5ex]
&\sin{\left(\frac{\varphi_0}{MN}{-}\frac{\varphi_{e}}{N}\right)} \\[0.5ex]
\end{aligned}\right\}_{l{+}p}\frac{J_{p}\left(\frac{\Tilde{\Pi}}{MN}\right)}{N^{2n+l-1}}e^{-\frac{{\varphi}_\mathrm{zpf}^2}{2N^2M^2}}\Bigg)$}   \\
&  { $M U_{N}(\varphi_s[\varphi])$ \\${+}\frac{1}{2}E_L(\varphi-M\varphi_s[\varphi])^2,$ \\ for $L\neq 0$}  & {given by Eq.~(\ref{eq:supercoefficiennt_as_sum}) with $c_n=-Mx_J\partial_\varphi^{n-1}\varphi_s[\varphi]|_{\Bar{\varphi}_{min}}$,\\
$\partial_\varphi\varphi_s=(M+\partial_{\varphi_s}^2U_N/{E_L})^{-1}$} \\ {\includegraphics[width=0.19\textwidth]{ 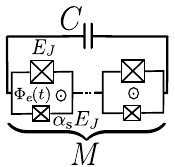}}& \SetCell{c,h}{${-}M\alpha_\mathrm{s}E_J\cos{\left(\frac{\varphi}{M}{-}r_a\varphi_{e}(t)\right)}$\\${-}{M}{E_J}\cos{\left(\frac{\varphi}{M}{+}r_b\varphi_{e}(t)\right)}$}&   \SetCell{r,h}{$\frac{({-}1)^{\left \lfloor\frac{2n{+}l{+}p}{2}\right \rfloor+1}E_J{\varphi}_\mathrm{zpf}^{2n{+}l}e^{{-}\frac{{\varphi}_\mathrm{zpf}^2}{2M^2}}}{n!(n+l)!M^{2n+l-1}}\mathcal{A}_p\left\{\!\begin{aligned}
&\cos{({\varphi_0}/{M}{-}\lambda^\prime_p)}\\[-0.5ex]
&\sin{({\varphi_0}/{M}{-}{\lambda^\prime_p})} \\[-0.5ex]
\end{aligned}\right\}_{l{+}p}$,\\{\footnotesize $\mathcal{A}_p{=}\sqrt{\alpha_\mathrm{s}^2J_p\left(\frac{\Pi_a}{M}\right)^2{+}J_p\left(\frac{\Pi_b}{M}\right)^2{+}2\alpha_\mathrm{s} J_p\left(\frac{\Pi_a}{M}\right)J_p\left(\frac{\Pi_b}{M}\right)\cos(\varphi_\mathrm{dc})}$,}\\ $\lambda^\prime_p=\arctan{\left(\frac{\alpha_\mathrm{s} J_p\left(\frac{\Pi_a}{M}\right)\sin{(r_a\varphi_\mathrm{dc})}{-}J_p\left(\frac{\Pi_b}{M}\right)\sin{(r_b\varphi_\mathrm{dc})}}{\alpha_\mathrm{s} J_p\left(\frac{\Pi_a}{M}\right)\cos{(r_a\varphi_\mathrm{dc})}{+}J_p\left(\frac{\Pi_b}{M}\right)\cos{(r_b\varphi_\mathrm{dc})}}\right)}$}\\ 
\end{tblr}
\caption{Supercoefficients for SNAIL-based and SQUID-based circuit designs. The geometric (stray) inductance, $L$, considered only for SNAIL-based designs (see \cite{fratini18} and Appendix~\ref{app:diff_potentials}), $x_J=E_L/E_J$, $N$ is the number of large JJs in the SNAIL (see Fig.~\ref{fig:SNAIL} in Appendix), $U_N$ is a potential of a single SNAIL;  for a SQUID-based circuits, the normalized external flux $\varphi_e(t)\equiv2\pi\Phi_e(t)/\Phi_0=\varphi_\mathrm{dc}+\varphi_\mathrm{ac}(t)$, where  $\Phi_0$ is the magnetic flux quantum and $\varphi_\mathrm{ac} (t)=2\varphi_\mathrm{ac0}\cos{\omega_dt}$, $\Pi_a$ and $\Pi_b$ are the linear-response partitions
between SQUID nodes with a convention $\Tilde{\Pi}\equiv\Pi_a-\Pi_b=2\varphi_\mathrm{ac0}M$, $r_a$ and  $r_b$ are gauge-dependent coefficients \cite{You2019,Riwar2022}, $r_a+r_b=1$; for all circuit designs, $M$ represents the number of single-loop nonlinear elements (SNAILs or SQUIDs) in the series array, $\alpha_\mathrm{s}$ is an asymmetry coefficient, and $C$ denotes the shunting capacitance (see Appendix~\ref{app:diff_potentials} for details on circuit designs).} 
\label{table:cEQD_schemes}
\end{table*}

The supercoefficients approach becomes even more advantageous when the potential can be expressed as a sum of periodic functions. To demonstrate this, we consider a single DOF potential for the special case of Josephson circuits without stray inductors.\ Specifically, circuits with permutation symmetry within each series array of nonlinear elements (JJs, SNAILs, etc.) \cite{ferguson2013,sandro2023,DiPaolo2021,Rymarz23,fratini18}, hereafter referred to as `symmetric' (see Appendix~\ref{app:diff_potentials}), and composed of individual single-loop nonlinear elements (\textit{e.g.}, SNAIL, SQUID, see Table \ref{table:cEQD_schemes}). In this case, the potential is given by
\begin{equation}
U(\varphi)=A\cos{(a_1\varphi{+}a_2\varphi_{e})}{+}B\cos{(b_1\varphi{+}b_2\varphi_{e})},
\label{eq:Uvarphi}
\end{equation}
where $A$, $B$, $a_1$, $b_1$, $a_2$, $b_2$ are determined by the circuit design, $\varphi_e$ is a normalized external flux (see details in Appendix~\ref{app:diff_potentials}). Consequently, the even- and odd-order nonlinearities, $c_n$, are always correspondingly proportional to cosine and sine functions of the equilibrium flux value, leading to a closed form for Eq.~(\ref{eq:supercoefficiennt_as_sum}) 
\begin{align}\label{eq:supercoefficients_joseph_circuit}
&C_{nl,p}=\frac{(-1)^{\left \lfloor\frac{2n+l+p}{2}\right \rfloor}{\varphi}_\mathrm{zpf}^{2n+l}}{n!(n+l)!}\times\\
&{\Bigg(Aa_1^{2n{+}l}J_p(a_1\Tilde{\Pi})e^{{-}\frac{{\varphi}_\mathrm{zpf}^2a_1^2}{2}}\left\{\!\small\begin{aligned}
&{-}\cos{(a_1\varphi_0{+}a_2\varphi_{e})}  \\[0.5ex]
&\sin{(a_1\varphi_0{+}a_2\varphi_{e})} \\[0.5ex]
\end{aligned}\right\}_{l{+}p}}\nonumber\\&{{+}Bb_1^{2n{+}l}J_p(b_1\Tilde{\Pi})e^{{-}\frac{{\varphi}_\mathrm{zpf}^2b_1^2}{2}}\left\{\!\small\begin{aligned}
&{-}\cos{(b_1\varphi_0{+}b_2\varphi_{e})}  \\[0.5ex]
&\sin{(b_1\varphi_0{+}b_2\varphi_{e})} \\[0.5ex]
\end{aligned}\right\}_{l{+}p}\Bigg)}\nonumber
\end{align}
for $2n+l+p\geq3$ and slightly modified form for $2n+l+p<3$ (see Appendix~\ref{app:general_pot_sym_circuits}) with even and odd $l+p$ invoking cosine or sine from curly brackets, respectively. This closed-form SC expression can be generalized for an arbitrary symmetric Josephson circuit, effectively providing the exact amplitudes of the parametric processes, which incorporate information about the drive via Bessel functions, about the non-commutativity of the bosonic operators via exponent terms, and about the circuit topology via $\varphi_0$ and the potential function. Notably, Eq.~(\ref{eq:supercoefficients_joseph_circuit}) allows one to avoid the slow convergence of the sum in Eq.~(\ref{eq:supercoefficiennt_as_sum}) for strong drives by collecting all drive terms in a Bessel function and retaining the periodic properties of the potential, in contrast to handling individual $c_n$'s. In particular, Eq.~(\ref{eq:supercoefficients_joseph_circuit}) reproduces the results for the specific cases of the undriven circuit and the strongly flux-driven SNAIL, as presented in Refs.~\cite{smith2016} and \cite{petrescu2023}, respectively.

\subsection{\added{Josephson junctions with higher harmonics}}

Recently, it was shown that the consideration of higher harmonics in JJs (\textit{e.g.}, $\cos(2\varphi)$ term in the Josephson energy) accounting for tunneling of multiple Copper pairs could have a significant effect on the dynamics of  superconducting circuits \cite{Willsch2024,wang2024systematicstudyhighejec}. Moreover, higher harmonics could be important for parametric processes \cite{féchant2025offsetchargedependence}. Here, we derive SCs for a higher-harmonics model of Josephson junctions. The general single DOF potential from Eq.~(\ref{eq:Uvarphi}) in this case is modified to
\begin{align}
U(\varphi)=\sum\limits_{m\ge1}A_m\cos{(ma_{1}\varphi{+}a_{2}\varphi_{e})}\\+\sum\limits_{m\ge1}B_m\cos{(mb_{1}\varphi{+}b_{2}\varphi_{e})},\nonumber
\end{align}
where $A_m$, $B_m$  are the amplitudes reflecting
coherent tunneling of groups of $m$ Cooper
pairs. The supercoefficients are derived straightforwardly, similar to Eq.~(\ref{eq:supercoefficients_joseph_circuit}),
\begin{widetext}
\begin{align}\label{eq:sc_joseph_circuit_harmonics}
{ C_{nl,p}=\frac{(-1)^{\left \lfloor\frac{2n+l+p}{2}\right \rfloor}{\varphi}_\mathrm{zpf}^{2n+l}}{n!(n+l)!}}
{ \Bigg(\sum\limits_{m\ge1}A_m(ma_{1})^{2n+l}J_p(ma_{1}\Tilde{\Pi})e^{{-}\frac{{\varphi}_\mathrm{zpf}^2a_1^2m^2}{2}}\left\{\!\begin{aligned}
&{-}\cos{(ma_{1}\varphi_0{+}a_2\varphi_{e})}  \\[0.5ex]
&\sin{(ma_1\varphi_0{+}a_2\varphi_{e})} \\[0.5ex]
\end{aligned}\right\}_{l{+}p}}\nonumber\\{{+}\sum\limits_{m\ge1}B_m(mb_1)^{2n+l}J_p(mb_1\Tilde{\Pi})e^{{-}\frac{{\varphi}_\mathrm{zpf}^2b_1^2m^2}{2}}\left\{\!\begin{aligned}
&{-}\cos{(mb_1\varphi_0{+}b_2\varphi_{e})}  \\[0.5ex]
&\sin{(mb_1\varphi_0{+}b_2\varphi_{e})} \\[0.5ex]
\end{aligned}\right\}_{l{+}p}\Bigg)}.
\end{align}
\end{widetext}
Regardless of the specific circuit design, the corresponding SCs allow us to spot some features of higher-order multi-photon processes for such a model.\ Particularly, from Eq.~\eqref{eq:sc_joseph_circuit_harmonics}, the modified terms $A_m(ma_{1})^{2n+l}J_p(ma_{1}\Tilde{\Pi})$ and $B_m(mb_1)^{2n+l}J_p(mb_1\Tilde{\Pi})$ for higher harmonics suggest that the large prefactors associated with higher indices $m$, $n$, $l$ could compensate for the smallness of $A_m$, $B_m$. In addition, higher-harmonic terms are  significantly more sensitive to increases in drive strength due to factor $m$ in the argument of Bessel functions.

\subsection{\added{Supercoefficients for an exact eigenstate basis}}

Notably, a similar closed-form expression for the SCs (\ref{eq:supercoefficients_joseph_circuit}) can be derived from an expansion of Hamiltonian (\ref{eq:full_cqed_ham}) expressed in the exact eigenstate basis. This expansion avoids some limitations of the harmonic oscillator approximation. Namely, it explicitly accounts for the charge offset $n_g$, potential periodicity, and multi-minima potentials. To preserve compactness of $\hat\varphi$, the drive can be incorporated into the potential function via a time-dependent unitary transformation \cite{Cohen23}
\begin{equation}    \hat{U}_D=\exp{\left(-i\hat{n}\frac{\Omega_d}{\omega_d}\sin{\omega_dt}\right)}.
\end{equation}
Here, for simplicity, we also omit drive phase ($\theta=0$).
The displaced-frame Hamiltonian is given by
\begin{align}
\hat{U}_{D}^\dag\hat H\hat{U}_{D}&-i\hat{U}_{D}^\dag\dot{\hat{U}}_{D}\rightarrow \\\hat H &= 4E_C(\hat{n}-n_g)^2+U\left(\hat{\varphi}+\frac{\Omega_d}{\omega_d}\sin{\omega_dt}\right).\nonumber
\end{align}
Unlike the harmonic oscillator representation, this expression preserves the periodicity of the potential. For symmetric Josephson circuits (see Eq.~(\ref{eq:Uvarphi})), we  
use trigonometric identities and the Jacobi-Anger expansions \cite{Arfken2013a, Cohen23} to
extract  and diagonalize the renormalized time-independent part of the Hamiltonian
\begin{align}
\hat{H}_0&= 4E_C(\hat{n}-n_g)^2{+}AJ_0\left(\frac{a_1\Omega_d}{\omega_d}\right)\cos{(a_1\hat\varphi{+}a_2\varphi_{e})}\nonumber\\\,&+BJ_0\left(\frac{b_1\Omega_d}{\omega_d}\right)\cos{(b_1\hat\varphi{+}b_2\varphi_{e})}=\sum\limits_j\varepsilon_j|j\rangle\langle j|.
\end{align}
Consequently,  the expansion of the full Hamiltonian in $\hat H_0$ eigenstate representation reads
\begin{align}\label{eq_app:ham_eig_expansion}
\hat H &=\sum\limits_j \varepsilon_j|j\rangle\langle j|\\&\,+\sum\limits_{\substack{n,l=0\\p=1}}C_{nl,p}^\mathrm{eig}|n\rangle\langle l|\left(e^{ip\omega_dt}+(-1)^p e^{-ip\omega_dt}\right)\nonumber
\end{align}
with the supercoefficients given by
\begin{align}\label{eq_app:supercoef_eig}
C_{nl,p}^\mathrm{eig} &= AJ_p\left(\frac{a_1\Omega_d}{\omega_d}\right)\left\{\!\begin{aligned}
&\langle n|\cos{(a_1\hat\varphi{+}a_2\varphi_{e})}|l\rangle \\[0.5ex]
&i\,\langle n|\sin{(a_1\hat\varphi{+}a_2\varphi_{e})}|l\rangle \\[0.5ex]
\end{aligned}\right\}_{p}\nonumber\\&+BJ_p\left(\frac{b_1\Omega_d}{\omega_d}\right)\left\{\!\begin{aligned}
&\langle n|\cos{(b_1\hat\varphi{+}b_2\varphi_{e})}|l\rangle \\[0.5ex]
&i\,\langle n|\sin{(b_1\hat\varphi{+}b_2\varphi_{e})}|l\rangle \\[0.5ex]
\end{aligned}\right\}_{p}.
\end{align}
The result of Eqs.~(\ref{eq_app:ham_eig_expansion}), (\ref{eq_app:supercoef_eig}) is easily generalized for the flux-driven case with $\varphi_{e}(t)=\varphi_\mathrm{dc}+\varphi_\mathrm{ac}(t)$ and $\varphi_\mathrm{ac}(t) = 2\varphi_\mathrm{ac0}\sin\omega_dt$ by substitutions ${a_1\Omega_d}/{\omega_d}\rightarrow 2\varphi_\mathrm{ac0}a_2$, ${b_1\Omega_d}/{\omega_d}\rightarrow 2\varphi_\mathrm{ac0}b_2$, and $\varphi_e\rightarrow\varphi_\mathrm{dc}$. The derivation of the SCs for a multiple-drive case is also straightforward. It introduces a product of Bessel functions and time-depended exponents originating from different drives (see Appendix \ref{app:multiple_drives}).

The expansion in Eq.~(\ref{eq_app:ham_eig_expansion}) formally presents the Hamiltonian applicable for an arbitrary drive with closed-form SCs for symmetric Josephson circuits. Additionally, the eigenstate representation rigorously accounts for the periodicity of potential and charge offset. It also could be better suited for circuits with $E_C/E_J\simeq1$ than the harmonic oscillator representation. However, the computational complexity of the eigenstate representation -- the diagonalization of the bare Hamiltonian, the tractability of different transition energies $\varepsilon_i-\varepsilon_j$  for models beyond rotating wave approximation, etc. -- makes it less practical for the description of low-energy dynamics for circuit designs with $E_C/E_J\ll1$, considered here. Fortunately, in this case, the system dynamics is less sensitive to charge offset and the periodicity properties of the potential [$U(\varphi)=U(\varphi+2\pi)$]. Moreover, unlike Eq.~(\ref{eq:ham_with_quad_term}), there is no easy way to generalize Eq.~(\ref{eq_app:ham_eig_expansion}) for the case of non-symmetric Josephson circuits (meaning the broken permutation symmetry within a series
an array of nonlinear elements or the presence of a stray inductance in the system) because of the transcendental functions in the definition of the single DOF potential. So, while this representation has its advantages and limitations compared to the harmonic oscillator approximation, the latter is better suited for the selection of circuit designs considered in this work.

\section{Applications}
\label{sec:applications}
To  demonstrate the versatility and flexibility of the SC approach,  we consider circuit designs with a capacitively driven array of superconducting nonlinear asymmetric
inductive elements (SNAIL)\footnote{\added{We use a generalized notion of a SNAIL having $N$ large Josephson junctions (e.g., see Ref.~\cite{Hillmann2020}). Such a definition covers a wider range of circuit designs. In particular, for $N \gg 1$, it describes the fluxonium.}}, with and without geometric inductance in the system, and will also consider flux-driven arrays of asymmetric dc superconducting quantum
interference devices (SQUID) (see Table \ref{table:cEQD_schemes}). These model circuits cover a wide spectrum of well-established designs in circuit QED which have been utilized for multiple tasks, such as readout and control of fluxonium qubits \cite{Zhu2013_fluxonium,Nguyen2022_fluxonium,Bao2022_fluxoniumm,Earnest2018_fluxonium}, Kerr-cat qubits \cite{frattini2022squeezed,Iyama2024,hajr2024,Kang2022}, engineering multi-wave couplers \cite{ChapmanStijn2023,degraaf2024,pietikäinen2024}, parametric amplifiers \cite{fratini18,ranadive2024,Sivak2020,kaufman2024}, universal gates \cite{Hillmann2020, Eriksson2024, yu2024}, etc.

We apply the SC approach for description and optimization of circuit designs for specific parametric processes with SCs derived from Eqs.~(\ref{eq:supercoefficiennt_as_sum}) and (\ref{eq:supercoefficients_joseph_circuit}). Circuit optimization involves deriving the effective Hamiltonian for a given circuit topology in terms of the supercoefficients and identifying optimal dynamic parameters within a specified range of circuit parameters (see the general guide in Appendix~\ref{app:circuit_opt}).  In particular, we consider the dynamics of a squeezed Kerr parametric oscillator (KPO) as a key element in schemes utilizing Kerr-cat qubits. \added{We also investigate a tunable beam-splitter interaction between two cavity modes coupled via a nonlinear element—such as an array of SNAILs or SQUIDs—which plays a crucial role in various quantum computing schemes \cite{Pietikainen:2022iqj,Lu2023,ChapmanStijn2023,Liu2024,tsunoda2023error,Teoh2023,Chou2024,degraaf2024}.}

\subsection{Kerr-cat dynamics}

\begin{figure}
\centering
\includegraphics[width=0.98\linewidth]{ 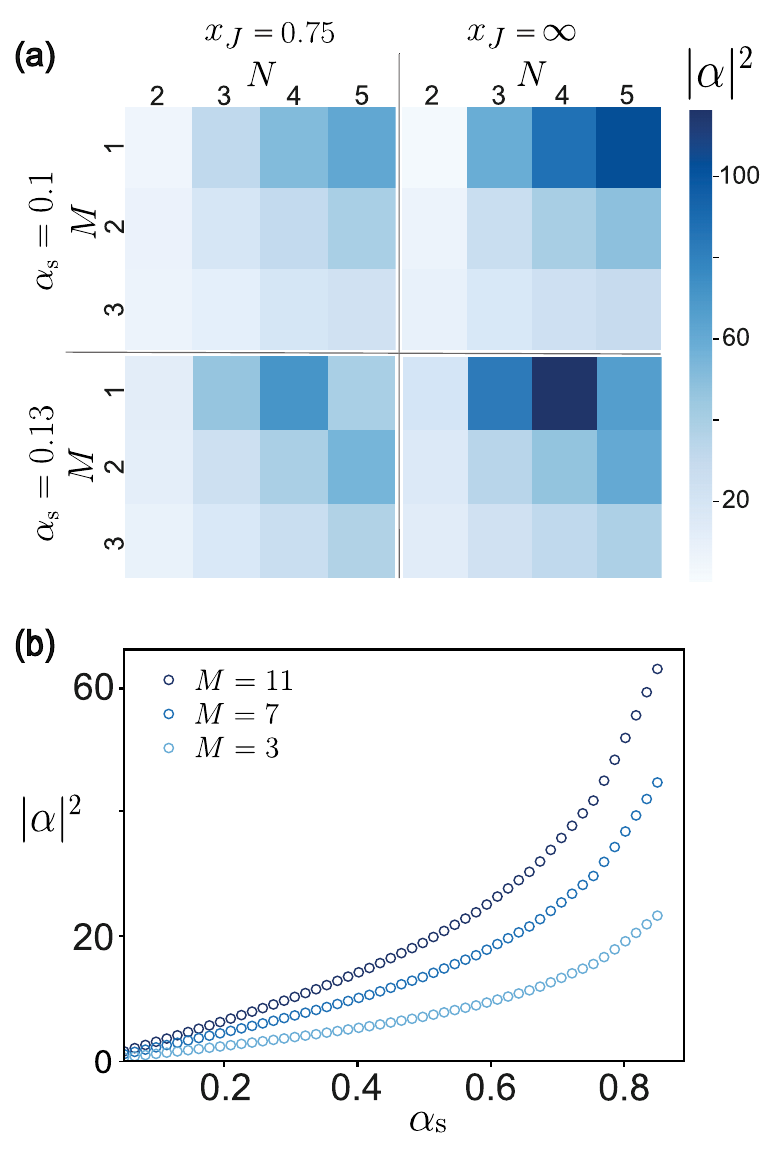}
\caption{(Color online) Maximum Kerr-cat size for different (a) SNAIL-based and (b) SQUID-based circuit designs (see Table \ref{table:cEQD_schemes}) with minimal Kerr nonlinearity  $K_\mathrm{lim}=\SI{1}{\mega\hertz}$, \added{effective drive amplitude} $\Pi=0.5$, \added{and combined index} $\mathrm{S}\leq8$ for $L\neq\infty$ (see Eq.~(\ref{eq:supercoefficiennt_as_sum})). For (a), (b) parameters are derived from \cite{Grimm2020,frattini2022squeezed}. More details on circuit optimization and simulation parameters are presented in Appendix~\ref{app:kerr_cat_scs}.} 
\label{fig:max_kc_size} 
\end{figure}
The effective Hamiltonian for Kerr-cat system is obtained for a KPO driven at $\omega_d\approx2\omega_q$ from Eq.~(\ref{eq:ham_supercoeff}), 
\begin{equation}
\begin{aligned}
\label{eq:Kerr_cat_ham}
\hat{\mathcal{H}}_{\mathrm{eff}}\approx -K\hat{a}^{\dagger 2}\hat{a}^2 +   \epsilon_2 (\hat{a}^{\dagger 2}+\hat{a}^2).
\end{aligned}
\end{equation}
The parameters are defined via SCs as
\begin{align}\label{eq:omega_q}
\omega_q &= \omega_0+{C}_{10,0}+{\Delta}^{(1)},\\
{K} &= -C_{20,0}+{K}^{(1)},\label{eq:kerr_nl}\\
{\epsilon_2}&=C_{02,1}+\epsilon_2^{(1)},\label{eq:eps2}
\end{align}
where $\omega_q$ is the Lamb- and Stark-shifted
small oscillation frequency and ${\Delta}^{(1)}$, ${K}^{(1)}$, $\epsilon_2^{(1)}$ are first-order corrections beyond the rotating wave approximation (RWA)(see Appendix~\ref{app:kerr_cat_scs}).

Here, for circuit optimization, we focus on two key dynamic parameters of the Kerr-cat: the Kerr nonlinearity, $K$, and the mean photon number in the squeezed KPO, $|\alpha|^2=\epsilon_2/K$ \cite{Grimm2020,Puri2017}. The former characterizes the gate times in corresponding quantum computing schemes and the latter sets the degree of bit-flip suppression for Kerr-qubits \cite{frattini2022squeezed,jaya2023interference}. To determine the optimal circuit design, we assume that circuits with the same Kerr nonlinearity but a larger Kerr-cat size, $|\alpha|^2$, for a fixed drive strength are preferable, as they allow for a faster increase in qubit lifetime and potentially mitigate the negative effects of the strong drive regime. An alternative optimization approach considers the dilution of nonlinearities, which suppresses higher-order multiphoton processes often associated with spurious decoherence of the Kerr-cat qubit \cite{Grimm2020,frattini2022squeezed,alex2024}. \added{However, this approach narrows the range of achievable Kerr nonlinearities and typically demands higher pump power.} 


The comparison of Kerr-cat sizes for different circuit topologies in Figures \hyperref[fig:max_kc_size]{\ref{fig:max_kc_size}(a)}, \hyperref[fig:max_kc_size]{\ref{fig:max_kc_size}(b)} allows ranking different circuit designs within a proposed range of the parameters (see 
Simulation parameters in Appendix~\ref{app:kerr_cat_scs})). Besides that, there are some pronounced features: the geometric inductance in the circuit generally impedes the growth of the Kerr-cat; for SQUID-based circuits, the  Kerr-cat growth rate strongly depends on the asymmetry of SQIUDs; even for almost symmetric SQUIDs,   Kerr-cat sizes are at least two times smaller than for the best SNAIL-based scheme. 

Although such a comparison of circuit configurations is somewhat limited because it does not consider, \textit{e.g.},  dissipation, higher-order parametric processes, chaotic behavior (see below), etc., it demonstrates the applicability of the SC approach as an optimization tool and benchmarks the circuit designs in the simplest non-dissipative case. However, a more comprehensive analysis of the optimal circuit designs should account for a wider set of considerations.


\subsection{\added{Onset of chaos in Kerr-cat qubit}} 

\begin{figure}
\centering
\includegraphics[width=0.98\linewidth]{ 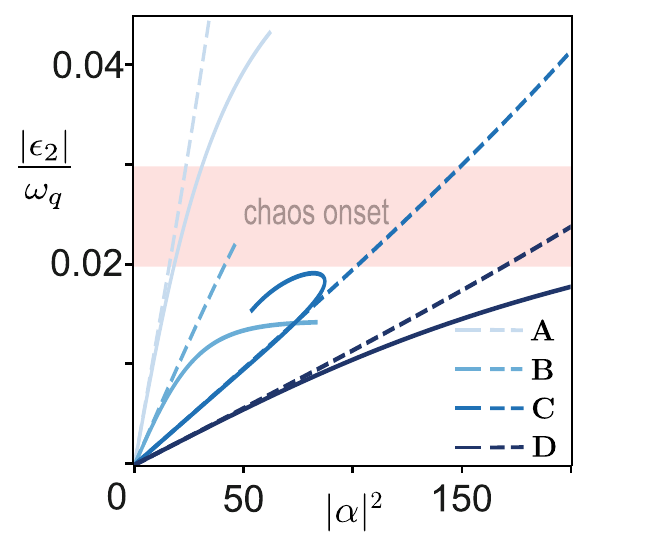}
\caption{(Color online) \added{Parametric curves  $|\epsilon_2(\Pi)|/\omega_q(\Pi)$ vs.\ $|\alpha|^2(\Pi)$ probes criteria for chaos onset for different SNAIL-based ($N=3$) circuit designs: \textbf{A} ($M=1$, $K=\SI{1.15}{\mega\hertz}$ for an undriven circuit), \textbf{B} ($M=2$, $K=\SI{-2.58}{\mega\hertz}$), \textbf{C} ($M=1$, $K=\SI{6.76}{\mega\hertz}$), \textbf{D} ($M=2$, $K=\SI{0.72}{\mega\hertz}$). Solid lines: SC approach with $\mathrm{S}\leq13$, dashed line: $\mathrm{S}\leq4$ (used in Ref.~\cite{chávezcarlos2024}). More details on the simulation parameters are presented in Table~\ref{tab:kerr_chaos_params} and Appendix~\ref{app:KC_chaos}.}} 
\label{fig:chaos_onset} 
\end{figure}
From classical chaos theory, it is well-known that driven nonlinear systems exhibit irregular dynamics near the separatrices, such as separatrix splitting and the emergence of a chaotic layer \cite{CHIRIKOV1979}.
The width of this chaotic layer increases for stronger drives. In superconducting circuits, this could lead to adverse effects, such as the introduction of an additional decoherence channel or even the destruction of the qubit if chaos extends to the computational states of the qubit \cite{Chepelianskii2002,Anand21,Cohen23,dumas24,Blais2021, chávezcarlos2024}. 
We examine the utility of the SC approach in gaining deeper insight into and predicting the emergence of chaotic behavior in Kerr-cat qubits. We utilize the results of Ref.~\cite{chávezcarlos2024} to predict the emergence of chaos in Kerr-cat qubits. In that work, the authors derive criteria for the onset of chaos and the transition to global chaos --- characterized by the merging of the chaotic layer originating from the separatrix of the full potential and the chaos on the parametric separatrix (the lemniscate of the Kerr-cat). This analysis is conducted by tracking the quasienergy spacing statistics of the Kerr-cat in Floquet simulations. The corresponding criterion is expressed in terms of the ratio $\epsilon_2/\omega_0$.

Here, we present a qualitative analysis of what the ratio $\epsilon_2/\omega_q$ (for $\Delta=0$ case) stands for by referring to the results for classical systems \cite{CHIRIKOV1979}. Physically, it represents the comparison of the level splitting of Kerr-cat quasienergies (approximately $4\epsilon_2$ \cite{jaya2023interference}) to the frequency of small oscillations, $\omega_q$ (see Eq.~\eqref{eq:omega_q}). To demonstrate the significance of this ratio, we check the Hamiltonian for Kerr-cat before time averaging in the Schrieffer-Wolff procedure (see Eq.~(\ref{eq_app:effH}) in Appendix) 
\begin{equation}
\begin{aligned}
\label{eq:Kerr_cat_ham}
\hat{H} = {-}K\hat{a}^{\dagger 2}\hat{a}^2 {+}   \epsilon_2 (\hat{a}^{\dagger 2}+\hat{a}^2){+}\hat{P}(\hat a^\dag,\hat a,\Tilde{\Pi}, e^{\pm i\omega_qt}),
\end{aligned}
\end{equation}
where $\hat P$ is a time-dependent polynomial operator with the slowest evolution frequency $\omega_q$. The last term is usually considered to contain all fast-oscillating ($4\epsilon_2\ll\omega_q$) and small-amplitude processes. However, since $\epsilon_2$ is a function of the amplitude of parametric drive, for some drive strength the quasienergy splitting $4\epsilon_2$ could become comparable to $\omega_q$. Additionally, the amplitudes of fast-oscillating terms in $\hat P$ can increase for stronger drive. Altogether, these changes in the amplitudes of the parametric processes can result in the splitting of the parametric separatrix, akin to the chaotic dynamics observed in a driven Duffing oscillator. In this scenario, the operator 
$\hat P$ in Eq.~\eqref{eq:Kerr_cat_ham} acts as the nonlinearly coupled drive. Although it is much more difficult to derive an explicit expression for the width of the chaotic layer at the Kerr-cat lemniscate  due to the complex form of the function $\hat P$, heuristic criteria can still be established \cite{CHIRIKOV1979,soskin2009,zaslavsky2007physics}
\begin{equation}
\Delta E\propto\frac{\omega_q}{4\epsilon_2}\exp{\left(-\frac{\omega_q}{4\epsilon_2}\right)}, 
\end{equation}
where the ratio $\epsilon_2/\omega_q$ is found to be a key parameter. Referring to the results of Ref.~\cite{chávezcarlos2024}, we set the criterion for the onset of chaos within the range  $\epsilon_2/\omega_q\in[0.02,0.03]$. Although somewhat coarse, this range for the chaos emergence criterion should adequately encompass the diverse set of considered circuit designs. Furthermore, we examine the applicability of the lower-order approximation employed in Ref.~\cite{chávezcarlos2024} using the SC approach. 

\begin{figure}
\centering
\includegraphics[width=0.98\linewidth]{ 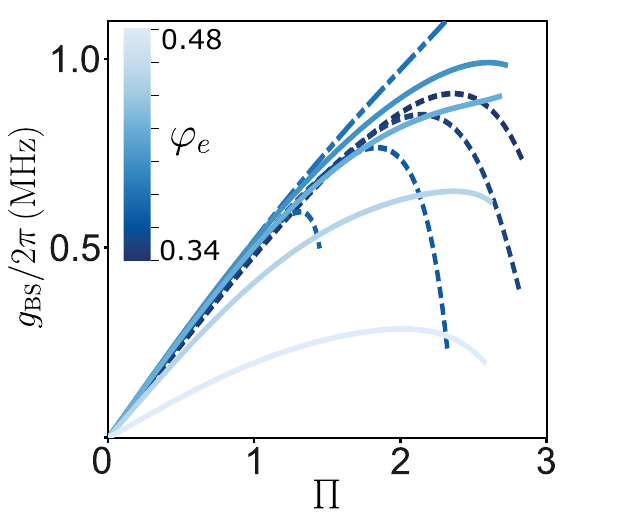}
    \caption{(Color online) \added{Beam-splitter interaction versus effective drive strength, showing sharp downturns due to multiphoton resonances for different flux biases with $\mathrm{S} \leq 13$ (see Eq.~\eqref{eq:supercoefficiennt_as_sum}). As a result of the Stark shift, the resonance condition $\tilde{\delta} = \delta + \Delta_a \approx 0$ is met at a particular drive strength (see Eq.~\eqref{eq:g_bs}), where $\delta = \omega_a^\prime + \omega_b^\prime - 2\omega_d$ and the circuit is driven at $\omega_d = \omega_c^\prime - \omega_b^\prime$. The presented curves demonstrate that multiphoton resonance manifests differently across various ranges of external flux: a sharp downturn shifting toward weaker drives for $\varphi_e \lesssim 0.37$ (dashed curves), no visible resonance within an intermediate range exemplified by $\varphi_e = 0.38$ (dash-dotted curve), and mild downturns for $\varphi_e \gtrsim 0.4$ (solid curves). The criteria for the values of $\Pi$ for each curve are derived from the constraints $g_{ab}/\tilde{\delta},\, g_{ac}/\tilde{\delta} \leq 0.25$, $\Tilde{\Pi}=\Pi/n_\mathrm{zpf}\lesssim2$ (see Eqs.~\eqref{eq:g_bs}, \eqref{eq_app:eff_ham_3modes}, and step 6 in Appendix \ref{app:circuit_opt}). The circuit parameters match the architecture described in Ref.~\cite{ChapmanStijn2023} (see details in Appendix~\ref{app:beam_splitter}).
}
} 
\label{fig:bs_downturn} 
\end{figure}
\begin{figure*}
\centering
\includegraphics[width=0.98\linewidth]{ 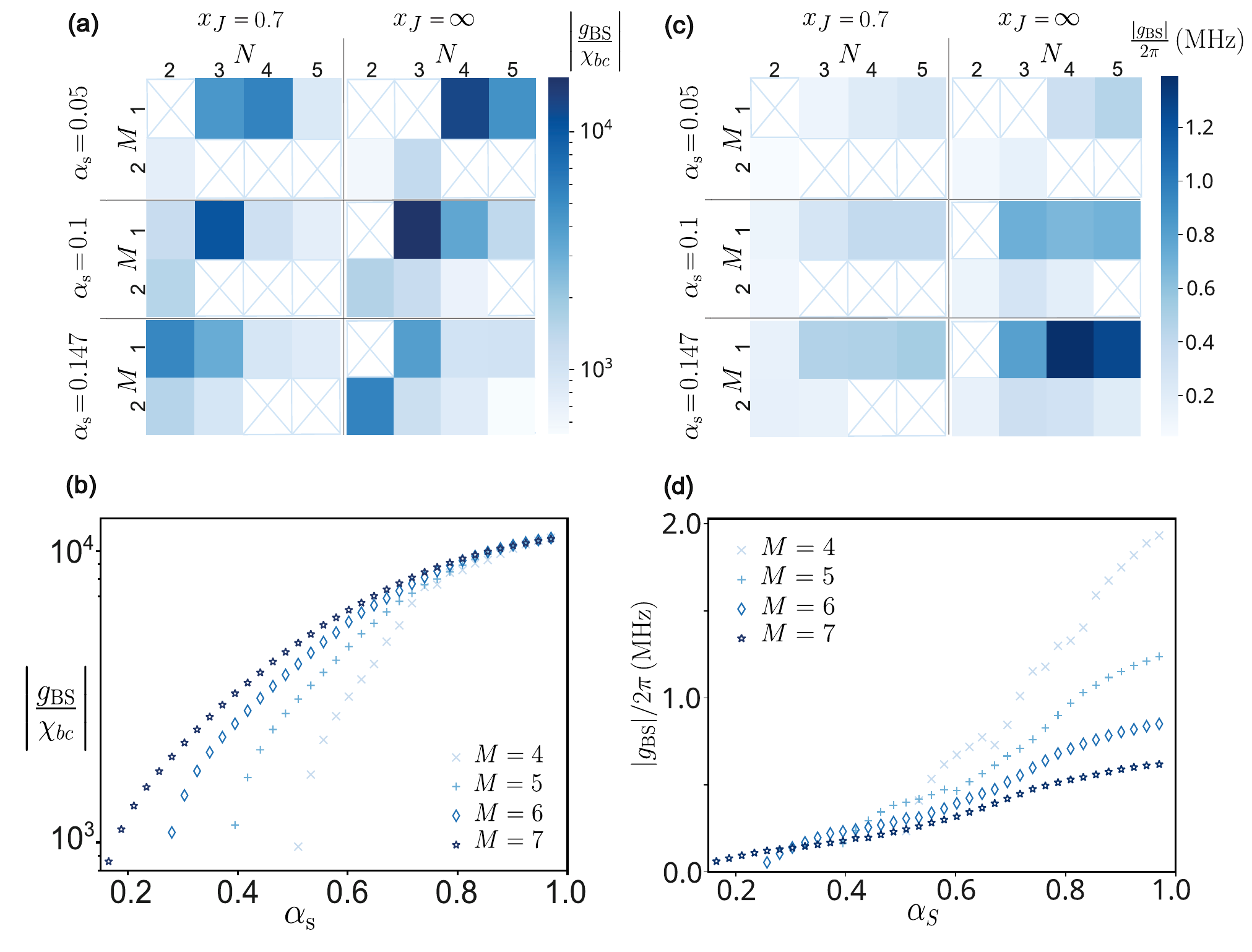}
\caption{(Color online) \added{Maximum on-off ratio between the beam-splitter interaction and the cavity-cavity cross-Kerr interaction, $\chi_{bc}$, for different (a) SNAIL-based and (b) SQUID-based circuit designs with $\chi_{bc} \geq \SI{30}{\hertz}$, effective drive amplitude $\Pi = 1$, and combined index $\mathrm{S} \leq 9$. Panels (c) and (d) show the corresponding beam-splitter interaction strengths. Crossed cells 
in (a) and (c) 
indicate circuits that fall outside the scope of the desired parameter range. Other simulation parameters follow Ref.~\cite{ChapmanStijn2023} and are detailed in Appendix~\ref{app:beam_splitter}. 
}}
\label{fig:on_off_ratio_and_bs_int} 
\end{figure*}
Figure~\ref{fig:chaos_onset} illustrates that, although a lower-order approximation predicts an inevitable onset of chaos—due to the linear scaling $|\alpha|^2 \propto \epsilon_2 \propto \Pi$—a more complete analysis reveals a richer and more intricate dynamical landscape. This complexity is shaped by the specifics of the circuit design and the nonlinear dependence of $|\alpha|^2$, $\epsilon_2$, and $\omega_q$ on drive strength (see Eqs.~\eqref{eq:omega_q}--\eqref{eq:eps2}, and Fig.~\ref{fig:chaos_suppl}). Notably, higher-order nonlinear corrections in SC approach show that, while configuration \textbf{D} enters a chaotic regime at stronger drives, configurations \textbf{B} and \textbf{C} can avoid chaos entirely. We also assess how the chaos criterion impacts the results in Figs.~\hyperref[fig:max_kc_size]{\ref{fig:max_kc_size}(a)} and~\hyperref[fig:max_kc_size]{\ref{fig:max_kc_size}(b)}. For the relatively weak drive $\Pi = 0.5$ considered, the corrections are minimal—for instance, only a slight deviation is observed in the $M=3$ curve of Fig.~\hyperref[fig:max_kc_size]{\ref{fig:max_kc_size}(b)} (see Fig.~\ref{fig:chaos_suppl} in Appendix~\ref{app:KC_chaos})

\subsection{Beam-splitter interaction} 

\added{The effective Hamiltonian for a parametrically tunable beam-splitter scheme, driven at frequency $\omega_d = \omega_c^\prime - \omega_b^\prime$, where $\omega_c^\prime$ and $\omega_b^\prime$ are the dressed frequencies of the coupled cavity modes, is given by (see Ref.~\cite{ChapmanStijn2023} and Appendix~\ref{app:beam_splitter}):
}
\begin{equation}
\hat{\mathcal{H}}_{\mathrm{eff}}\approx g_\mathrm{BS}(\hat{b}^\dag\hat{c}+\hat{b}\hat{c}^\dag)+\chi_{bc}\hat b^\dag\hat b\hat c^\dag\hat c.
\end{equation}
Here, the beam-splitter  and cavity-cavity cross-Kerr interactions are given by 
\begin{align}\label{eq:g_bs}
g_\mathrm{BS}&=C_{00,01,01,1}-{2g_{ab}g_{ac}}/{\Tilde{\delta}}, \\
\chi_{bc}&=C_{00,10,10,0}+\chi_{bc}^{(1)},
\end{align}
where $C_{n_al_a,n_bl_b,n_cl_c,p}$ are generalized SCs for the case of multiple DOF for nonlinearly coupled cavities with the indices $n_a$, $l_a$ corresponding to the coupler modes and $n_b$, $l_b$, $n_c$, $l_c$ to different cavity modes, $g_{ab}$ and $g_{ac}$ are parametric two-mode squeezing amplitudes, and the second terms in the above equations represent first-order corrections beyond the RWA (see Appendix~\ref{app:beam_splitter}).

In particular, the correction to $g_\mathrm{BS}$ \added{attributed to two-mode squeezing processes} explains the abrupt decrease of the beam-spitter interaction for stronger drives observed in Ref.~\cite{ChapmanStijn2023} that is due to the higher-order resonance at $\Tilde{\delta}=\delta+\Delta_a=0$, where $\delta=\omega_a^\prime+\omega_b^\prime-n\omega_d=\omega_a^\prime+\omega_c^\prime-(n+1)\omega_d$ with $n=2$ \added{and $\omega_a^\prime$ being the dressed frequency of the coupler mode}. The resonant condition strongly depends on the flux bias in the circuit and the detuning $\Delta_a = C_{10,00,00,0}+\Delta_a^{(1)}$ encapsulating Lamb and Stark shifts of the coupler mode (see Fig.~\ref{fig:bs_downturn}).

Figures \hyperref[fig:on_off_ratio_and_bs_int]{\ref{fig:on_off_ratio_and_bs_int}(a)}, \hyperref[fig:on_off_ratio_and_bs_int]{\ref{fig:on_off_ratio_and_bs_int}(b)}  benchmark different circuit designs in terms of the achievable on-off-ratio for the beam-splitter interaction. Unlike Kerr-cat optimization, we see rather diverse results for the variation of circuit parameters which reflects the larger number of limitations we chose to impose on the optimization routine (see Appendix~\ref{app:applications} in order to compare it with the circuit scheme employed in Ref.~\cite{ChapmanStijn2023}). Therefore, there are optimal windows in the configuration parameters, rather than a smooth dependence on specific parameters of the circuit topology for SNAIL-based schemes. For SQUID-based schemes, we again  observe the advantage of more symmetric SQUIDs. Additionally, for $\alpha_\mathrm{s}\approx 1$ the on-off ratio \deleted{does not} \added{only weakly depends} on the number of SQUIDs in the array, \added{since the dominant contributions to both the beam-splitter and cavity-cavity cross-Kerr interactions, $C_{00,01,01,1}$ and $C_{00,10,10,0}$, approximately scale as $1/M^3$ (see Eqs.~\eqref{eq_app:supercoeff_closed_form_3DOF} and \eqref{eq_app:A_p})}.
Generally, we conclude that for $\chi_{bc}\geq\SI{30}{\hertz}$, $\Pi=1$ there are circuit configurations that generate on-off ratios almost an order of magnitude larger than in the original scheme \cite{ChapmanStijn2023}. \added{However, to be practically useful, a high on-off ratio should be accompanied by sufficiently large values of $g_\mathrm{BS}$. From Figs.~\hyperref[fig:on_off_ratio_and_bs_int]{\ref{fig:on_off_ratio_and_bs_int}(c)} and~\hyperref[fig:on_off_ratio_and_bs_int]{\ref{fig:on_off_ratio_and_bs_int}(d)}, we observe that, at least in case of non-dissipative dynamics, an array of SQUIDs appears to be best suited for implementing a tunable beam-splitter interaction, simultaneously offering a high on-off ratio and a large interaction strength.
}

\section{Summary}

\added{We have developed a comprehensive analytical framework for evaluating and optimizing parametric processes in driven  Josephson circuits. By introducing the concept of supercoefficients—exact amplitudes that encode all relevant aspects of circuit topology, nonlinearity, and external driving—we provide a powerful and generalizable tool for the analysis of parametric interactions. This framework is based on a systematic normal-ordered expansion of the effective Hamiltonian and is applicable across a wide range of circuit architectures, including those with strong drives and complex configurations.}

Although the applications considered above provide only a glimpse of the versatility of the SC approach, 
it is important to emphasize explicitly the universality of the proposed approach for study of arbitrary parametric processes within single DOF circuits including multiple drives, multiple-loop circuits, higher-order harmonic models of JJs, etc., and multiple DOF circuits with a single nonlinear mode weakly coupled to the linear modes.  

We anticipate that the normal-ordered Hamiltonian expansion with the supercoefficients given in Eqs.~(\ref{eq:ham_supercoeff}), 
(\ref{eq:supercoefficiennt_as_sum}) could become a standard tool replacing the usual expansion in Eq.~(\ref{eq:ham_gn}) involving individual $g_n$'s since the former enables a higher level of abstraction regardless of the circuit design and the drive strength, effectively gathering all relevant expansion terms required to calculate the amplitude of various parametric processes. This is especially pertinent to the circuits with  strong drive or slowly diminishing higher-order nonlinear coefficients,  $g_n$, and significantly simplifies the study of system dynamics beyond the RWA. The flexibility and simplicity of the SC approach could be used for simulation of the optimal circuit designs for a particular task. \added{This can aid in selecting optimal circuit architectures prior to fabrication and help mitigate unwanted effects such as irregular dynamics, unintended multiphoton resonances, among others.} Furthermore, the proposed approach could be a stepping stone towards an even more general approach overcoming current limitations, \textit{i.e.}, the use of a single DOF approximation and the restrictions  on the geometric inductance  and the symmetry of the circuits needed to obtain closed-form supercoefficients.

\section{Acknowledgments}  We thank Alessandro Miano, Rodrigo Cortiñas and Jayameenakshi Venkatraman for their insightful discussions, and Stijn de Graaf and Yaxing Zhang for their valuable insights on the beam-splitter interaction.\ RB thanks Nataliia Okhremchuk and Yurii  Kozlov for their input during the early stages. This research was sponsored by the Army Research Office (ARO), and was accomplished under Grant Number W911NF-23-1-0051. The views and conclusions contained in this document are those of the authors and should not be interpreted as representing the official policies, either expressed or implied, of the Army Research Office (ARO), or the U.S. Government. 
The U.S. Government is authorized to reproduce and distribute reprints for Government purposes notwithstanding any copyright notation herein.

External Interests Disclosure: SMG is a consultant for, and equity holder in, Quantum Circuits, Inc.
\appendix

\section{Normal-ordered Hamiltonian for a single-degree-of-freedom circuit}
\label{app:eff_ham_derivaiton}

In this section, we derive a normal-ordered expansion of the Hamiltonian for an arbitrary potential with a single degree of freedom. We begin by considering a capacitively driven circuit and then demonstrate how the result can be generalized to flux-driven circuits and to scenarios involving multiple drives of different types (\textit{i.e.}, charge and flux drives).

\subsection{Capacitively driven circuits}

We begin with the Hamiltonian for a capacitively driven circuit with charge offset $n_g=0$, where the potential $U$ is an arbitrary smooth function
\begin{equation}
    \hat H = 4E_C\hat{n}^2+U(\hat{\varphi})+\hat{n}\Omega_d\cos{(\omega_d t+\theta)}.
\end{equation}
Utilizing the bosonic basis for the charge and phase operators, ${\hat{{n}}=-i{n}_\mathrm{zpf}{(\hat{a}-\hat{a}^\dag)}}$, ${\hat{{\varphi}}={\varphi}_\mathrm{zpf}{(\hat{a}+\hat{a}^\dag)}+\varphi_0}$, we rewrite the Hamiltonian as
\begin{align}   
\begin{split}
\hat{H}&=\omega_0\hat{a}^\dag\hat{a}
\\&{+}{{U}\left({{\varphi}_\mathrm{zpf}}\left(\hat{a}+\hat{a}^\dag\right)+\varphi_0\right)}{-}\frac{E_Jc_2\left({{\varphi}_\mathrm{zpf}}\left(\hat{a}+\hat{a}^\dag\right)\right)^2}{2}\\&\qquad{-}i{{n}_\mathrm{zpf}{\Omega}_{d}}(\hat{a}-\hat{a}^\dag)\cos{(\omega_d t+\theta)}.
\end{split}
\end{align}
The quadratic term of the potential is explicitly subtracted to avoid double counting. Additionally, we have defined dimensionless nonlinear coefficients $c_n = {\partial_{\varphi}^2{U}(\varphi)}|_{\varphi_0}/E_J$, where $\varphi_0$ is a minimum of the function $U(\varphi)$. Next, we account for the drive term via a linear displacement by applying the unitary transformation 
\begin{align}\label{eq:disp_trans}
\hat{U}_{D}=\exp{\left(\beta(t)\hat{a}^\dag-\beta^{*}(t)\hat{a}\right)}
\end{align} 
with
\begin{align}
\beta(t)= \frac{i{\Omega}_d{n}_\mathrm{zpf}}{2}\frac{e^{-i(\omega_d t+\theta)}}{({\omega}_d-\omega_0)}-\frac{i{\Omega}_d{n}_\mathrm{zpf}}{2}\frac{e^{i(\omega_d t+\theta)}}{({\omega}_d+\omega_0)},
\end{align}
to the Hamiltonian, $\hat{U}_{D}^\dag\hat H\hat{U}_{D}-i\hat{U}_{D}^\dag\dot{\hat{U}}_{D}\rightarrow\hat H$, which results in 
\begin{align}\label{eq_app:ham_afer_disps} 
\begin{split}
\hat{{H}}&=\omega_0\hat{a}^\dag\hat{a}
{+}{{U}\left({{\varphi}_\mathrm{zpf}}\left(\hat{a}{+}\hat{a}^\dag\right){+}\Tilde{\Pi}\cos{(\omega_d{t}{+}\gamma)}{+}\varphi_0\right)}\\&-\frac{E_Jc_2}{2}{\left({{\varphi}_\mathrm{zpf}}\left(\hat{a}{+}\hat{a}^\dag\right)+\Tilde{\Pi}\cos{(\omega_d{t}{+}\gamma)}\right)^2}.
\end{split}
\end{align}
We have introduced notation for the
the effective drive amplitude $\Tilde{\Pi} = \Omega_d{\omega}_d/(\omega_d^2-\omega_0^2)$ as well as the phase of the effective drive $\gamma=\theta{-}{\pi}/2$. Further, instead of the usual expansion in orders of ${\varphi}_\mathrm{zpf}({\hat{a}+\hat{a}^\dag})+\Tilde{\Pi}\cos{(\omega_d{t}+\gamma)}$, we focus on deriving a normal-ordered expansion of 
$U$ for bosonic operators. This approach enables us to distinguish individual parametric processes with their corresponding amplitudes, incorporating contributions from all orders of the nonlinearity. Formally, we can write down the potential as an expansion around $\varphi_0$ as\footnote{We use notations ${\partial_\varphi}$ and $\frac{\partial}{\partial\varphi}$ interchangeably.}
\begin{widetext}
\begin{align}
\begin{split}
{U}({\varphi}_\mathrm{zpf}({\hat{a}+\hat{a}^\dag})+\Tilde{\Pi}\cos{(\omega_d{t}+\gamma)}+\varphi_0) &=
\sum\limits_{n=0}\frac{1}{n!}\left.\frac{\partial^n {U}(\varphi)}{\partial \varphi^n}\right|_{\varphi=\varphi_0}\left({\varphi}_\mathrm{zpf}({\hat{a}+\hat{a}^\dag})+\Tilde{\Pi}\cos{(\omega_d{t}+\gamma)}\right)^n \\&=e^{\left({\varphi}_\mathrm{zpf}({\hat{a}+\hat{a}^\dag})+\Tilde{\Pi}\cos{(\omega_d{t}+\gamma)}\right)\frac{\partial}{\partial \varphi}}\left.{U}(\varphi)\right|_{\varphi=\varphi_0}.
\end{split}
\end{align}
\end{widetext}
where we have made one additional step reformulating the expansion as an action of the translation operator $(\hat{T}(x)f(r))|_{r=x} = f(r+x)$, $\hat{T}(x) = \exp(x{\partial_r})$ (a similar approach was applied to multiple DOF circuit in Ref.~\cite{weiss2021}). In such an expression, the substitution $\varphi=\varphi_0$ is applied after differentiation is performed. It is also important to emphasize that the bosonic operators $\hat a^\dag$, $\hat a$ commute with the derivative ${\partial_\varphi}$, since the latter acts only on the classical function $U(\varphi)$.
We then utilize the normal-ordered expansion \cite{wilcox1967,marcos2013} 
\begin{widetext}
\begin{align}\label{eq:exp_op}
\begin{split}
&\exp\left(\alpha\left({\hat{a}+\hat{a}^\dag}\right)\frac{\partial}{\partial\varphi}\right) =\sum\limits_{\substack{n=0\\ m=0}}^{\infty}\frac{e^{\frac{\alpha^2}{2}\frac{\partial^2}{\partial \varphi^2}}\left(\alpha\frac{\partial}{\partial \varphi}\right)^{n+m}}{n!m!}\hat{a}^{\dag n}\hat{a}^m\\
&=\sum\limits_{n=0}\frac{e^{\frac{\alpha^2}{2}\frac{\partial^2}{\partial \varphi^2}}\left({\alpha}\frac{\partial}{\partial \varphi}\right)^{2n}}{(n!)^2}\hat{a}^{\dag n}\hat{a}^{n}
+\sum\limits_{\substack{n=0\\ l\neq 0}}\frac{e^{\frac{\alpha^2}{2}\frac{\partial^2}{\partial \varphi^2}}\left({\alpha}\frac{\partial}{\partial \varphi}\right)^{2n+l}}{n!(n+l)!}\left(\hat{a}^{\dag n}\hat{a}^{n+l}+\hat{a}^{\dag n+l}\hat{a}^{n}\right),
\end{split}
\end{align}
\end{widetext}
where the exponent under the sum originates from the non-commutation of $\hat{a}^\dag$ and $\hat{a}$. Using the Jacobi-Anger expansion
\begin{align}
\exp(a\cos{\psi})=I_0(a)+\sum\limits_{n=1}^{\infty}I_{n}(a)(e^{in\psi}+e^{-in\psi}),
\end{align}
where $I_n(x)$ is a modified Bessel function of the first kind,
we can write the Hamiltonian as
\begin{align}\label{eq:ham_supercoeff_with_quad} 
\begin{split}
\hat{{H}}&=\omega_0\hat{a}^\dag\hat{a}{-}\frac{E_Jc_2}{2}{\left({{\varphi}_\mathrm{zpf}}\left(\hat{a}{+}\hat{a}^\dag\right){+}\Tilde{\Pi}\cos{(\omega_d{t}{+}\gamma)}\right)^2}\\&{+}\sum\limits_{n,l,p}^{\{l,p\}^\prime}\Tilde{C}_{nl,p}(\hat{a}^{\dag n}\hat{a}^{n{+}l}{+}\hat{a}^{\dag n{+}l}\hat{a}^{n})\\&\qquad\qquad\times(e^{ip(\omega_dt{+}\gamma)}{+}e^{-ip(\omega_dt{+}\gamma)}).
\end{split}
\end{align}
The prime superscription of the sum means that for terms with $l=0$ and $p=0$, an additional $1/2$ multiplier should be applied. We have defined the coefficients 
\begin{equation}\label{eq:supercoeff_diff_form}
\Tilde{C}_{nl,p}{=}\frac{e^{\frac{{\varphi}_\mathrm{zpf}^2}{2}\frac{\partial^2}{\partial \varphi^2}}({{\varphi}_\mathrm{zpf}}\frac{\partial}{\partial \varphi})^{{2n{+}l}}I_{p}(\Tilde{\Pi}\frac{\partial}{\partial \varphi})}{n!(n{+}l)!}\left.{U}(\varphi)\right|_{\varphi{=}\varphi_0},
\end{equation}
collecting the quadratic term into the sum.
The final compact normal-ordered Hamiltonian for an arbitrary potential is
\begin{align}\label{eq:ham_supercoeff_without_quad}  
\begin{split}
\hat{{H}}=\omega_0\hat{a}^\dag\hat{a}&+\sum\limits_{n,l,p}^{\{l,p\}^\prime}{C}_{nl,p}(\hat{a}^{\dag n}\hat{a}^{n+l}+\hat{a}^{\dag n+l}\hat{a}^{n})\\&\qquad\times(e^{ip(\omega_dt+\gamma)}+e^{-ip(\omega_dt+\gamma)}),
\end{split}
\end{align}
where we utilized the expansion for the Bessel and exponential functions in definition (\ref{eq:supercoeff_diff_form}) and introduced the supercoefficients (SC)
\begin{align}\label{eq_app:supercoeff_sum_form}
\begin{split}
C_{nl,p}&=\sum\limits_{k,m=0}^{\mathrm{S}\geq3}
\frac{c_{2n{+}l{+}2k{+}p{+}2m}E_J{{\varphi}_\mathrm{zpf}^{2n{+}l}}}{m!k!(k{+}p)!n!(n{+}l)!}
\\&\qquad\qquad\times\left(\frac{{\varphi}_\mathrm{zpf}^2}2\right)^m{}\left(\frac{{\Tilde{\Pi}}}2\right)^{2k{+}p}.
\end{split}
\end{align}
\added{where $\mathrm{S}=2n+l+p+2m+2k$.}

Notice that the linear term $(\hat a+\hat a^\dag)$ has a non-zero supercoefficient $C_{01,0}$ consisting of odd-ordered nonlinear coefficients starting with $c_3$. This term is a small linear displacement of the phase, and arises from the non-commutativity of $\hat a^\dag$ and $\hat a$. In principle, such a displacement could be accounted for via redefinition of the minimum condition,
\begin{align}
\partial_\varphi U(\varphi)+\sum\limits_{m=0}^{2m+1\geq3}
\frac{{\varphi}_\mathrm{zpf}^{2m}{\partial_{\varphi}^{2m+1}{U}(\varphi)}}{2^m\,m!}
=0,    
\end{align}
where the sum accounts for additional displacement due to the linear term with the supercoefficient $C_{01,0}$.
However, it results only in a minor correction to $\varphi_0$ due to the smallness of $\varphi_\mathrm{zpf}$. We therefore keep the initial minimum condition and treat the linear term explicitly. 

\subsection{Flux-driven circuits}

Let us separately consider the case of the flux-driven circuit to address its specifics. For flux-driven circuits, the Hamiltonian has the  general form
\begin{equation}
\hat H_\mathrm{f} = 4E_C\hat{n}^2+U(\hat{\varphi},\varphi_{e}(t)).
\end{equation}
For circuits with time-dependent external flux, the drive itself contains information about the circuit topology \cite{You2019,Riwar2022}. In that regard, we have to consider the more explicit form of the potential and try to preserve the generality of the approach simultaneously. For an arbitrary single DOF approximation of a Josephson circuit composed of individual single-loop nonlinear
elements (\textit{e.g.}, a SQUID \cite{Iyama2024}, see next Section), the potential can be decomposed as
\begin{align}
\begin{split}
&U(\hat{\varphi},\varphi_{e}(t))=U^{(a)}(\hat \varphi,\varphi_{e}(t))+U^{(b)}( \hat\varphi,\varphi_{e}(t))\\&=A\cos{(a_1\hat\varphi{+}a_2\varphi_{e}(t))}{+}B\cos{(b_1\hat\varphi{+}b_2\varphi_{e}(t))},
\end{split}
\end{align}
where $A$, $B$, and $a_i$, $b_i$ ($i=1,2$) are the coefficients reflecting the topology of the particular circuit design. In particular, the values of $a_2$ and $b_2$ are defined by the irrotational condition in the case of flux drive \cite{You2019,Riwar2022} and constrained by $b_2-a_2=1$. For a flux-driven circuit, we distinguish the static and fluctuating part of the external flux $\varphi_e(t)=\varphi_\mathrm{dc}+\varphi_\mathrm{ac}(t)$, where we specialize to the case of a harmonic flux drive $\varphi_\mathrm{ac}(t)=2\varphi_\mathrm{ac0}\cos{(\omega_dt+\gamma)}$ (the drive phase $\gamma$ is defined by modulation properties of the flux oscillation). We rewrite the Hamiltonian by explicitly extracting the flux drive in the form of linear displacement of the phase
\begin{align}
\begin{split}
\hat H_\mathrm{f} &= 4E_C\hat{n}^2+U^{(a)}(\hat{\varphi}\\&\quad+\frac{a_2}{a_1}\varphi_\mathrm{ac}(t))+U^{(b)}(\hat{\varphi}+\frac{b_2}{b_1}\varphi_\mathrm{ac}(t)),
\end{split}
\end{align}
where $U^{(i)}(\hat{\varphi}+\frac{i_2}{i_1}\varphi_\mathrm{ac}(t))\equiv U^{(i)}(\hat{\varphi}+\frac{i_2}{i_1}\varphi_\mathrm{ac}(t),\varphi_\mathrm{dc})$, $i=a\,,b$. This Hamiltonian resembles the one obtained for a capacitive drive (see Eq.~(\ref{eq_app:ham_afer_disps})).
We again explicitly extract the quadratic terms 
\begin{align}
\begin{split}
&\hat H_\mathrm{f} = 4E_C\hat{n}^2\\&+\frac{E_Jc_{2}^{(a)}}{2}{\left(\hat{\varphi}{+}\frac{a_2}{a_1}\varphi_\mathrm{ac}(t)\right)^2}
{+}\frac{E_Jc_{2}^{(b)}}{2}\left(\hat{\varphi}{+}\frac{b_2}{b_1}\varphi_\mathrm{ac}(t)\right)^2\\
&\quad+U_{nl}^{(a)}(\hat{\varphi}+\frac{a_2}{a_1}\varphi_\mathrm{ac}(t))+U_{nl}^{(b)}(\hat{\varphi}+\frac{b_2}{b_1}\varphi_\mathrm{ac}(t)),
\end{split}
\end{align}
where $U_{nl}^{(i)}(\varphi)=U^{(i)}(\varphi)-\partial_\varphi^2U^{(i)}(\varphi)|_{\varphi_0}\varphi^2/2$, $i=a\,,b$, comprises the nonlinear parts of the potential, $c_{n}^{(i)}={\partial_{\varphi}^n{U}_0^{(i)}(\varphi)}|_{\varphi_0}/E_J$ and $\varphi_0$ is a minimum of the undriven potential derived from condition ${\partial_{\varphi}{U}_0(\varphi)}|_{\varphi_0}=0$, with ${U}_0(\varphi)\equiv{U}(\varphi,\varphi_\mathrm{dc})$. Since $c_2 = c_{2}^{(a)}+c_{2}^{(b)}$ the definition of $\varphi_\mathrm{zpf} = \sqrt[\leftroot{-2}\uproot{2}4]{{2E_C}/{E_Jc_2}}$ remains the same. Using the harmonic oscillator basis, we rewrite the Hamiltonian as
\begin{align}\label{eq:ham_flux_dr_add_term}
\begin{split}
\hat H_\mathrm{f} &= \omega_0\hat{a}^\dag\hat{a}\\&+\frac{E_J{{\varphi}_\mathrm{zpf}}\left(\hat{a}+\hat{a}^\dag\right)}{2}\left(c_{2}^{(a)}\frac{a_2}{a_1}+c_{2}^{(b)}\frac{b_2}{b_1}\right)\varphi_\mathrm{ac}(t)
\\&\qquad+U_{nl}^{(a)}(\varphi_\mathrm{zpf}\left(\hat{a}+\hat{a}^\dag\right)+\frac{a_2}{a_1}\varphi_\mathrm{ac}(t))\\&\qquad+U_{nl}^{(b)}(\varphi_\mathrm{zpf}\left(\hat{a}+\hat{a}^\dag\right)+\frac{b_2}{b_1}\varphi_\mathrm{ac}(t)),
\end{split}
\end{align}
where we omit the constant term. Compared to the case of the charge-driven circuit, we have the second term in Eq.~\eqref{eq:ham_flux_dr_add_term}, an additional drive term that contributes to the linear response of the system to the drive. To eliminate this drive, we perform a displacement transformation similar to Eq.~(\ref{eq:disp_trans}) with
\begin{align}
\beta(t)= \frac{-{\Omega}_d^{lin}e^{-i({\omega}_d{t}+\gamma)}}{2({\omega}_d-\omega_0)}+\frac{{\Omega}_d^{lin}e^{i({\omega}_d{t}+\gamma)}}{2({\omega}_d+\omega_0)}
\end{align}
and
\begin{align}
\Omega_d^{lin} = {E_J{{\varphi}_\mathrm{zpf}}}\left(c_{2}^{(a)}\frac{a_2}{a_1}+c_{2}^{(b)}\frac{b_2}{b_1}\right)\varphi_\mathrm{ac0}, 
\end{align}
to account for these additional drive terms. The displaced-frame Hamiltonian then given by
\begin{align}\label{eq:ham_flux_drive}
\begin{split}
\hat H_\mathrm{f} &= \omega_0\hat{a}^\dag\hat{a}
\\&+U_{nl}^{(a)}(\varphi_\mathrm{zpf}\left(\hat{a}+\hat{a}^\dag\right)+\Pi_a\cos{(\omega_dt+\gamma)})\\&+U_{nl}^{(b)}(\varphi_\mathrm{zpf}\left(\hat{a}+\hat{a}^\dag\right)+\Pi_b\cos{(\omega_dt+\gamma)}),
\end{split}
\end{align}
with the effective drive amplitudes defined as
\begin{widetext}
\begin{align}
&\Pi_a {=} 2\varphi_\mathrm{ac0}\left(\frac{a_2}{a_1}\left(1{-}\frac{E_Jc_{2}^{(a)}\varphi_\mathrm{zpf}^2\omega_0}{2(\omega_d^2-\omega_0^2)}\right){-}\frac{b_2}{b_1}\frac{E_Jc_{2}^{(b)}\varphi_\mathrm{zpf}^2\omega_0}{2(\omega_d^2-\omega_0^2)}\right),\\
&\Pi_b {=}2\varphi_\mathrm{ac0}\left(\frac{b_2}{b_1}\left(1{-}\frac{E_Jc_{2}^{(b)}\varphi_\mathrm{zpf}^2\omega_0}{2(\omega_d^2-\omega_0^2)}\right){-}\frac{a_2}{a_1}\frac{E_Jc_{2}^{(a)}\varphi_\mathrm{zpf}^2\omega_0}{2(\omega_d^2-\omega_0^2)}\right).
\end{align}
\end{widetext}
It might be useful to estimate the contributions from additional drive term in $\Pi_a$ and $\Pi_b$.  By using the definition $\omega_0=2E_Jc_2\varphi_\mathrm{zpf}^2$, we conclude that 
\begin{equation}
\left|\frac{E_Jc_{2}^{(i)}\varphi_\mathrm{zpf}^2\omega_0}{2(\omega_d^2-\omega_0^2)}\right|\leq\left|\frac{1}{4((\omega_d/\omega_0)^2-1)}\right|,\quad i=a,\,b.
\end{equation}
Therefore, the maximum magnitude of additional contribution is fully defined by the ratio $\omega_d/\omega_0$. For example, for two-photon squeezing drives, $\omega_d=2\omega_0$, the correction is not larger than $1/12$. However, for $\omega_d\simeq\omega_0$ it could become a dominant input to the effective amplitude of the flux drive.
Next, from Eq.~(\ref{eq:ham_flux_drive}), we can write the normal-ordered form of Hamiltonian  (\ref{eq:ham_supercoeff_without_quad}) with supercoefficients consisting of two terms
\begin{equation}
C_{nl,p}=C^{(a)}_{nl,p}+C^{(b)}_{nl,p},
\end{equation}
where each term is defined by
\begin{align}
\begin{split}
C^{(i)}_{nl,p}&=\sum\limits^{\mathrm{S}\geq3}_{k,m}
\frac{c_{2n{+}l{+}2k{+}p{+}2m}^{(i)}E_J{{\varphi}_\mathrm{zpf}^{2n{+}l}}}{m!k!(k{+}p)!n!(n{+}l)!}\\&\qquad\qquad\times
\left(\frac{{\varphi}_\mathrm{zpf}^2}2\right)^m{}\left(\frac{{{\Pi_i}}}2\right)^{2k{+}p}.\label{eq:super_coef_flux_drive}
\end{split}
\end{align}

\paragraph*{Alternative gauge for a flux-driven circuit}It was shown in \cite{You2019, Riwar2022} that the irrotational condition allows
to avoid the inconsistency in the description of a flux-driven circuit related to the treatment of time-dependent
fluxoid quantization. However, for the sake of completeness, we consider an alternative gauge \cite{You2019,Riwar2022,Lu2023} where we account for time-dependent external flux via additional capacitive terms in the Hamiltonian (see Eq.~(14) in \cite{You2019}).  In this case, the system potential can be written in a more symmetric form as
\begin{align}
\begin{split}
U&(\hat{\varphi},\varphi_{e}(t))=\\&A\cos{(a_1\hat\varphi{-}\frac{1}{2}\varphi_{e}(t))}+B\cos{(b_1\hat\varphi{+}\frac{1}{2}\varphi_{e}(t))},
\end{split}
\end{align}
and the Hamiltonian becomes
\begin{align}
\begin{split}
\hat H_\mathrm{f} &= \omega_0\hat{a}^\dag\hat{a}-\frac{ien_\mathrm{zpf}(a_2+b_2)}{2}(\hat a-\hat a^\dag)\Dot{\varphi}_{ac}(t)
\\&+U_{nl}^{(a)}(\hat{\varphi}-\frac{1}{2a_1}\varphi_\mathrm{ac}(t))+U_{nl}^{(b)}(\hat{\varphi}+\frac{1}{2b_1}\varphi_\mathrm{ac}(t)),
\end{split}
\end{align}
where the capacitive drive is defined by the time derivative of the external flux oscillations.
Again, we perform the displacement transformation to account for a capacitive drive in the nonlinear part of the potential, so the displaced-frame Hamiltonian and the effective drive amplitudes are defined by 
\begin{align}
\begin{split}
\hat H_\mathrm{f} &= \omega_0\hat{a}^\dag\hat{a}
+U_{nl}^{(a)}(\varphi_\mathrm{zpf}\left(\hat{a}+\hat{a}^\dag\right)+\Pi_a(t))\\&\qquad\qquad+U_{nl}^{(b)}(\varphi_\mathrm{zpf}\left(\hat{a}+\hat{a}^\dag\right)+\Pi_b(t))),
\end{split}\\
\begin{split}
\Pi_a(t) &= -\frac{\varphi_\mathrm{ac0}}{a_1}\cos{(\omega_dt+\gamma)}\\&\qquad\quad+\frac{e\varphi_\mathrm{ac0}(a_2+b_2)\omega_d^2}{(\omega_d^2-\omega_0^2)}\sin{(\omega_dt+\gamma)},
\end{split}\\
\begin{split}
\Pi_b(t) &= \frac{\varphi_\mathrm{ac0}}{b_1}\cos{(\omega_dt+\gamma)}\\&\qquad\quad+\frac{e\varphi_\mathrm{ac0}(a_2+b_2)\omega_d^2}{(\omega_d^2-\omega_0^2)}\sin{(\omega_dt+\gamma)},
\end{split}
\end{align}
correspondingly. We also used $n_\mathrm{zpf}\varphi_\mathrm{zpf}=1/2$. At this point, we should keep in mind that we have a different potential, variable space, and, consequently, different $\varphi_\mathrm{zpf}$, $\omega_0$ and other gauge-dependent parameters than in the case where the irrotational condition was applied. Hence, it is not trivial to show analytically the equivalence between additional drive terms in both cases. However, we can spot some key properties for both results. For highly symmetric circuits (capacitances and inductive energies of all Josephson junctions in the circuit are equal) with $A=B$, $a_2=-1/2$, $b_2=1/2$, $a_1=b_1$, $U_a(x)=U_b(x)$, additional contributions to the displacement
vanish. Since we have different signs for displacements, $\Pi_a=-\Pi_b$, $U_a+U_b$ contains only even-order terms for bosonic operators yielding a parity-protected Hamiltonian \cite{Lu2023}.

\subsection{Multiple drives}
\label{app:multiple_drives}

Since in the proposed derivation of the effective Hamiltonian, the effect of the drive is accounted for via linear displacement of the mode, it is easy to generalize the result for the case of multiple drives on the system. If there are $K$ capacitive  drives and a single flux drive we obtain the normal-ordered expansion
\begin{widetext}
\begin{equation}\label{eq_app:multi_dr}
\hat{{H}}=\omega_0\hat{a}^\dag\hat{a}+\sum\limits_{n,l,\{p\}}^{\{l,\{p\}\}^\prime}C_{nl,\{p\}}(\hat{a}^{\dag n}\hat{a}^{n+l}+\hat{a}^{\dag n+l}\hat{a}^{n})\prod\limits_{i}^{K+1}(e^{ip_i(\omega_{di}t+\gamma_i)}+e^{-ip_i(\omega_{di}t+\gamma_i)}),
\end{equation}
where we have a set of indices $\{p\}$, drive frequencies $\{\omega_d\}$ and phases $\{\gamma\}$ with corresponding SCs defined as
\begin{align}
C_{nl,\{p\}}&=C^{(a)}_{nl,\{p\}}+C^{(b)}_{nl,\{p\}},\\
C^{(j)}_{nl,\{p\}}&=\sum\limits^{\mathrm{S}\geq3}_{\{k\},\,m}
\frac{c_{2n{+}l{+}\sum_i{(2k_i{+}p_i)}{+}2m}^{(j)}E_J{{\varphi}_\mathrm{zpf}^{2n{+}l}}}{m!n!(n{+}l)!\prod_i(k_i!(k_i{+}p_i)!)}
\left(\frac{{\varphi}_\mathrm{zpf}^2}2\right)^m{}\left(\frac{{{\Pi_j}}}2\right)^{2k_0{+}p_0}\prod\limits_{i=1}^{K}\left(\frac{{{\Tilde{\Pi}_i}}}2\right)^{2k_i{+}p_i},\quad j=a,b.
\end{align}
\end{widetext}
We redefined $\mathrm{S}={2n{+}l{+}\sum_{i}{(2k_i{+}p_i)}{+}2m}$ and reserved indices $k_0$ and $p_0$ for the flux drive terms. The effective drive amplitudes $\{\Tilde{\Pi}\}$ are defined by corresponding parameters of classical capacitive drives ($\{{\Omega}\}$, $\{\omega_d\}$ and  $\{\gamma\}$). As we show in the next Section, in a closed-form expression for the supercoefficients, multiple drives are accounted for via a product of  Bessel functions $\prod_iJ_{p_i}(x_i)$ with corresponding arguments proportional to the effective drive amplitudes $\{\Tilde{\Pi}\}$. Although the generalization to an arbitrary number of flux drives is also possible, for the sake of brevity, we consider only a single flux drive.

\section{Supercoefficients for different circuit QED schemes}\label{app:diff_potentials}

In this Section, we derive the supercoefficients for various superconducting circuit topologies (refer to Table \ref{table:cEQD_schemes} in the main text), enabling the study of arbitrary parametric processes. In the main text, we employ these SCs to estimate the dynamical parameters of a squeezed Kerr parametric oscillator and a tunable beam-splitter interaction between two cavities.

\subsection{General potential for single degree of freedom approximation for an arbitrary symmetric Josephson circuit}
\label{app:general_pot_sym_circuits}

\begin{figure}[!t]
    \centering    \includegraphics{ 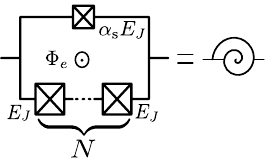}
    \caption{Superconducting nonlinear asymmetric
inductive elements (SNAIL) with $N$ large Josephson junctions.}
    \label{fig:SNAIL}
\end{figure}

In this subsection, we outline the derivation of the supercoefficients for a general potential of symmetric Josephson circuits with no stray inductors which are composed of identical single-loop nonlinear elements (SNAIL, SQUIDs, etc.) in the single DOF approximation (for example, see circuit designs in Table I of the main text). Our definition of symmetric circuits coincides with the one in Ref.~\cite{ferguson2013} for a fluxonium and in Ref.~\cite{fratini18} for an array of SNAILs. That is, we consider circuits that obey a permutation symmetry for a series array of Josephson junctions in a particular nonlinear element  (\textit{e.g.}, all $N$ array JJs in a SNAIL are identical, see Fig.~\ref{fig:SNAIL}) and for a series array of nonlinear elements (\textit{e.g.}, all $M$ SQUIDs, SNAILs, etc. are identical, see Table I in the main text). Such a symmetry leads to an approximation that assumes an equal phase drop for each circuit element. In this case, we can write the potential as
\begin{equation}\label{eq_app:gen_pot}
U(\varphi)=A\cos{(a_1\varphi+a_2\varphi_{e})}+B\cos{(b_1\varphi+b_2\varphi_{e})},
\end{equation}
where all coefficients are defined by the circuit topology, the capacitive and inductive energy of individual elements, and $\varphi_e$ is a normalized external flux \cite{ferguson2013,DiPaolo2021,sandro2023,Iyama2024, Rymarz23}. For example, for an array of $M$ SNAILs: $A=-\alpha_\mathrm{s}ME_J$, $B=-MNE_J$ $a_1=1/M$, $b_1=1/(MN)$ (see next subsection), $a_2$ and $b_2$ reflect the gauge choice for electromagnetic potential \cite{Riwar2022} with $b_2-a_2=1$. In the absence of stray inductors, this potential encompasses all the circuit designs discussed in the main text.  To write down SCs defined by Eq.~(\ref{eq_app:supercoeff_sum_form}), we obtain the corresponding nonlinear coefficients
\begin{widetext}
\begin{align}\label{eq_app:general_nl_coeff}
&c_{2n{+}l{+}2k{+}p{+}2m}=(-1)^{\left \lfloor\frac{2n{+}l{+}2k{+}p{+}2m}{2}\right \rfloor}\times\\&\left(Aa_1^{2n{+}l{+}2k{+}p{+}2m}\left\{\!\begin{aligned}
&-\cos{(a_1\varphi_0{+}a_2\varphi_{e})}\\[0.5ex]
&\sin{(a_1\varphi_0{+}a_2\varphi_{e})}\\[0.5ex]
\end{aligned}\right\}_{l+p}{+}Bb_1^{2n{+}l{+}2k{+}p{+}2m}\left\{\!\begin{aligned}
&-\cos{(b_1\varphi_0{+}b_2\varphi_{e})} \\[0.5ex]
&\sin{(b_1\varphi_0{+}b_2\varphi_{e})} \\[0.5ex]
\end{aligned}\right\}_{l+p}\right)\nonumber,
\end{align}
where 
\begin{equation}
\left\{\!\begin{aligned}
&a  \\[0.5ex]
&b \\[0.5ex]
\end{aligned}\right\}_{k}\equiv\left\{\!\begin{aligned}
&a & \text{if } & k &\text{is even}  \\[0.5ex]
&b & \text{if } & k &\text{is odd}\\[0.5ex]
\end{aligned}\right\}
\end{equation}
and  $\left \lfloor{x}\right \rfloor$ is the floor function meaning the greatest integer less than or equal to $x$. Next, using the property $\left \lfloor{x+n}\right \rfloor=\left \lfloor{x}\right \rfloor+n$ for $n\in\mathbb{N}$ and substituting (\ref{eq_app:general_nl_coeff}) into the definition of SC, we can rewrite the SC in a closed form
\begin{align}\label{eq_app:supercoefficients_joseph_circuit}
C_{nl,p}&=\frac{(-1)^{\left \lfloor\frac{2n+l+p}{2}\right \rfloor}{\varphi}_\mathrm{zpf}^{2n+l}}{n!(n+l)!} \Bigg(Aa_1^{2n+l}J_p(a_1\Tilde{\Pi})e^{{-}\frac{{\varphi}_\mathrm{zpf}^2a_1^2}{2}}\left\{\!\begin{aligned}
&{-}\cos{(a_1\varphi_0{+}a_2\varphi_{e})}  \\[0.5ex]
&\sin{(a_1\varphi_0{+}a_2\varphi_{e})} \\[0.5ex]
\end{aligned}\right\}_{l{+}p}\\&\quad{+}Bb_1^{2n+l}J_p(b_1\Tilde{\Pi})e^{{-}\frac{{\varphi}_\mathrm{zpf}^2b_1^2}{2}}\left\{\!\begin{aligned}
&{-}\cos{(b_1\varphi_0{+}b_2\varphi_{e})}  \\[0.5ex]
&\sin{(b_1\varphi_0{+}b_2\varphi_{e})} \\[0.5ex]
\end{aligned}\right\}_{l{+}p}\Bigg),\quad \text{for}\quad2n+l+p>2,\nonumber\\
C_{nl,p}&=\frac{(-1)^{\left \lfloor\frac{2n+l+p}{2}\right \rfloor}{\varphi}_\mathrm{zpf}^{2n+l}}{n!(n+l)!}
{ \Bigg(Aa_1^{2n+l}\left(J_p(a_1\Tilde{\Pi})e^{{-}\frac{{\varphi}_\mathrm{zpf}^2a_1^2}{2}}-\left(\frac{a_1\Tilde{\Pi}}{2}\right)^p\right)\left\{\!\begin{aligned}
&{-}\cos{(a_1\varphi_0{+}a_2\varphi_{e})}  \\[0.5ex]
&\sin{(a_1\varphi_0{+}a_2\varphi_{e})} \\[0.5ex]
\end{aligned}\right\}_{l{+}p}}\label{eq_app:supercoefficients_joseph_circuit2}\\&\quad+Bb_1^{2n+l}\left(J_p(b_1\Tilde{\Pi})e^{{-}\frac{{\varphi}_\mathrm{zpf}^2b_1^2}{2}}-\left(\frac{b_1\Tilde{\Pi}}{2}\right)^p\right)\left\{\!\begin{aligned}
&{-}\cos{(b_1\varphi_0{+}b_2\varphi_{e})}  \\[0.5ex]
&\sin{(b_1\varphi_0{+}b_2\varphi_{e})} \\[0.5ex]
\end{aligned}\right\}_{l{+}p}\Bigg),\quad \text{for}\quad2n+l+p\leq2.\nonumber
\end{align}
\end{widetext}
We have employed the expansion of the Bessel and exponent functions and accounted for the condition $\mathrm{S}\geq3$ in Eq.~\eqref{eq_app:supercoeff_sum_form} for the second expression excluding terms with $k=0$ and $m=0$. Hereafter, for the sake of brevity, we obtain supercoefficients for specific circuit designs only for $2n+l+p>2$ in the form of Eq.~\eqref{eq_app:supercoefficients_joseph_circuit}, keeping in mind that for SCs with $2n+l+p\leq2$, the substitution $J_p(x)e^{{-}y}\rightarrow J_p(x)e^{{-}y}-({x}/{2})^p$ should be applied, with corresponding $p$, $x$ and $y$, to reproduce Eq.~\eqref{eq_app:supercoefficients_joseph_circuit2}.

The above considerations facilitate the generalization of the results to nonlinear circuits without any loops (\textit{e.g.}, for a transmon consisting of a capacitively shunted JJ, with $B=0$, $b_1=0$, $a_1=1$, $\varphi_e=0$, and all terms with odd $l+p$ are zero), as well as to circuits composed of multi-loop nonlinear elements, which typically introduce additional cosine terms to the potential \cite{bhandari2024symmetricallythreadedsquidsgeneration, You2019}. Circuit without permutation symmetry lead to the single DOF potential defined via transcendental functions. This makes it impossible to derive a closed form for the SCs since there is no compact expression for the $c_n$'s \cite{fratini18,sandro2023}. In this case, the sum representation of SC defined in Eq.~(\ref{eq_app:supercoeff_sum_form}) should be used instead.

\subsection{Capacitively-driven array of SNAILs}

We now proceed to apply this theory for a specific circuit designs, starting with the simple example of the circuit scheme composed of a single SNAIL element (see Fig.~2 in the main text). For such a scheme, the potential reads as
\begin{equation}\label{eq:fluxonium_pot}U_N(\varphi)=-\alpha_\mathrm{s} E_J\cos{\varphi}-NE_J\cos\left(\frac{\varphi-\varphi_{e}}{N}\right),
\end{equation}
where $\alpha_\mathrm{s}$ is an asymmetry coefficient reflecting the  inductance ratio between the large Josephson junctions and the ``black sheep'' JJ, $E_J$ is the Josephson energy of the large JJs, $\varphi_{e}=2\pi\Phi_{e}/\Phi_0$, $\Phi_{e}$ is the external magnetic flux, $\Phi_0$ is the magnetic flux quantum, $N$ is the number of large JJs. For simplicity, throughout this study, we focus on circuits with $\alpha_\mathrm{s}<1/N$, which ensures the presence of a single potential minimum \cite{Frattini3wm}. However, formally, the approach can also be applied to a multi-well potential, provided the focus is on small oscillations of the phase around a specific minimum of the potential.

Using the result of the previous subsection, the SC can be written in a closed form as
\begin{widetext}
\begin{align}\label{supercoefficients_fluxonium}
&C_{nl,p}=\frac{(-1)^{\left \lfloor\frac{2n+l+p}{2}\right \rfloor}E_J{\varphi}_\mathrm{zpf}^{2n+l}}{n!(n+l)!}\times\nonumber\\
&\Bigg({\alpha_\mathrm{s}}\left\{\!\begin{aligned}
&-\cos{\varphi_0} \\[0.5ex]
&\sin{\varphi_0} \\[0.5ex]
\end{aligned}\right\}_{l+p}J_{p}(\Pi)e^{-\frac{{\varphi}_\mathrm{zpf}^2}{2}}{+}\left\{\!\begin{aligned}
&-\cos{\left((\varphi_0{-}\varphi_{e})/{N}\right)} \\[0.5ex]
&\sin{\left((\varphi_0{-}\varphi_{e})/{N}\right)} \\[0.5ex]
\end{aligned}\right\}_{l+p}\left(\frac{1}{N}\right)^{2n+l-1}J_{p}\left(\frac{\Pi}{N}\right)e^{-\frac{{\varphi}_\mathrm{zpf}^2}{2N^2}}\Bigg).
\end{align}
To generalize this result for the case of an array of SNAILs, we assume an equal phase drop over each SNAIL yielding the potential \cite{fratini18} 
\begin{align}\label{eq:array_fluxonium_pot_no_geom_ind}    U(\varphi)=-M\alpha_\mathrm{s} E_J\cos{\frac{\varphi}{M}}-MNE_J\cos\left(\frac{\varphi}{MN}-\frac{\varphi_{e}}{N}\right),
\end{align}
with the corresponding SCs  
\begin{align}\label{eq_app:sc_snail_arr}
&C_{nl,p}=\frac{(-1)^{\left \lfloor\frac{2n+l+p}{2}\right \rfloor}E_J{\varphi}_\mathrm{zpf}^{2n+l}}{n!(n+l)!M^{2n+l-1}}\times\nonumber\\
&\Bigg({\alpha_\mathrm{s}}\left\{\!\begin{aligned}
&-\cos{\frac{\varphi_0}{M}} \\[0.5ex]
&\sin{\frac{\varphi_0}{M}} & \\[0.5ex]
\end{aligned}\right\}_{l+p}J_{p}\left(\frac{\Tilde{\Pi}}{M}\right)e^{-\frac{{\varphi}_\mathrm{zpf}^2}{2M^2}}{+}\left\{\!\begin{aligned}
&-\cos{\left(\frac{\varphi_0}{MN}{-}\frac{\varphi_{e}}{N}\right)}  \\[0.5ex]
&\sin{\left(\frac{\varphi_0}{MN}{-}\frac{\varphi_{e}}{N}\right)} \\[0.5ex]
\end{aligned}\right\}_{l+p}\left(\frac{1}{N}\right)^{2n+l-1}J_{p}\left(\frac{\Tilde{\Pi}}{MN}\right)e^{-\frac{{\varphi}_\mathrm{zpf}^2}{2N^2M^2}}\Bigg).
\end{align}
\end{widetext}

\subsubsection*{Geometric inductance in circuit}

Since there are cases where the geometric (parasitic) inductance in a circuit (\textit{e.g.}, originating from nanowires) \cite{frattini2022squeezed,ChapmanStijn2023,Sivak2020} could be substantial, we separately analyze circuits composed of an array of SNAILs in the presence of a geometric inductance. As it was proposed for an array of SNAILs  in Ref.~\cite{fratini18,Frattini2021} and analyzed  in depth in Ref.~\cite{Rymarz23}  for circuits where the internal capacitance of the JJs and parasitic capacitance of the circuit are much smaller than shunting capacitance, the potential of the system could be reduced to a single DOF approximation
\begin{equation}
U(\varphi) = M U_{N}(\varphi_s[\varphi]) +\frac{1}{2}E_L(\varphi-M\varphi_s[\varphi])^2,
\end{equation}
where $U_{N}(\varphi)$ is the potential for a single SNAIL (see the previous subsection), $E_L$ is the energy parameter of the linear (geometric) inductance in the system. The transcendental function $\varphi_s[\varphi]$ is derived from 
\begin{equation}\label{eq:current conservation}
    \alpha \sin{\varphi_s} + \sin{\frac{\varphi_s- \varphi_{e}}{3}}+x_J (M\varphi_s-\varphi)=0,
\end{equation}
which enforces current conservation in the circuit.
The participation ratio $x_J = L_J/L$ is defined by the inductance of large Josephson junctions, $L_J\propto1/E_J$, in SNAIL and the geometric inductance $L$. 

Nonlinear coefficients for this potential ${c}_{n}={\partial_{\varphi}^n{U}(\varphi)}|_{\overline{\varphi}_{\min}}/E_J$,
where $\overline{\varphi}_{{\rm min}}=M\varphi_\mathrm{s,min}$ are determined from the condition ${\partial_{\varphi}{U}(\varphi)}=0$ and the equality $\varphi_s[\Bar{\varphi}_\mathrm{min}]=\varphi_\mathrm{s,min}$. The minimum value for an individual SNAIL potential $\varphi_\mathrm{s,min}$ is calculated from the condition ${\partial_\varphi U_N(\varphi)}=0$ (see details in Ref.~\cite{fratini18}). Using the current conservation condition \eqref{eq:current conservation}, we can formally obtain the nonlinear coefficients of an arbitrary order. For example, the first four coefficients read as
\begin{align}
{c}_{1}&=x_{J}(\overline{\varphi}_{\rm min}-M\varphi_{s}[\overline{\varphi}_{\rm min}]),\\ 
{c}_{2}&=x_{J}(1-M\frac{\partial\varphi_{s}}{\partial\varphi}[\overline{\varphi}_{\rm min}]),\\
{c}_{3}&=-Mx_{J}\frac{\partial^{2}\varphi_{s}}{\partial^{2}\varphi}[\overline{\varphi}_{\rm min}],\\ 
{c}_{4}&=-Mx_{J}\frac{\partial^{3}\varphi_{s}}{\partial^{3}\varphi}[\overline{\varphi}_{\rm min}],
\end{align}
where the derivatives of the implicit function $\varphi_{s}[\varphi]$ can be found by differentiating Eq.~(\ref{eq:current conservation}). The first derivative of $\varphi_s$, for example, is given by
\begin{equation}
\frac{\partial\varphi_s}{\partial\varphi}=\left(M+\left.\frac{1}{E_L}\frac{\partial^2U_N}{\partial\varphi_s^2}\right|_{\varphi_s[\Bar{\varphi}_\mathrm{min}]}\right)^{-1}.
\end{equation}
The higher-order coefficients could be obtained analytically or via symbolic calculation of higher-order derivatives of the function $\varphi_s[\varphi]$ \cite{fratini18}.  

Unfortunately, whenever there is a substantial geometric inductance in the circuit, it is impossible to derive a closed-form expression for the supercoefficients due to the presence of transcendental function $\varphi_{s}[\varphi]$ in the potential. However, we can calculate the SCs $C_{nl,p}$ following the definition in Eq.~(\ref{eq_app:supercoeff_sum_form}) up to a required order by calculating the corresponding ${c}_n$'s. For the asymptotic case $L\ll L_J$ the result reduces to Eq.~(\ref{eq_app:sc_snail_arr}). Additionally,  for Josephson circuits without geometric inductance, one also has to use SCs from Eq.~(\ref{eq_app:supercoeff_sum_form}) if a series array of nonlinear inductance elements is not permutation symmetric. In this case, the potential also contains transcendental functions due to the unequal phase drops across different Josephson junctions \cite{sandro2023,Rymarz23}.

\subsection{Array of DC SQUIDs with flux drive}

For a flux-driven circuit composed of an array of asymmetric DC SQUIDs, the single DOF approximation for the potential reads \cite{Iyama2024}
\begin{align}
\begin{split}
    U(\varphi,t)=-M\alpha_\mathrm{s}E_J\cos{\left(\frac{\varphi}{M}-r_a\varphi_{e}(t)\right)}\\-ME_J\cos{\left(\frac{\varphi}{M}+r_b\varphi_{e}(t)\right)},
\end{split}
\end{align}
where the external flux can be decomposed into static and oscillating parts, $\varphi_{e}(t)=\varphi_\mathrm{dc}+\varphi_\mathrm{ac}(t)$ with $\varphi_\mathrm{ac}(t) = 2\varphi_\mathrm{ac0}\cos\omega_dt$, $r_a$ and $r_b$ are obtained utilizing the irrotational gauge and defined by the capacitance of the circuit elements \cite{You2019} with $r_a+r_b=1$.

Here, we use SC definition from Eq.~(\ref{eq:super_coef_flux_drive}) for flux-driven circuits with the coefficients $a_1=b_1=1/M$, $a_2=-r_a$, $b_2=r_b$, the effective drive amplitudes are given by 
\begin{widetext}
\begin{align}
\Pi_a &{=} 2\varphi_\mathrm{ac0}M\left(-r_a\left(1{-}\frac{E_Jc_{2}^{(a)}\varphi_\mathrm{zpf}^2\omega_0}{2(\omega_d^2-\omega_0^2)}\right){-}r_b\frac{E_Jc_{2}^{(b)}\varphi_\mathrm{zpf}^2\omega_0}{2(\omega_d^2-\omega_0^2)}\right),\\
\Pi_b &{=} 2\varphi_\mathrm{ac0}M\left(r_b\left(1{-}\frac{E_Jc_{2}^{(b)}\varphi_\mathrm{zpf}^2\omega_0}{2(\omega_d^2-\omega_0^2)}\right){+}r_a\frac{E_Jc_{2}^{(a)}\varphi_\mathrm{zpf}^2\omega_0}{2(\omega_d^2-\omega_0^2)}\right),
\end{align}
and Eq.~(\ref{eq_app:supercoefficients_joseph_circuit}) to obtain the supercoefficients 
\begin{align}
C_{nl,p} = \frac{(-1)^{\left \lfloor\frac{2n+l+p}{2}\right \rfloor}E_J{\varphi}_\mathrm{zpf}^{2n+l}e^{-\frac{{\varphi}_\mathrm{zpf}^2}{2M^2}}}{n!(n+l)!M^{2n+l-1}}\times\hspace{40mm}\\
\left(\alpha_\mathrm{s}J_p\left(\frac{\Pi_a}{M}\right)\left\{\!\begin{aligned}
&-\cos{\left(\frac{\varphi_0}{M}-r_a\varphi_\mathrm{dc}\right)} \\[0.5ex]
&\sin{\left(\frac{\varphi_0}{M}-r_a\varphi_\mathrm{dc}\right)} \\[0.5ex]
\end{aligned}\right\}_{l+p}+J_p\left(\frac{\Pi_b}{M}\right)\left\{\!\begin{aligned}
&-\cos{\left(\frac{\varphi_0}{M}+r_b\varphi_\mathrm{dc}\right)} \\[0.5ex]
&\sin{\left(\frac{\varphi_0}{M}+r_b\varphi_\mathrm{dc}\right)} \\[0.5ex]
\end{aligned}\right\}_{l+p}\right).\nonumber
\end{align}
Using the trigonometric identities, $A\cos{x} + B\sin{x} = R\cos{(x- \lambda)}$, where $R=A^2+B^2$ and $\tan{\lambda}=B/A$, this can be written in a compact form as
\begin{align}
C_{nl,p} = \frac{(-1)^{\left \lfloor\frac{2n+l+p}{2}\right \rfloor+1}E_J{\varphi}_\mathrm{zpf}^{2n+l}e^{-\frac{{\varphi}_\mathrm{zpf}^2}{2M^2}}}{n!(n+l)!M^{2n+l-1}}\mathcal{A}_p\left\{\!\begin{aligned}
&\cos{\left(\frac{\varphi_0}{M}-\lambda^\prime_p\right)} \\[0.5ex]
&\sin{\left(\frac{\varphi_0}{M}-{\lambda^\prime_p}\right)} \\[0.5ex]
\end{aligned}\right\}_{l+p},
\end{align}
where we introduce the functions
\begin{align}
\mathcal{A}_p(\Pi_a,\Pi_b,\alpha_\mathrm{s},\varphi_\mathrm{dc})&=\sqrt{\alpha_\mathrm{s}^2J_p\left(\frac{\Pi_a}{M}\right)^2+J_p\left(\frac{\Pi_b}{M}\right)^2+2\alpha_\mathrm{s}J_p\left(\frac{\Pi_a}{M}\right)J_p\left(\frac{\Pi_b}{M}\right)\cos(\varphi_\mathrm{dc})},\label{eq_app:A_p}\\\lambda^\prime_p(\Pi_a,\Pi_b,r_a,r_b,\alpha_\mathrm{s},\varphi_\mathrm{dc})&=\arctan{\left(\frac{\alpha_\mathrm{s}J_p\left(\frac{\Pi_a}{M}\right)\sin{(r_a\varphi_\mathrm{dc})}-J_p\left(\frac{\Pi_b}{M}\right)\sin{(r_b\varphi_\mathrm{dc})}}{\alpha_\mathrm{s}J_p\left(\frac{\Pi_a}{M}\right)\cos{(r_a\varphi_\mathrm{dc})}+J_p\left(\frac{\Pi_b}{M}\right)\cos{(r_b\varphi_\mathrm{dc})}}\right)}\label{eq_app:lambda_p} 
\end{align}
\end{widetext}
and use the properties of the arctangent function, $\arctan{(1/x)} = \pi/2- \arctan{(x)}$. The minimum $\varphi_0=\lambda M+2\pi n$, $n\in\mathbb{Z}$ is found from the condition ${\partial_\varphi U_0}=0$  for a potential $U_0(\varphi)\equiv U(\varphi,\varphi_\mathrm{ac0}= 0)$ which determines the topology of the circuit 
\begin{equation}\label{eq:undriven_squid_arr}
U_0(\varphi)=-M\Tilde{E}_J\cos{\left(\frac{\varphi}{M}-\lambda\right)},
\end{equation}
where
\begin{align}
\Tilde{E}_J &=E_J\sqrt{1+\alpha_\mathrm{s}^2+2\alpha_\mathrm{s}\cos{\varphi_\mathrm{dc}}},\\
\lambda &=\arctan{\left(\frac{\alpha_\mathrm{s}\sin{(r_a\varphi_\mathrm{dc})}-\sin{(r_b\varphi_\mathrm{dc})}}{\alpha_\mathrm{s}\cos{(r_a\varphi_\mathrm{dc})}+\cos{(r_b\varphi_\mathrm{dc})}}\right)}.
\end{align}
Furthermore, to compare the amplitudes of the parametric processes in capacitively and flux-driven circuits, we assume that the two types of driving schemes are equivalent if we define an effective drive amplitude $\Tilde{\Pi}\equiv\Pi_a-\Pi_b=2\varphi_\mathrm{ac0}M$, with $\Pi=n_\mathrm{zpf}\Tilde{\Pi}$, for flux-driven circuits. In this case, $\Pi_a$ and $\Pi_b$ would represent the partitions of the effective drive amplitude between nodes of the SQUID loop attributed to the irrotational condition for the time-dependent external flux.

\paragraph*{Weak drive approximation and  Kerr-cat parameters}
To examine the closed-form SCs for a circuit composed of an array of DC SQUIDs, we reproduce results from Ref.~\cite{Iyama2024} for the Kerr parametric oscillator within the weak drive approximation. Specifically, we rederive the expressions for the Kerr nonlinearity $K$ and the squeezing drive amplitude $\epsilon_2$ in the effective Hamiltonian from Ref.~\cite{Iyama2024} for the case of a weak drive $\varphi_\mathrm{ac0}\ll2\pi$. Since Ref.~\cite{Iyama2024} considers only the rotating wave approximation, we limit ourselves to definitions (see next Section for details)
\begin{align}
\hat{\mathcal{H}}_{\mathrm{eff}} &= -K\hat{a}^{\dagger 2}\hat{a}^2 +   \epsilon_2 (\hat{a}^{\dagger 2}+\hat{a}^2),\nonumber\\
{K} &=-C_{20,0},\nonumber\\ {\epsilon}_2 &=C_{02,1}.\nonumber
\end{align}

Using the explicit expression for SC
\begin{align}\label{eq_app:K_sqid_weak_dr}
\begin{split}
K =\frac{E_J{\varphi}_\mathrm{zpf}^4M}{2!2!}\left(\frac{1}{M}\right)^{4}&e^{-\frac{{\varphi}_\mathrm{zpf}^2}{2M^2}}\mathcal{A}_0
\cos{(\frac{\varphi_0}{M}-\lambda^\prime_0)}\\&=\frac{E_Ce^{-\frac{{\varphi}_\mathrm{zpf}^2}{2M^2}}}{2M^2},
\end{split}
\end{align}
where $\varphi_\mathrm{zpf}=\sqrt[4]{2E_CM/\Tilde{E}_J}$ with $c_2E_J=\Tilde{E}_J/M$ (see Eq.~(\ref{eq:undriven_squid_arr})), $\varphi_0=\lambda M$. We have used the approximation
\begin{align}
\mathcal{A}_0 &\approx \Tilde{E}_J/E_J\\
\lambda^\prime_0 &\approx\lambda
\end{align}
for small $\varphi_\mathrm{ac0}$, utilizing the asymptotic expression $J_0(x)\xrightarrow[]{x\ll1}1$. 
The explicit  expression for the squeezing drive amplitude $\epsilon_2$ for weak drive is approximated by
\begin{widetext}
\begin{align}\label{eq_app:eps2_sqid_weak_dr}
\epsilon_2&{=} -\frac{E_J{{\varphi}_\mathrm{zpf}^2}}{2!M}e^{-\frac{{\varphi}_\mathrm{zpf}^2}{2M^2}}
\left(\alpha_\mathrm{s}J_1\left(\frac{\Pi_a}{M}\right)
\sin{(\frac{\varphi_0}{M}-r_a\varphi_\mathrm{dc})}
+J_1\left(\frac{\Pi_b}{M}\right)\sin{(\frac{\varphi_0}{M}+r_b\varphi_\mathrm{dc})}\right)\\
&\approx-\frac{E_J{{\varphi}_\mathrm{zpf}^2}}{2!M}e^{-\frac{{\varphi}_\mathrm{zpf}^2}{2M^2}}
\left(\alpha_\mathrm{s}\frac{\Pi_a}{2M}
\sin{(\frac{\varphi_0}{M}-r_a\varphi_\mathrm{dc})}
+\frac{\Pi_b}{2M}\sin{(\frac{\varphi_0}{M}+r_b\varphi_\mathrm{dc})}\right)\nonumber\\
&=-\frac{E_J{{\varphi}_\mathrm{zpf}^2}}{2!M}e^{-\frac{{\varphi}_\mathrm{zpf}^2}{2M^2}}
\left(\alpha_\mathrm{s}(-r_a\varphi_\mathrm{ac0})
\sin{(\frac{\varphi_0}{M}-r_a\varphi_\mathrm{dc})}
+(r_b\varphi_\mathrm{ac0})\sin{(\frac{\varphi_0}{M}+r_b\varphi_\mathrm{dc})}\right)\nonumber\\
&=\frac{\varphi_\mathrm{ac0}{{\varphi}_\mathrm{zpf}^2}}{2!M}e^{-\frac{{\varphi}_\mathrm{zpf}^2}{2M^2}}\frac{\partial}{\partial{\varphi_\mathrm{dc}}}\left(\Tilde{E}_J\cos{\left(\frac{\varphi_0}{M}-\lambda\right)}\right)\nonumber\\&=\frac{\varphi_\mathrm{ac0}\sqrt{\frac{2E_c}{M\Tilde{E}_J}}}{2}e^{-\frac{{\varphi}_\mathrm{zpf}^2}{2M^2}}\frac{\partial\Tilde{E}_J}{\partial{\varphi_\mathrm{dc}}},\nonumber
\end{align}
\end{widetext}
where we used the approximation $J_1(x)\xrightarrow[]{x\ll1}x/2$ and the current conservation condition $\alpha_\mathrm{s}
\sin{({\varphi_0}/{M}-r_a\varphi_\mathrm{dc})}
=-\sin{({\varphi_0}/{M}+r_b\varphi_\mathrm{dc})}$.   Up to a non-commutativity exponent multiplier and Stark shift in the definition of inductive energy, $\Tilde{E}_J$, coming from the averaging of higher-order drive terms in Ref.~\cite{Iyama2024}, both Eqs.~(\ref{eq_app:K_sqid_weak_dr}) and (\ref{eq_app:eps2_sqid_weak_dr}) reproduce the results of Ref.~\cite{Iyama2024}. Additionally, we note that the authors do not consider additional displacement terms in $\Pi_a$, $\Pi_b$ (see subsection on SCs for the flux-driven circuits). For the chosen approximation, this looks justified for the amplitudes of the parametric processes in terms of the obtained expressions. As we demonstrate in the following sections, in the case of a strong drive, it is generally necessary to use the SC approach and consider orders beyond the RWA.

\section{Optimization of circuit designs}
\label{app:applications}

\added{In this Section, we present a comprehensive guide for optimizing circuit designs tailored to specific parametric processes using the supercoefficient (SC) framework. This methodology encompasses formulating the problem within a rotating frame, deriving static effective Hamiltonians via perturbative techniques like the Schrieffer-Wolff transformation, and calculating SCs that encapsulate the circuit's topology and nonlinear characteristics. We also discuss strategies for parameter selection, simulation constraints, and convergence monitoring, ensuring that the optimization process remains tractable even under strong parametric drives. This structured approach facilitates the systematic design and analysis of advanced superconducting circuits, such as Kerr-cat qubits and beam-splitter interactions.}

\subsection{Study of optimal circuit designs for parametric processes}
\label{app:circuit_opt}

Here, we present a general guide for the optimization of circuit designs for specific parametric processes of interest within the SC approach and discuss its application for the problems considered in the main text:
\begin{enumerate}[label=\arabic*., align=left, left=0pt, labelsep=1em]
\item Formulate the concrete problem involving parametric drives, and move into a rotating frame   $\hat{U}_R=\exp{(i\omega^\prime\hat{a}^\dag\hat{a}t)}$, $\hat{U}_{R}^\dag\hat H\hat{U}_{R}-i\hat{U}_{R}^\dag\dot{\hat{U}}_{R}\rightarrow \hat H$, where $\omega^\prime\approx\frac{q}{r}\omega_d$,  $q,\,r\in \mathcal{Z}$, and  $\omega_0-\omega^\prime\ll\omega_0$, so 
    \begin{align}   
            \hat{H}(t)=(\omega_0-\omega^\prime)\hat{a}^{\dag}\hat{a}&+\\\,\sum\limits_{\substack{n=0\\ l=0\\p=0}}^{\{l,p\}^\prime}{C}_{nl,p}\left(\hat{a}^{\dag n}\hat{a}^{n+l}\right.&\left.e^{-il\omega^\prime{t}}(e^{ip{\omega}_d{t}}+\text{c.c.})+\text{H.c.}\right),\nonumber
    \end{align}
where we take $\gamma=0$ for the sake of brevity.
At this point, depending on the type of wave mixing involved, one chooses the order up to which the Hamiltonian expansion is taken. In the main text, we limit the expansion with $2n+l+p\leq4$ and $p\leq2$ for a Kerr parametric oscillator dealing with three- and four-wave mixing as the dominant processes and with $2n+l+p\leq4$ for the beam-splitter interaction problem involving additionally five-wave mixing. Such an approximation, in particular, is justified by the smallness of $\varphi_\mathrm{zpf}$ and $C_{nl,p}\propto\varphi_\mathrm{zpf}^{2n+l}$.

\item Calculate the static effective Hamiltonian up to a required order using the Schrieffer-Wolff (SW) procedure or any alternative averaging method. In the main text, we employ the procedure introduced in Ref.~\cite{jaya2022} computing the unitary generator $\hat S(t)$ for the canonical transformation and  obtaining a static effective Hamiltonian
via 
\begin{align}\label{eq_app:effH}
\begin{split}
\hat{\mathcal{H}}_{\mathrm{eff}} \equiv \overline{e^{\hat{S} / i \hbar} \hat{H}(t) e^{-\hat{S} / i \hbar}}-i \hbar\overline{ e^{\hat{S} / i \hbar} \partial_{t} e^{-\hat{S} / i \hbar}}\\
= \overline{\hat{{H}}(t)}+ \frac{1}{i\hbar}\overline{[\hat S, \hat{{H}}(t)]}+\frac{1}{2!(i\hbar)^2}\overline{[\hat S,[\hat S, \hat{{H}}(t)]]} \\+ \cdots  + \overline{\partial_t\hat S}  + \frac{1}{2! i\hbar} \overline{[\hat S, \partial_t {\hat S}]} + \cdots
\end{split}
\end{align}
where the  Baker-Campbell-Hausdorff formula\footnote{$e^{-\hat A} \hat B e^{\hat A} = \hat B + [\hat B, \hat A] + \frac{1}{2!} [[\hat B, \hat A],\hat A] + \ldots$} was applied and the time averaging is given by $\overline{f}(t)\equiv\frac{1}{T}\int_0^T{f(t)}dt$, where $T=2\pi/\omega$. Then, both the generator $\hat S$ and the effective Hamiltonian $\hat{\mathcal{H}}_{\mathrm{eff}}$ can be calculated perturbatively with a recursive procedure up to the required order of $1/\omega$ considering the expansion of the time-dependent Hamiltonian 
\begin{equation}
\hat{{H}}(t)=\sum\limits_{m\in \mathbb{Z}}\hat{H}_me^{im\omega t},
\end{equation}
so 
\begin{align}    \hat{\mathcal{H}}_{\mathrm{eff}} = \sum_{n\in\mathbb{N}} \hat{\mathcal{H}}_{\mathrm{eff}}^{(n)}, \quad \hat S = \sum_{n\in\mathbb{N}} \hat S^{(n)}.
\end{align}
The RWA for the Hamiltonian is defined as $\hat{\mathcal{H}}_{\mathrm{eff}}^{(0)}=\overline{\hat{H}(t)}$ with $\hat S^{(0)}=0$, while the first-order correction is obtained from
\begin{align}
\frac{\hat S^{(1)}}{\hbar}&=- \int dt\; \textbf{osc}\hat{H}(t),\\
\hat{\mathcal{H}}_{\mathrm{eff}}^{(1)}&=\frac{1}{i\hbar}\overline{\left([\hat S^{(1)}, \hat{{H}}(t)]+\frac{1}{2!} [\hat S^{(1)}, \partial_t {\hat S^{(1)}}]\right)},
\end{align}
where $\textbf{osc}f= f-\overline{f}$. The next order of perturbations can be calculated via the recursive procedure presented in the original paper \cite{jaya2022}. Although it is possible to obtain corrections up to the required order, for simplicity, the main text and next sections focus only on the first-order correction beyond the RWA.


\item Obtain the supercoefficients for circuit designs of interest by using  Eqs.\  (\ref{eq_app:supercoefficients_joseph_circuit}), (\ref{eq_app:supercoefficients_joseph_circuit2}) for symmetric Josephson circuits or Eq.\  (\ref{eq_app:supercoeff_sum_form}) for circuits with the geometric inductance or broken permutation symmetry. This is the first step where the specific potential function must be considered. In the main text, we describe circuits consisting of an array of SNAILs or asymmetric SQUIDs (see Table I in the main text).

\item Define a list of free parameters to consider (in the main text, for example, $\varphi_e$, $N$, $M$, $\alpha_\mathrm{s}$, $x_J$) and fixed parameters ($E_C$, $E_J$). In the main text, we employ an effective drive amplitude $\Pi=n_\mathrm{zpf}\Tilde{\Pi}$ often used to describe the input from the capacitive parametric drive \cite{frattini2022squeezed,ChapmanStijn2023,jaya2022lind}. For flux-driven circuits, we use $\Tilde{\Pi}\equiv\Pi_a-\Pi_b=2\varphi_\mathrm{ac0}M$ (see details above).

\item Define the limitations of a simulation. This step is highly problem-specific and is primarily determined by the intended utility of the parametric processes. For the Kerr-cat problem, we introduce the minimal value of Kerr nonlinearity considered. For the beam-splitter interaction problem, we ensure the validity of the dispersive approximation and introduce the minimal value of cavity-cavity cross-Kerr, $\chi_{ab}$, considered (see next subsection). So, for example, to find the maximum  Kerr-cat size, for each circuit design we fix parameters $N$, $M$, $\alpha_\mathrm{s}$, $x_J$, $\Pi$, require $|K({\Pi})|>K_\mathrm{lim}=\SI{1}{\mega\hertz}$ and scan all values of external flux, $\varphi_e$. 
     
\item  Monitor the convergence of the expansion of $C_{nl,p}$ for strongly driven circuits with the linear inductance or broken permutation symmetry using Eq.~(\ref{eq_app:supercoeff_sum_form}). In the main text, the convergence is ensured by using only drive strengths with $\Tilde{\Pi}\lesssim 2$ (except for the \textbf{D} configuration for KPO where diluted higher-order $c_n$'s compensate for a stronger drive, see Table~\ref{tab:kerr_chaos_params} below). This is further supported by the smallness of the zero-point fluctuations of the phase and the factorial functions in the denominator of the sum defining the supercoefficients (\ref{eq_app:supercoeff_sum_form}).
\end{enumerate}

\subsection{Supercoefficients and Kerr-cat parameters}
\label{app:kerr_cat_scs}

\begin{table*}[!t]
	\centering
\begin{tabular}{l|c|c|c|c}
\textbf{Parameter} & \textbf{A} & \textbf{B} & \textbf{C} & \textbf{D}\\ \hline
Nonlinear resonator frequency, $\omega_0/2\pi$  & \SI{5.6}{\giga\hertz}\,  & \SI{5.2}{\giga\hertz}\,  & \SI{5.9}{\giga\hertz}\, & \SI{6.3}{\giga\hertz}\, \\ \hline
Kerr nonlinearity, $K/2\pi$ at $\Pi=0$  & \SI{6.76}{\mega\hertz}\, & \SI{-2.58}{\mega\hertz}\, & \SI{1.15}{\mega\hertz}\, & \SI{0.72}{\mega\hertz}\, \\ \hline
Number of SNAILs in an array, $M$    &  1 & 2& 1& 2 \\ \hline
Number of large JJs in SNAIL, $N$   & 3 & 3 & 3 & 3\\ \hline
Nonlinear inductance of large JJs, $L_J$    & \SI{0.8}{\nano\henry} & \SI{0.6}{\nano\henry}& \SI{0.35}{\nano\henry}& \SI{0.39}{\nano\henry} \\ \hline
Inductance participation ratio, $x_J$   & 100 & 1 & 10 & 0.27 \\ \hline
Shunting capacitance, $C$   & \SI{0.32}{\pico\farad} & \SI{0.16}{\pico\farad}& \SI{0.62}{\pico\farad}& \SI{0.17}{\pico\farad} \\ \hline
Asymmetry coefficient, $\alpha_\mathrm{s}$   & 0.11 & 0.11 &  0.05 & 0.0739  \\ \hline
External flux, $\varphi_e$ & 0.32 & 0.46  & 0.34 & 0.25 \\ \hline
Maximum effective drive amplitude, $\Tilde{\Pi} (\Pi)$ & 1.5 (3.5)  & 3. (5.) & 2.9  (10.)&  5.3 (10) \\ \hline		
\end{tabular}  
\caption{Parameters of the circuit configurations considered in studying the onset of chaos in Kerr-cat qubits in the main text.}
\label{tab:kerr_chaos_params}
\end{table*}
We apply the SC approach to study a parametrically squeezed Kerr nonlinear oscillator (KPO) which, in particular, is used to engineer Kerr-cat qubits \cite{fratini18,frattini2022squeezed,jaya2022lind,Iyama2024}. By applying a classical drive at frequency $\omega_d\approx2\omega_0$ and limiting the Hamiltonian expansion with conditions $2n+l\leq4$ and $p\leq2$ in the RWA approximation one obtains 
\begin{equation}
\begin{split}
\label{eq:Kerr_cat_ham}
\hat{\mathcal{H}}_{\mathrm{RWA}} &= \left(\omega_0-\frac{\omega_d}{2}+{C}_{10,0}\right) \hat{a}^\dagger \hat{a}\\&\qquad+C_{20,0}\hat{a}^{\dagger 2}\hat{a}^2 +   C_{02,1} (\hat{a}^{\dagger 2}+\hat{a}^2).
\end{split}
\end{equation}
To account for terms beyond RWA we apply the Schrieffer-Wolff procedure up to the first order to obtain
\begin{equation}
\begin{aligned}
\label{eq:Kerr_cat_ham}
\hat{\mathcal{H}}_{\mathrm{eff}} = \Delta \hat{a}^\dagger \hat{a}-K\hat{a}^{\dagger 2}\hat{a}^2 +   \epsilon_2 (\hat{a}^{\dagger 2}+\hat{a}^2),
\end{aligned}
\end{equation}
with Kerr-cat parameters defined as
\begin{align}
{\Delta} &= \omega_0-\frac{\omega_d}{2}+{C}_{10,0}+{\Delta}^{(1)},\\
{K} &= -C_{20,0}+{K}^{(1)},\\
{\epsilon_2}&=C_{02,1}+\epsilon_2^{(1)},
\end{align}
and first-order corrections beyond RWA
\begin{widetext}
\begin{align}
\epsilon_2^{(1)}&=(-2C_{01,1}C_{03,0}-6C_{01,0}C_{03,1}-\frac{6}{5}C_{01,2}C_{03,1}-6C_{02,1}C_{04,0}-2C_{02,0}C_{10,1}+2C_{02,2}C_{10,1}\nonumber\\&\quad-C_{02,1}C_{10,2}+2C_{01,1}C_{11,0}-12C_{03,1}C_{11,0}-2C_{01,0}C_{11,1}+\frac{2}{3}C_{01,2}C_{11,1}-4C_{03,0}C_{11,1})/\omega_d,\\
K^{(1)}&=-(-6C_{03,0}^2-\frac{108}{5}C_{03,1}^2-36C_{04,0}^2-6C_{11,0}^2+4C_{11,1}^2-12C_{02,0}C_{12,0}-18C_{12,0}^2)/\omega_d,\\
\Delta^{(1)}&=(-4C_{02,0}^2-2C_{02,1}^2+\frac{8}{3}C_{02,2}^2-12C_{03,0}^2-\frac{216}{5}C_{03,1}^2-48C_{04,0}^2-8C_{01,0}C_{11,0}-4C_{11,0}^2\nonumber\\&\quad+\frac{16}{3}C_{01,1}C_{11,1}+\frac{8}{3}C_{11,1}^2-12C_{02,0}C_{12,0}-6C_{12,0}^2)/\omega_d.
\end{align} 
\end{widetext}
The above expressions reproduce the result for the dynamic parameters of the Kerr-cat in terms of the nonlinear coefficients, $g_n$, presented in \cite{frattini2022squeezed} (see Appendix A therein) for the corresponding order of the Schrieffer-Wolff procedure and nonlinear coefficients. For example, if we use the expansion of $C_{nl,p}$ only up to $\mathrm{S}\leq4$, $p\leq2$ (see Eq.~(\ref{eq_app:supercoeff_sum_form})), use $\Tilde\Pi=2\varphi_\mathrm{zpf}{\Pi}$ and omit terms $\mathcal{O}(\varphi_{\mathrm{zpf}}^{5})$ (corresponds to order 2 in Ref.~\cite{frattini2022squeezed}), we obtain the usual expressions for the Kerr-cat parameters 
\begin{align}
\begin{split}
{\Delta} &= \omega_0-\frac{\omega_d}{2}+{C}_{10,0}+{\Delta}^{(1)}\rightarrow\\ &\quad\omega_0-\frac{\omega_d}{2}+6g_4|\Pi|^2 - \frac{18g_3^2|\Pi|^2}{\omega_d} + 2K,
\end{split}\\
{K} &= -C_{20,0}+{K}^{(1)}\rightarrow-\frac{3g_4}{2} +  \frac{20g_3^2}{3\omega_d},\\
{\epsilon_2} &=C_{02,1}+\epsilon_2^{(1)}\rightarrow g_3\Pi.
\end{align}
However, the SCs approach offers a more flexible framework for addressing higher-order contributions to the amplitudes of the corresponding parametric processes. Specifically, closed-form SCs for symmetric Josephson circuits enable us to address circuit topology and drive power exactly, rather than relying on a perturbative approach for $g_n$'s and orders of $\Pi$. This proves advantageous for describing more complex circuit topologies and strong drives, and, in particular, it enables a more efficient Schrieffer-Wolff procedure. Notably, it eliminates the need to include higher-order terms from the expansion \cite{frattini2022squeezed,GarciaMata2024effectiveversus}
\begin{align}
\hat{{H}}=\omega_0\hat{a}^\dag\hat{a}+\sum\limits_{ n\geq3}{g_n}\left(\hat{a}{+}\hat{a}^\dag{+}\Pi e^{-i\omega_d{t}}{+}\Pi^{\ast} e^{i\omega_d{t}}\right)^n
\end{align}
to account for corrections to $\Delta$, $K$, $\epsilon_2$ arising from higher-order multiphoton processes.    

\paragraph*{Simulation parameters} In the main text, we consider a Kerr-cat with zero detuning ($\Delta=0$) by driving the KPO at $\omega_d=2\omega_q$, where Lamb- and Stark-shifted small oscillation frequency $\omega_q = \omega_0+{C}_{10,0}+{\Delta}^{(1)}$. For Kerr-cat qubit optimization, the energy parameters of the simulated SNAIL-based circuit designs are derived from experiments \cite{Grimm2020,frattini2022squeezed}: the inductance of the large Josephson junction  $L_J=\SI{0.391}{\nano\henry}$, the shunting capacitance $C=\SI{0.172}{\pico\farad}$. So, for SNAIL-based circuits, the plasma frequency, $\omega_0/2\pi$, varies from $\SI{4}{\giga\hertz}$ to $\SI{12}{\giga\hertz}$. For SQUID-based circuit designs, we use the same inductance of large JJs but $2.5$ times larger shunting capacitance to ensure close frequency range $\omega_0/2\pi\in[\SI{5}{\giga\hertz},\SI{17}{\giga\hertz}]$, and $r_a=0.9$, $r_b=0.1$. Other parameters of the simulation: $\Pi=0.5$, $\mathrm{S}\leq8$ for circuits with finite geometric inductance, $\gamma=0$ and $\gamma=\pi$ for positive and negative values of $K$, respectively, to ensure a double-well potential of the Kerr-cat \cite{jaya2023interference}.

\subsection{Onset of chaotic behavior in Kerr-cat qubit}
\label{app:KC_chaos}

\begin{figure*}[t!]
\centering
\includegraphics[width=0.9\linewidth]{ 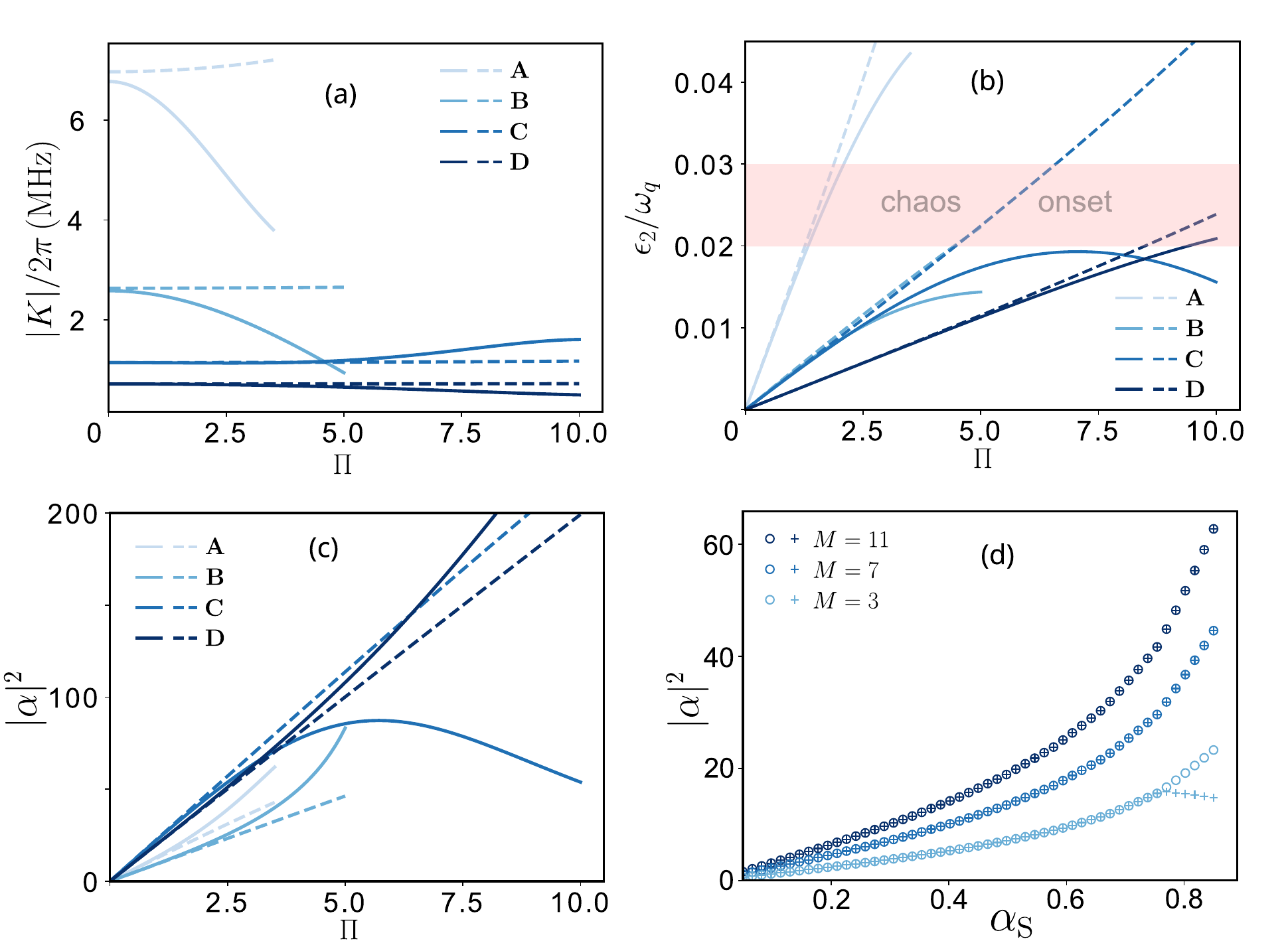}
\caption{(Color online) Chaos in the Kerr-cat qubit.  (a) Dependence of Kerr-nonlinearity $K$ on effective drive strength, $\Pi$. Dashed lines represent result for SC approximated with $\mathrm{S}\leq4$, solid lines calculated for $\mathrm{S}\leq13$ (see Eq.~(\ref{eq_app:supercoeff_sum_form})). (b) Chaos emergence criterion for dependence $\epsilon_2(\Pi)/\omega_q(\Pi)$ vs.~$\Pi$. (c) Dependence of Kerr-cat size on $\Pi$. (d) Dependence of Kerr-cat size on JJ asymmetry coefficient $\alpha_\mathrm{s}$ in SQUID-based circuits (see Table I in the main text) without (circles) and with (crosses) chaos emergence criterion enforced. Parameters of the circuit configurations \textbf{A}-\textbf{D} for (a)-(c) are presented in Table \ref{tab:kerr_chaos_params}.} 
\label{fig:chaos_suppl} 
\end{figure*}

\added{The simulations are performed for the circuit configurations presented in Table~\ref{tab:kerr_chaos_params}.} We demonstrate (see Fig.~\ref{fig:chaos_onset} in the main text) that the susceptibility of Kerr-cat qubits to chaotic behavior is highly dependent on the circuit topology. It also determines the achievable range of dynamic parameters constrained by chaos. Notably, the circuit topology plays a crucial role in dictating whether the system is affected by chaos and the drive strength at which this occurs. Additionally, we illustrate that the lower-order approximation for Kerr-cat parameters (dashed lines in Fig.~\ref{fig:chaos_onset} in the main text and Figs.~\ref{fig:chaos_suppl}(a)-(c)) is only valid for weaker drives and smaller Kerr-cat sizes. For stronger drives, a higher-order approximation or a closed-form expression for the supercoefficients is necessary to accurately capture topology-related specifics. Here, we also depict the dependence of dynamic parameters on drive strength to better understand the dynamics of considered circuits (see  Figs.~\ref{fig:chaos_suppl}(a)-(c)). We also examined the influence of the criteria for chaos emergence by applying an additional limitation, 
$\epsilon_2/\omega_q\leq0.025$ (see step 5 in the guide for circuit design optimization), to SNAIL-based and SQUID-based circuit designs of the Kerr-cat problems discussed in the main text (Figs.~\hyperref[fig:max_kc_size]{\ref{fig:max_kc_size}(a)} and \hyperref[fig:max_kc_size]{\ref{fig:max_kc_size}(b)} therein). Based on this criterion, only the SQUID-based circuit configuration with $M=3$ is minimally affected (see Fig.~\ref{fig:chaos_suppl}(d)).

\paragraph*{Simulation parameters} The circuit configurations presented in Table \ref{tab:kerr_chaos_params} are considered. The configurations \textbf{A} and \textbf{D} are referring to the experimental setups utilized in \cite{Grimm2020} and \cite{frattini2022squeezed}, correspondingly. Circuit designs \textbf{B} and \textbf{C} are chosen to demonstrate more complex topology-dependent dynamics. We compare different approximations for the SCs with $\mathrm{S}\leq4$ corresponding to \cite{chávezcarlos2024} and $\mathrm{S}\leq13$.

\subsection{Supercoefficients for two cavities coupled by nonlinear coupler}\label{app:beam_splitter}

The normal-ordered expansion of a single DOF Hamiltonian for Josephson circuits could be easily generalized for circuits with weakly capacitively coupled linear oscillators. We can demonstrate it for the case of two linear oscillators, each capacitively coupled to a shared nonlinear coupling element \cite{Pietikainen:2022iqj,Lu2023,ChapmanStijn2023}. In this case, the Hamiltonian is given by
\begin{align}\label{eq:3modes}\begin{split}
\hat{{H}}&=\omega_a\hat{a}^\dag\hat{a}+\omega_b\hat{b}^\dag\hat{b}+\omega_c\hat{c}^\dag\hat{c}
\\&+{{U}_{nl}\left({{\varphi}_\mathrm{zpf}}\left(\hat{a}{+}\hat{a}^\dag\right){+}\varphi_0\right)}\\&-i{{n}_\mathrm{zpf}{\Omega}_{d}}(\hat{a}-\hat{a}^\dag)\cos{({\omega}_d{t}+
\theta)}+\hat{H}_I,
\end{split}\\
\hat{H}_I&=-g_b(\hat{a}-\hat{a}^\dag)(\hat{b}-\hat{b}^\dag)-g_c(\hat{a}-\hat{a}^\dag)(\hat{c}-\hat{c}^\dag),
\end{align}
where the nonlinear part of the potential $U_{nl}$ is fully determined by the coupler, $\hat a^\dag$ and $\hat a$ are bosonic operators of the coupler mode, $\hat b^\dag$, $\hat b$, $\hat c^\dag$, $\hat c$ are  bosonic operators of linear oscillators.
By employing the dispersive approximation and the displacement transformation for the drive term (as detailed in Ref.~\cite{ChapmanStijn2023} and in the first section on deriving a normal-ordered Hamiltonian for a driven circuit), we can express the Hamiltonian in terms of dressed modes as
\begin{align}\label{eq:3dressed_modes}
\begin{split}
\hat{{H}}&=\omega_a^\prime\hat{a}^\dag\hat{a}+\omega_b^\prime\hat{b}^\dag\hat{b}+\omega_c^\prime\hat{c}^\dag\hat{c}
\\&+{U}_{nl}\left({{\varphi}_\mathrm{zpf}}\left(\hat{a}+\hat{a}^\dag+\xi_b(\hat{b}+\hat{b}^\dag)+\xi_c(\hat{c}+\hat{c}^\dag)\right)\right.\\&\qquad\qquad\qquad\left.+{\Tilde{\Pi}}\cos{({\omega}_d{t}+\gamma)}{+}\varphi_0\right),
\end{split}
\end{align}
where the coefficients $\xi_{b,c}=2g_{b,c}\omega_{b,c}/(\omega_{b,c}^2-\omega_{a}^2)$, with $g_{b,c}/(\omega_{b,c}-\omega_{a})\ll1$, the effective drive amplitude $\Tilde{\Pi} = \Omega_d{\omega}_d/(\omega_d^2-\omega_a^{\prime2})$,  the phase $\gamma=\theta-{\pi}/2$, and the dressed frequencies are
\begin{align}
\omega_a^\prime &=\omega_a + \frac{g_b^2}{\omega_a-\omega_b}+ \frac{g_c^2}{\omega_a-\omega_c},\\
\omega_b^\prime &=\omega_b - \frac{g_b^2}{\omega_a-\omega_b},\\
\omega_c^\prime &=\omega_c - \frac{g_c^2}{\omega_a-\omega_c}.
\end{align}
Applying the procedure from Appendix~\ref{app:eff_ham_derivaiton} to the Hamiltonian in Eq.~(\ref{eq:3dressed_modes}), we can write its general normal-ordered expansion for the three-mode case. Namely, the corresponding supercoefficients are derived from 
\begin{widetext}
\begin{equation}\label{eq:supercoeff_diff_form3modes}
\Tilde{C}_{n_al_a,n_bl_b,n_cl_c,p}{=}\frac{e^{\frac{{\varphi}_\mathrm{zpf}^2}{2}(1+\xi_b^2+\xi_c^2)\frac{\partial^2}{\partial \varphi^2}}({{\varphi}_\mathrm{zpf}^2}\frac{\partial^2}{\partial \varphi^2})^{\frac{2(n_a{+}n_b{+}n_c){+}(l_a{+}l_b{+}l_c)}{2}}\xi_b^{2n_b+l_b}\xi_c^{2n_c+l_c}I_{p}({\Tilde{\Pi}}\frac{\partial}{\partial \varphi})}{n_a!(n_a{+}l_a)!n_b!(n_b{+}l_b)!n_c!(n_c{+}l_c)!}\left.{U}(\varphi)\right|_{\varphi{=}\varphi_0},
\end{equation}
where each pair of the indices $n$ and $l$ correspond to a particular mode in the system, and is defined by
\begin{align}\label{eq_app:supercoeff_sum_form_3DOF}
C_{n_al_a,n_bl_b,n_cl_c,p}&=\sum\limits_{k,m}^{\mathrm{S}\geq3}
\frac{c_{2(n_a{+}n_b{+}n_c){+}(l_a{+}l_b{+}l_c){+}2k{+}p{+}2m}E_J{{\varphi}_\mathrm{zpf}^{2(n_a{+}n_b{+}n_c){+}(l_a{+}l_b{+}l_c)}}\xi_b^{2n_b+l_b}\xi_c^{2n_c+l_c}}{m!k!(k{+}p)!n_a!(n_a{+}l_a)!n_b!(n_b{+}l_b)!n_c!(n_c{+}l_c)!}\\
&\quad\times
\left(\frac{{\varphi}_\mathrm{zpf}^2(1{+}\xi_b^2{+}\xi_c^2)}{2}\right)^m{}\left(\frac{{\Tilde{\Pi}}}2\right)^{2k{+}p},\nonumber
\end{align}
where  $\mathrm{S}=2(n_a{+}n_b{+}n_c){+}(l_a{+}l_b{+}l_c){+}2k{+}p{+}2m$. Then, the normal-ordered Hamiltonian is given by
\begin{align}\label{eq:ham_supercoeff_with_quad_3DOF}    \hat{{H}}=\omega_a^\prime\hat{a}^\dag\hat{a}+\omega_b^\prime\hat{b}^\dag\hat{b}+\omega_c^\prime\hat{c}^\dag\hat{c}
+\sum\limits_{ \substack{n_a,n_b,n_c,\\l_a,l_b,l_c,p=0}}^{\{l_a,l_b,l_c,p\}^\prime}C_{n_al_a,n_bl_b,n_cl_c,p}(\hat{a}^{\dag n_a}\hat{a}^{n_a+l_a}+\hat{a}^{\dag n_a+l_a}\hat{a}^{n})(\hat{b}^{\dag n_b}\hat{b}^{n_b+l_b}+\hat{b}^{\dag n_b+l_b}\hat{b}^{n_b})\nonumber\\
\times(\hat{c}^{\dag n_c}\hat{c}^{n_c+l_c}+\hat{c}^{\dag n_c+l_c}\hat{c}^{n_c})(e^{ip(\omega_dt+\gamma)}+e^{-ip(\omega_dt+\gamma)}).
\end{align}
The superscription of the sum means that for terms with $l_a=0$, $l_b=0$, $l_c=0$, and $p=0$, an additional $1/2$ multiplier should be applied. The derivation of the closed form of the SC for symmetric Josephson circuits is straightforward (\textit{e.g.}, see subsection \textit{General potential for single degree of freedom approximation for an arbitrary symmetric Josephson circuit}). For example, for SQUID-based circuit designs, the supercoefficients are
\begin{align}\label{eq_app:supercoeff_closed_form_3DOF}
&C_{n_al_a,n_bl_b,n_cl_c,p}={(-1)^{\left \lfloor\frac{2(n_a{+}n_b{+}n_c){+}(l_a{+}l_b{+}l_c){+}p}{2}\right \rfloor+1}\added{ME_J{\left(\frac{{\varphi}_\mathrm{zpf}}{M}\right)^{2(n_a{+}n_b{+}n_c){+}(l_a{+}l_b{+}l_c)}}}\xi_b^{2n_b+l_b}\xi_c^{2n_c+l_c}}\\
&\times{\left(n_a!(n_a{+}l_a)!n_b!(n_b{+}l_b)!n_c!(n_c{+}l_c)!\right)^{-1}}\added{\exp{\left(-\frac{{\varphi}_\mathrm{zpf}^2(1{+}\xi_b^2{+}\xi_c^2)}{2M^2}\right)}}\mathcal{A}_p\left\{\!\begin{aligned}
&\cos{(\frac{\varphi_0}{M}{-}\lambda^\prime_p)}\\[-0.5ex]
&\sin{(\frac{\varphi_0}{M}{-}{\lambda^\prime_p})} \\[-0.5ex]
\end{aligned}\right\}_{\added{l_a{+}l_b{+}l_c}{+}p},\nonumber
\end{align}
\end{widetext}
with $\mathcal{A}_p$ and $\lambda^\prime_p$ defined in Eqs.~(\ref{eq_app:A_p}) and (\ref{eq_app:lambda_p}).

\subsubsection*{Beam-splitter interaction}

The above description can be applied to a parametrically driven beam-splitter interaction, which is useful in multiple applications in quantum computing designs \cite{Liu2024,tsunoda2023error,Teoh2023,Chou2024,degraaf2024}. In the main text, we consider the scheme 
of programmable beam-splitter interaction between linear cavities considered in Ref.~\cite{ChapmanStijn2023}. For drive frequency $\omega_d  = \omega_c^\prime-\omega_b^\prime$, the Hamiltonian within the RWA approximation reads as
\begin{align}
\begin{split}
\hat{\mathcal{H}}^{\mathrm{RWA}}&\approx\Delta_a^{(0)}\hat{a}^\dag\hat{a}+\delta\hat{b}^\dag\hat{b}+\delta\hat{c}^\dag\hat{c}+\left(g_{bc}e^{-i\gamma}\hat{b}^\dag\hat{c}\right.\\&\left.+g_{ac}^{(0)}e^{-in\gamma}\hat{c}^\dag\hat{a}^\dag+g_{ab}^{(0)}e^{-i(n+1)\gamma}\hat{b}^\dag\hat{a}^\dag+\text{H.c.}\right)\\&\hspace{20mm}+\chi_{bc}^{(0)}\hat b^\dag\hat b\hat c^\dag\hat c,
\end{split}
\end{align}
where we omit self-Kerr terms for the cavities but consider two-mode squeezing terms with amplitudes $g_{ac}$ and  $g_{ab}$ to account for possible slow oscillating processes with frequency $\delta=\omega_a^\prime+\omega_b^\prime-n\omega_d=\omega_a^\prime+\omega_c^\prime-(n+1)\omega_d$ for $n\in\mathcal{N}$, $n>0$. The corresponding amplitudes of parametric processes are given by 
\begin{align}
g_{bc} &= C_{00,01,01,1},\\  
g_{ab}^{(0)} &= C_{01,01,00,n},\\
g_{ac}^{(0)} &= C_{01,00,01,n+1},\\
\Delta_a^{(0)} &= C_{10,00,00,0},\\
\chi_{bc}^{(0)}&=C_{00,10,10,0},
\end{align}
with amplitude $g_{bc}$ reproducing the result for $g_\mathrm{BS}$ considered in Ref.~\cite{ChapmanStijn2023}. The specific values of the index $n$ depend on the particular circuit parameters. For microwave cavity modes with resonant frequencies of $3$ and $\SI{7}{\giga\hertz}$, and a coupler frequency between $4.5$ and $\SI{6}{\giga\hertz}$ as discussed in Ref.~\cite{ChapmanStijn2023}, we will consider only the case $n=2$. Additionally, for the sake of simplicity, we omit the phase $\gamma=0$.
From the SW procedure, the first-order corrections beyond the RWA for the couplings between different normal modes, the detuning, and cavity-cavity cross-Kerr are given by
\begin{widetext}
\begin{align}  
g_{ab}^{(1)} &= -\frac{C_{01,01,00,2} C_{02,00,00,0}}{\omega_a^\prime}-\frac{2C_{01,01,00,1} C_{02,00,00,1}}{2\omega_a^\prime-\omega_d}-\frac{C_{01,01,00,0} C_{02,00,00,2}}{\omega_a^\prime-2\omega_d}-\frac{2C_{01,00,00,2} C_{02,01,00,0}}{\omega_a^\prime+2\omega_d}\label{eq:ga_corr}\\
&\quad-\frac{2C_{01,00,00,1} C_{02,01,00,1}}{\omega_a^\prime+\omega_d}-\frac{C_{01,01,00,1} C_{10,00,00,1}}{\omega_d}-\frac{C_{01,00,00,2} C_{10,01,00,0}}{\omega_a^\prime-2\omega_d}-\frac{C_{01,00,00,1} C_{10,01,00,1}}{\omega_a^\prime-\omega_d},\nonumber\\
g_{ac}^{(1)} &=-\frac{2C_{01,00,01,2} C_{02,00,00,1}}{2\omega_a^\prime-\omega_d}-\frac{C_{01,00,01,1} C_{02,00,00,2}}{\omega_a^\prime-\omega_d}-\frac{2C_{01,00,00,3} C_{02,00,01,0}}{\omega_a^\prime+3\omega_d}-\frac{2C_{01,00,00,2} C_{02,00,01,1}}{\omega_a^\prime+2\omega_d}\\
&\quad-\frac{C_{01,00,01,2} C_{10,00,00,1}}{\omega_d}-\frac{C_{01,00,01,1} C_{10,00,00,2}}{2\omega_d}-\frac{C_{01,00,00,3} C_{10,00,01,0}}{\omega_a^\prime-3\omega_d}-\frac{C_{01,00,00,2} C_{10,00,01,1}}{\omega_a^\prime-2\omega_d},\nonumber
\end{align}
\begin{align}
\begin{split}
&\Delta_a^{(1)}{=} 
{-}\frac{4C_{01,00,00,0}C_{11,00,00,0}+2C_{11,00,00,0}^2}{\omega_a^\prime}{-}{\left(4C_{01,00,00,1}C_{11,00,00,1}{+}2C_{11,00,00,1}^2\right)}\left(\frac{1}{\omega_a^\prime-\omega_d}{+}\frac{1}{\omega_a^\prime+\omega_d}\right),
\end{split}\\
\begin{split}\label{eq:chi_corr}
&\chi_{bc}^{(1)}{=}C_{01,01,01,0}^2\left(\frac{1}{\omega_a^\prime{-}5\omega_d}{-}\frac{1}{3\omega_a^\prime{-}5\omega_d}{-}\frac{1}{\omega_a^\prime-\omega_d}{-}\frac{1}{\omega_a^\prime{+}\omega_d}\right){-}2C_{01,00,01,0}C_{01,10,01,0}\left(\frac{1}{2\omega_a^\prime{-}3\omega_d}{+}\frac{1}{3\omega_d}\right)\\&+
C_{01,01,01,1}^2\left(\frac{1}{\omega_a^\prime-6\omega_d}-\frac{2}{\omega_a^\prime}+\frac{1}{\omega_a^\prime-4\omega_d}-\frac{1}{3\omega_a^\prime-4\omega_d}-\frac{4}{3\omega_a^\prime-6\omega_d}-\frac{1}{\omega_a^\prime+2\omega_d}\right)\\&{-}C_{01,01,00,0}C_{01,01,10,0}\left(\frac{1}{\omega_a^\prime{-}\omega_d}{+}\frac{1}{\omega_d}\right){-}\frac{2C_{01,00,10,0}C_{01,10,00,0}}{\omega_a^\prime}
{-}2C_{01,00,10,1}C_{01,10,00,1}\left(\frac{1}{\omega_a^\prime{-}\omega_d}{+}\frac{1}{\omega_a^\prime{+}\omega_d}\right).
\end{split}
\end{align}
\end{widetext}
There is a renormalized explicit expression only for  $\chi_{bc}=\chi_{bc}^{(0)}+\chi_{bc}^{(1)}$ in Ref.~\cite{ChapmanStijn2023} which can be reproduced with a lower-order expansion of the SCs ($\mathrm{S}\leq4$).  Switching to the rotating frame, so $\hat a\rightarrow\hat ae^{-i\Delta_a t}$, $\hat b\rightarrow\hat be^{-i\delta t}$, and $\hat c\rightarrow\hat ce^{-i\delta t}$, we obtain
\begin{align}
\begin{split}
\hat{H}(t)=\left(g_{bc}\hat{b}^\dag\hat{c}+g_{ac}\hat{c}^\dag\hat{a}^\dag e^{i\Tilde{\delta}t}\right.&\left.+g_{ab}\hat{b}^\dag\hat{a}^\dag e^{i\Tilde{\delta}t}+\text{H.c.}\right)\\&+\chi_{bc}\hat b^\dag\hat b\hat c^\dag\hat c,
\end{split}
\end{align}
where $\Tilde{\delta}=\delta+\Delta_a$. Then, by applying a separate Schrieffer-Wolff procedure for time scales $1/\Tilde{\delta}$, the effective Hamiltonian reads
\begin{align}\label{eq_app:eff_ham_3modes}
\begin{split}
\hat{\mathcal{H}}_{\mathrm{eff}}&\approx-\frac{2g_{ab}^2}{\Tilde{\delta}}\hat{b}^\dag\hat{b}-\frac{2g_{ac}^2}{\Tilde{\delta}}\hat{c}^\dag\hat{c}-\frac{2(g_{ac}^2+g_{ab}^2)}{\Tilde{\delta}}\hat{a}^\dag\hat{a}\\&\quad+g_\mathrm{BS}(\hat{b}^\dag\hat{c}+\hat{b}\hat{c}^\dag)+\chi_{bc}\hat b^\dag\hat b\hat c^\dag\hat c
\end{split}
\end{align}
with $g_{ab}/\Tilde{\delta},\, g_{ab}/\Tilde{\delta}\ll1$, with renormalized beam-splitter interaction $g_\mathrm{BS}=g_{bc}-2g_{ab}g_{ac}/\Tilde{\delta}$, two-mode squeezing amplitudes $g_{ab}=g_{ab}^{(0)}+g_{ab}^{(1)}$, $g_{ac}=g_{ac}^{(0)}+g_{ac}^{(1)}$, and detuning of the coupler mode $\Delta_a=\Delta_a^{(0)}+\Delta_a^{(1)}$. Notably, amplitudes $g_{ab}^{(1)}\propto\xi_b$ and $g_{ac}^{(1)}\propto\xi_c$ (indices $n_b=n_c=0$ and $l_b+l_c=1$) ensuring the correction to $g_\mathrm{BS}$ is of the same order in terms of $\xi_c$, $\xi_b$.

\paragraph*{Simulation parameter} In the optimization routine, we impose the following constraints: $g_{ab}/\Tilde{\delta},\, \added{g_{ac}/\Tilde{\delta}}\leq0.25$ to ensure the validity of the effective Hamiltonian in terms of Schrieffer-Wolff procedure for time scales $1/\Tilde{\delta}$; $\chi_{bc}\geq\SI{30}{\hertz}$ to prevent divergence of the algorithm near Kerr-free point. The energy parameters for SNAIL-based circuit designs are derived from Ref.~\cite{ChapmanStijn2023}: $E_J=\SI{86}{\giga\hertz}$, $E_C=\SI{177}{\mega\hertz}$. For SQUID-based circuits, we choose the same capacitive energy and $E_J=\SI{186}{\giga\hertz}$, $r_a=0.9$, and $r_b=0.1$. Common parameters are bare cavity frequencies $\omega_b/2\pi= \SI{2.976}{\giga\hertz}$, $\omega_c/2\pi= \SI{6.915}{\giga\hertz}$ and capacitive couplings
$g_b/2\pi= \SI{75.6}{\mega\hertz}$, $g_c/2\pi= \SI{134.9}{\mega\hertz}$. Additional limits for the parameters range: a bare coupler frequency $\omega_a/2\pi\in[\SI{4.5}{\giga\hertz}, \SI{6}{\giga\hertz}]$ and exclusion of flux-bias where multiphoton resonances are possible (defined by the denominators in the expressions for $g_\mathrm{BS}$ and $\xi_{bc}$, Eqs. (\ref{eq:ga_corr})-(\ref{eq:chi_corr})).

\bibliography{refs}

\begin{thebibliography}{73}%
\makeatletter
\providecommand \@ifxundefined [1]{%
 \@ifx{#1\undefined}
}%
\providecommand \@ifnum [1]{%
 \ifnum #1\expandafter \@firstoftwo
 \else \expandafter \@secondoftwo
 \fi
}%
\providecommand \@ifx [1]{%
 \ifx #1\expandafter \@firstoftwo
 \else \expandafter \@secondoftwo
 \fi
}%
\providecommand \natexlab [1]{#1}%
\providecommand \enquote  [1]{``#1''}%
\providecommand \bibnamefont  [1]{#1}%
\providecommand \bibfnamefont [1]{#1}%
\providecommand \citenamefont [1]{#1}%
\providecommand \href@noop [0]{\@secondoftwo}%
\providecommand \href [0]{\begingroup \@sanitize@url \@href}%
\providecommand \@href[1]{\@@startlink{#1}\@@href}%
\providecommand \@@href[1]{\endgroup#1\@@endlink}%
\providecommand \@sanitize@url [0]{\catcode `\\12\catcode `\$12\catcode `\&12\catcode `\#12\catcode `\^12\catcode `\_12\catcode `\%12\relax}%
\providecommand \@@startlink[1]{}%
\providecommand \@@endlink[0]{}%
\providecommand \url  [0]{\begingroup\@sanitize@url \@url }%
\providecommand \@url [1]{\endgroup\@href {#1}{\urlprefix }}%
\providecommand \urlprefix  [0]{URL }%
\providecommand \Eprint [0]{\href }%
\providecommand \doibase [0]{https://doi.org/}%
\providecommand \selectlanguage [0]{\@gobble}%
\providecommand \bibinfo  [0]{\@secondoftwo}%
\providecommand \bibfield  [0]{\@secondoftwo}%
\providecommand \translation [1]{[#1]}%
\providecommand \BibitemOpen [0]{}%
\providecommand \bibitemStop [0]{}%
\providecommand \bibitemNoStop [0]{.\EOS\space}%
\providecommand \EOS [0]{\spacefactor3000\relax}%
\providecommand \BibitemShut  [1]{\csname bibitem#1\endcsname}%
\let\auto@bib@innerbib\@empty
\bibitem [{\citenamefont {Blais}\ \emph {et~al.}(2004)\citenamefont {Blais}, \citenamefont {Huang}, \citenamefont {Wallraff}, \citenamefont {Girvin},\ and\ \citenamefont {Schoelkopf}}]{Blais2004}%
  \BibitemOpen
  \bibfield  {author} {\bibinfo {author} {\bibfnamefont {A.}~\bibnamefont {Blais}}, \bibinfo {author} {\bibfnamefont {R.-S.}\ \bibnamefont {Huang}}, \bibinfo {author} {\bibfnamefont {A.}~\bibnamefont {Wallraff}}, \bibinfo {author} {\bibfnamefont {S.~M.}\ \bibnamefont {Girvin}},\ and\ \bibinfo {author} {\bibfnamefont {R.~J.}\ \bibnamefont {Schoelkopf}},\ }\bibfield  {title} {\bibinfo {title} {Cavity quantum electrodynamics for superconducting electrical circuits: An architecture for quantum computation},\ }\href {https://doi.org/10.1103/PhysRevA.69.062320} {\bibfield  {journal} {\bibinfo  {journal} {Phys. Rev. A}\ }\textbf {\bibinfo {volume} {69}},\ \bibinfo {pages} {062320} (\bibinfo {year} {2004})}\BibitemShut {NoStop}%
\bibitem [{\citenamefont {Wallraff}\ \emph {et~al.}(2004)\citenamefont {Wallraff}, \citenamefont {Schuster}, \citenamefont {Blais}, \citenamefont {Frunzio}, \citenamefont {Huang}, \citenamefont {Majer}, \citenamefont {Kumar}, \citenamefont {Girvin},\ and\ \citenamefont {Schoelkopf}}]{Wallraff_cQED_2004}%
  \BibitemOpen
  \bibfield  {author} {\bibinfo {author} {\bibfnamefont {A.}~\bibnamefont {Wallraff}}, \bibinfo {author} {\bibfnamefont {D.~I.}\ \bibnamefont {Schuster}}, \bibinfo {author} {\bibfnamefont {A.}~\bibnamefont {Blais}}, \bibinfo {author} {\bibfnamefont {L.}~\bibnamefont {Frunzio}}, \bibinfo {author} {\bibfnamefont {R.~S.}\ \bibnamefont {Huang}}, \bibinfo {author} {\bibfnamefont {J.}~\bibnamefont {Majer}}, \bibinfo {author} {\bibfnamefont {S.}~\bibnamefont {Kumar}}, \bibinfo {author} {\bibfnamefont {S.~M.}\ \bibnamefont {Girvin}},\ and\ \bibinfo {author} {\bibfnamefont {R.~J.}\ \bibnamefont {Schoelkopf}},\ }\bibfield  {title} {\bibinfo {title} {Strong coupling of a single photon to a superconducting qubit using circuit quantum electrodynamics},\ }\href {https://doi.org/10.1038/nature02851} {\bibfield  {journal} {\bibinfo  {journal} {Nature}\ }\textbf {\bibinfo {volume} {431}},\ \bibinfo {pages} {162} (\bibinfo {year} {2004})}\BibitemShut {NoStop}%
\bibitem [{\citenamefont {Blais}\ \emph {et~al.}(2021)\citenamefont {Blais}, \citenamefont {Grimsmo}, \citenamefont {Girvin},\ and\ \citenamefont {Wallraff}}]{Blais2021}%
  \BibitemOpen
  \bibfield  {author} {\bibinfo {author} {\bibfnamefont {A.}~\bibnamefont {Blais}}, \bibinfo {author} {\bibfnamefont {A.~L.}\ \bibnamefont {Grimsmo}}, \bibinfo {author} {\bibfnamefont {S.~M.}\ \bibnamefont {Girvin}},\ and\ \bibinfo {author} {\bibfnamefont {A.}~\bibnamefont {Wallraff}},\ }\bibfield  {title} {\bibinfo {title} {Circuit quantum electrodynamics},\ }\href {https://doi.org/10.1103/RevModPhys.93.025005} {\bibfield  {journal} {\bibinfo  {journal} {Rev. Mod. Phys.}\ }\textbf {\bibinfo {volume} {93}},\ \bibinfo {pages} {025005} (\bibinfo {year} {2021})}\BibitemShut {NoStop}%
\bibitem [{\citenamefont {Aoki}\ \emph {et~al.}(2024)\citenamefont {Aoki}, \citenamefont {Kanao}, \citenamefont {Goto}, \citenamefont {Kawabata},\ and\ \citenamefont {Masuda}}]{Aoki2024}%
  \BibitemOpen
  \bibfield  {author} {\bibinfo {author} {\bibfnamefont {T.}~\bibnamefont {Aoki}}, \bibinfo {author} {\bibfnamefont {T.}~\bibnamefont {Kanao}}, \bibinfo {author} {\bibfnamefont {H.}~\bibnamefont {Goto}}, \bibinfo {author} {\bibfnamefont {S.}~\bibnamefont {Kawabata}},\ and\ \bibinfo {author} {\bibfnamefont {S.}~\bibnamefont {Masuda}},\ }\bibfield  {title} {\bibinfo {title} {Control of the $zz$ coupling between kerr cat qubits via transmon couplers},\ }\href {https://doi.org/10.1103/PhysRevApplied.21.014030} {\bibfield  {journal} {\bibinfo  {journal} {Phys. Rev. Appl.}\ }\textbf {\bibinfo {volume} {21}},\ \bibinfo {pages} {014030} (\bibinfo {year} {2024})}\BibitemShut {NoStop}%
\bibitem [{\citenamefont {Hua}\ \emph {et~al.}(2024)\citenamefont {Hua}, \citenamefont {Xu}, \citenamefont {Wang}, \citenamefont {Ma}, \citenamefont {Zhou}, \citenamefont {Cai}, \citenamefont {Ai}, \citenamefont {xi~Liu}, \citenamefont {Li}, \citenamefont {Zou},\ and\ \citenamefont {Sun}}]{hua2024}%
  \BibitemOpen
  \bibfield  {author} {\bibinfo {author} {\bibfnamefont {Z.}~\bibnamefont {Hua}}, \bibinfo {author} {\bibfnamefont {Y.}~\bibnamefont {Xu}}, \bibinfo {author} {\bibfnamefont {W.}~\bibnamefont {Wang}}, \bibinfo {author} {\bibfnamefont {Y.}~\bibnamefont {Ma}}, \bibinfo {author} {\bibfnamefont {J.}~\bibnamefont {Zhou}}, \bibinfo {author} {\bibfnamefont {W.}~\bibnamefont {Cai}}, \bibinfo {author} {\bibfnamefont {H.}~\bibnamefont {Ai}}, \bibinfo {author} {\bibfnamefont {Y.}~\bibnamefont {xi~Liu}}, \bibinfo {author} {\bibfnamefont {M.}~\bibnamefont {Li}}, \bibinfo {author} {\bibfnamefont {C.-L.}\ \bibnamefont {Zou}},\ and\ \bibinfo {author} {\bibfnamefont {L.}~\bibnamefont {Sun}},\ }\href {https://arxiv.org/abs/2410.06904} {\bibinfo {title} {Engineering the nonlinearity of bosonic modes with a multi-loop squid}} (\bibinfo {year} {2024}),\ \Eprint {https://arxiv.org/abs/2410.06904} {arXiv:2410.06904 [quant-ph]} \BibitemShut {NoStop}%
\bibitem [{\citenamefont {Bhandari}\ \emph {et~al.}(2024)\citenamefont {Bhandari}, \citenamefont {Huang}, \citenamefont {Hajr}, \citenamefont {Yanik}, \citenamefont {Qing}, \citenamefont {Wang}, \citenamefont {Santiago}, \citenamefont {Dressel}, \citenamefont {Siddiqi},\ and\ \citenamefont {Jordan}}]{bhandari2024symmetricallythreadedsquidsgeneration}%
  \BibitemOpen
  \bibfield  {author} {\bibinfo {author} {\bibfnamefont {B.}~\bibnamefont {Bhandari}}, \bibinfo {author} {\bibfnamefont {I.}~\bibnamefont {Huang}}, \bibinfo {author} {\bibfnamefont {A.}~\bibnamefont {Hajr}}, \bibinfo {author} {\bibfnamefont {K.}~\bibnamefont {Yanik}}, \bibinfo {author} {\bibfnamefont {B.}~\bibnamefont {Qing}}, \bibinfo {author} {\bibfnamefont {K.}~\bibnamefont {Wang}}, \bibinfo {author} {\bibfnamefont {D.~I.}\ \bibnamefont {Santiago}}, \bibinfo {author} {\bibfnamefont {J.}~\bibnamefont {Dressel}}, \bibinfo {author} {\bibfnamefont {I.}~\bibnamefont {Siddiqi}},\ and\ \bibinfo {author} {\bibfnamefont {A.~N.}\ \bibnamefont {Jordan}},\ }\href {https://arxiv.org/abs/2405.11375} {\bibinfo {title} {Symmetrically threaded squids as next generation kerr-cat qubits}} (\bibinfo {year} {2024}),\ \Eprint {https://arxiv.org/abs/2405.11375} {arXiv:2405.11375 [quant-ph]} \BibitemShut {NoStop}%
\bibitem [{\citenamefont {Royer}\ \emph {et~al.}(2018)\citenamefont {Royer}, \citenamefont {Puri},\ and\ \citenamefont {Blais}}]{Royer2018}%
  \BibitemOpen
  \bibfield  {author} {\bibinfo {author} {\bibfnamefont {B.}~\bibnamefont {Royer}}, \bibinfo {author} {\bibfnamefont {S.}~\bibnamefont {Puri}},\ and\ \bibinfo {author} {\bibfnamefont {A.}~\bibnamefont {Blais}},\ }\bibfield  {title} {\bibinfo {title} {Qubit parity measurement by parametric driving in circuit qed},\ }\href {https://doi.org/10.1126/sciadv.aau1695} {\bibfield  {journal} {\bibinfo  {journal} {Science Advances}\ }\textbf {\bibinfo {volume} {4}},\ \bibinfo {pages} {eaau1695} (\bibinfo {year} {2018})}\BibitemShut {NoStop}%
\bibitem [{\citenamefont {Lu}\ \emph {et~al.}(2023)\citenamefont {Lu}, \citenamefont {Maiti}, \citenamefont {Garmon}, \citenamefont {Ganjam}, \citenamefont {Zhang}, \citenamefont {Claes}, \citenamefont {Frunzio}, \citenamefont {Girvin},\ and\ \citenamefont {Schoelkopf}}]{Lu2023}%
  \BibitemOpen
  \bibfield  {author} {\bibinfo {author} {\bibfnamefont {Y.}~\bibnamefont {Lu}}, \bibinfo {author} {\bibfnamefont {A.}~\bibnamefont {Maiti}}, \bibinfo {author} {\bibfnamefont {J.~W.~O.}\ \bibnamefont {Garmon}}, \bibinfo {author} {\bibfnamefont {S.}~\bibnamefont {Ganjam}}, \bibinfo {author} {\bibfnamefont {Y.}~\bibnamefont {Zhang}}, \bibinfo {author} {\bibfnamefont {J.}~\bibnamefont {Claes}}, \bibinfo {author} {\bibfnamefont {L.}~\bibnamefont {Frunzio}}, \bibinfo {author} {\bibfnamefont {S.~M.}\ \bibnamefont {Girvin}},\ and\ \bibinfo {author} {\bibfnamefont {R.~J.}\ \bibnamefont {Schoelkopf}},\ }\bibfield  {title} {\bibinfo {title} {High-fidelity parametric beamsplitting with a parity-protected converter},\ }\href {https://doi.org/10.1038/s41467-023-41104-0} {\bibfield  {journal} {\bibinfo  {journal} {Nature Communications}\ }\textbf {\bibinfo {volume} {14}},\ \bibinfo {pages} {5767} (\bibinfo {year} {2023})}\BibitemShut {NoStop}%
\bibitem [{\citenamefont {Leghtas}\ \emph {et~al.}(2015)\citenamefont {Leghtas}, \citenamefont {Touzard}, \citenamefont {Pop}, \citenamefont {Kou}, \citenamefont {Vlastakis}, \citenamefont {Petrenko}, \citenamefont {Sliwa}, \citenamefont {Narla}, \citenamefont {Shankar}, \citenamefont {Hatridge}, \citenamefont {Reagor}, \citenamefont {Frunzio}, \citenamefont {Schoelkopf}, \citenamefont {Mirrahimi},\ and\ \citenamefont {Devoret}}]{Leghtas2015}%
  \BibitemOpen
  \bibfield  {author} {\bibinfo {author} {\bibfnamefont {Z.}~\bibnamefont {Leghtas}}, \bibinfo {author} {\bibfnamefont {S.}~\bibnamefont {Touzard}}, \bibinfo {author} {\bibfnamefont {I.~M.}\ \bibnamefont {Pop}}, \bibinfo {author} {\bibfnamefont {A.}~\bibnamefont {Kou}}, \bibinfo {author} {\bibfnamefont {B.}~\bibnamefont {Vlastakis}}, \bibinfo {author} {\bibfnamefont {A.}~\bibnamefont {Petrenko}}, \bibinfo {author} {\bibfnamefont {K.~M.}\ \bibnamefont {Sliwa}}, \bibinfo {author} {\bibfnamefont {A.}~\bibnamefont {Narla}}, \bibinfo {author} {\bibfnamefont {S.}~\bibnamefont {Shankar}}, \bibinfo {author} {\bibfnamefont {M.~J.}\ \bibnamefont {Hatridge}}, \bibinfo {author} {\bibfnamefont {M.}~\bibnamefont {Reagor}}, \bibinfo {author} {\bibfnamefont {L.}~\bibnamefont {Frunzio}}, \bibinfo {author} {\bibfnamefont {R.~J.}\ \bibnamefont {Schoelkopf}}, \bibinfo {author} {\bibfnamefont {M.}~\bibnamefont {Mirrahimi}},\ and\ \bibinfo {author} {\bibfnamefont {M.~H.}\ \bibnamefont {Devoret}},\ }\bibfield  {title} {\bibinfo
  {title} {Confining the state of light to a quantum manifold by engineered two-photon loss},\ }\href {https://doi.org/10.1126/science.aaa2085} {\bibfield  {journal} {\bibinfo  {journal} {Science}\ }\textbf {\bibinfo {volume} {347}},\ \bibinfo {pages} {853} (\bibinfo {year} {2015})}\BibitemShut {NoStop}%
\bibitem [{\citenamefont {Xiao}\ \emph {et~al.}(2023)\citenamefont {Xiao}, \citenamefont {Venkatraman}, \citenamefont {Cortiñas}, \citenamefont {Chowdhury},\ and\ \citenamefont {Devoret}}]{xiao2023diagrammatic}%
  \BibitemOpen
  \bibfield  {author} {\bibinfo {author} {\bibfnamefont {X.}~\bibnamefont {Xiao}}, \bibinfo {author} {\bibfnamefont {J.}~\bibnamefont {Venkatraman}}, \bibinfo {author} {\bibfnamefont {R.~G.}\ \bibnamefont {Cortiñas}}, \bibinfo {author} {\bibfnamefont {S.}~\bibnamefont {Chowdhury}},\ and\ \bibinfo {author} {\bibfnamefont {M.~H.}\ \bibnamefont {Devoret}},\ }\href@noop {} {\bibinfo {title} {A diagrammatic method to compute the effective hamiltonian of driven nonlinear oscillators}} (\bibinfo {year} {2023}),\ \Eprint {https://arxiv.org/abs/2304.13656} {arXiv:2304.13656 [quant-ph]} \BibitemShut {NoStop}%
\bibitem [{\citenamefont {Hajr}\ \emph {et~al.}(2024)\citenamefont {Hajr}, \citenamefont {Qing}, \citenamefont {Wang}, \citenamefont {Koolstra}, \citenamefont {Pedramrazi}, \citenamefont {Kang}, \citenamefont {Chen}, \citenamefont {Nguyen}, \citenamefont {Junger}, \citenamefont {Goss}, \citenamefont {Huang}, \citenamefont {Bhandari}, \citenamefont {Frattini}, \citenamefont {Puri}, \citenamefont {Dressel}, \citenamefont {Jordan}, \citenamefont {Santiago},\ and\ \citenamefont {Siddiqi}}]{hajr2024}%
  \BibitemOpen
  \bibfield  {author} {\bibinfo {author} {\bibfnamefont {A.}~\bibnamefont {Hajr}}, \bibinfo {author} {\bibfnamefont {B.}~\bibnamefont {Qing}}, \bibinfo {author} {\bibfnamefont {K.}~\bibnamefont {Wang}}, \bibinfo {author} {\bibfnamefont {G.}~\bibnamefont {Koolstra}}, \bibinfo {author} {\bibfnamefont {Z.}~\bibnamefont {Pedramrazi}}, \bibinfo {author} {\bibfnamefont {Z.}~\bibnamefont {Kang}}, \bibinfo {author} {\bibfnamefont {L.}~\bibnamefont {Chen}}, \bibinfo {author} {\bibfnamefont {L.~B.}\ \bibnamefont {Nguyen}}, \bibinfo {author} {\bibfnamefont {C.}~\bibnamefont {Junger}}, \bibinfo {author} {\bibfnamefont {N.}~\bibnamefont {Goss}}, \bibinfo {author} {\bibfnamefont {I.}~\bibnamefont {Huang}}, \bibinfo {author} {\bibfnamefont {B.}~\bibnamefont {Bhandari}}, \bibinfo {author} {\bibfnamefont {N.~E.}\ \bibnamefont {Frattini}}, \bibinfo {author} {\bibfnamefont {S.}~\bibnamefont {Puri}}, \bibinfo {author} {\bibfnamefont {J.}~\bibnamefont {Dressel}}, \bibinfo {author} {\bibfnamefont {A.~N.}\ \bibnamefont {Jordan}},
  \bibinfo {author} {\bibfnamefont {D.}~\bibnamefont {Santiago}},\ and\ \bibinfo {author} {\bibfnamefont {I.}~\bibnamefont {Siddiqi}},\ }\href {https://arxiv.org/abs/2404.16697} {\bibinfo {title} {High-coherence kerr-cat qubit in 2d architecture}} (\bibinfo {year} {2024}),\ \Eprint {https://arxiv.org/abs/2404.16697} {arXiv:2404.16697 [quant-ph]} \BibitemShut {NoStop}%
\bibitem [{\citenamefont {Shillito}\ \emph {et~al.}(2022)\citenamefont {Shillito}, \citenamefont {Petrescu}, \citenamefont {Cohen}, \citenamefont {Beall}, \citenamefont {Hauru}, \citenamefont {Ganahl}, \citenamefont {Lewis}, \citenamefont {Vidal},\ and\ \citenamefont {Blais}}]{Shillito2022}%
  \BibitemOpen
  \bibfield  {author} {\bibinfo {author} {\bibfnamefont {R.}~\bibnamefont {Shillito}}, \bibinfo {author} {\bibfnamefont {A.}~\bibnamefont {Petrescu}}, \bibinfo {author} {\bibfnamefont {J.}~\bibnamefont {Cohen}}, \bibinfo {author} {\bibfnamefont {J.}~\bibnamefont {Beall}}, \bibinfo {author} {\bibfnamefont {M.}~\bibnamefont {Hauru}}, \bibinfo {author} {\bibfnamefont {M.}~\bibnamefont {Ganahl}}, \bibinfo {author} {\bibfnamefont {A.~G.}\ \bibnamefont {Lewis}}, \bibinfo {author} {\bibfnamefont {G.}~\bibnamefont {Vidal}},\ and\ \bibinfo {author} {\bibfnamefont {A.}~\bibnamefont {Blais}},\ }\bibfield  {title} {\bibinfo {title} {Dynamics of transmon ionization},\ }\href {https://doi.org/10.1103/PhysRevApplied.18.034031} {\bibfield  {journal} {\bibinfo  {journal} {Phys. Rev. Appl.}\ }\textbf {\bibinfo {volume} {18}},\ \bibinfo {pages} {034031} (\bibinfo {year} {2022})}\BibitemShut {NoStop}%
\bibitem [{\citenamefont {Khezri}\ \emph {et~al.}(2023)\citenamefont {Khezri}, \citenamefont {Opremcak}, \citenamefont {Chen}, \citenamefont {Miao}, \citenamefont {McEwen}, \citenamefont {Bengtsson}, \citenamefont {White}, \citenamefont {Naaman}, \citenamefont {Sank}, \citenamefont {Korotkov}, \citenamefont {Chen},\ and\ \citenamefont {Smelyanskiy}}]{googleRWA}%
  \BibitemOpen
  \bibfield  {author} {\bibinfo {author} {\bibfnamefont {M.}~\bibnamefont {Khezri}}, \bibinfo {author} {\bibfnamefont {A.}~\bibnamefont {Opremcak}}, \bibinfo {author} {\bibfnamefont {Z.}~\bibnamefont {Chen}}, \bibinfo {author} {\bibfnamefont {K.~C.}\ \bibnamefont {Miao}}, \bibinfo {author} {\bibfnamefont {M.}~\bibnamefont {McEwen}}, \bibinfo {author} {\bibfnamefont {A.}~\bibnamefont {Bengtsson}}, \bibinfo {author} {\bibfnamefont {T.}~\bibnamefont {White}}, \bibinfo {author} {\bibfnamefont {O.}~\bibnamefont {Naaman}}, \bibinfo {author} {\bibfnamefont {D.}~\bibnamefont {Sank}}, \bibinfo {author} {\bibfnamefont {A.~N.}\ \bibnamefont {Korotkov}}, \bibinfo {author} {\bibfnamefont {Y.}~\bibnamefont {Chen}},\ and\ \bibinfo {author} {\bibfnamefont {V.}~\bibnamefont {Smelyanskiy}},\ }\bibfield  {title} {\bibinfo {title} {Measurement-induced state transitions in a superconducting qubit: Within the rotating-wave approximation},\ }\href {https://doi.org/10.1103/PhysRevApplied.20.054008} {\bibfield  {journal} {\bibinfo
  {journal} {Phys. Rev. Appl.}\ }\textbf {\bibinfo {volume} {20}},\ \bibinfo {pages} {054008} (\bibinfo {year} {2023})}\BibitemShut {NoStop}%
\bibitem [{\citenamefont {Sank}\ \emph {et~al.}(2016)\citenamefont {Sank}, \citenamefont {Chen}, \citenamefont {Khezri}, \citenamefont {Kelly}, \citenamefont {Barends}, \citenamefont {Campbell}, \citenamefont {Chen}, \citenamefont {Chiaro}, \citenamefont {Dunsworth}, \citenamefont {Fowler}, \citenamefont {Jeffrey}, \citenamefont {Lucero}, \citenamefont {Megrant}, \citenamefont {Mutus}, \citenamefont {Neeley}, \citenamefont {Neill}, \citenamefont {O'Malley}, \citenamefont {Quintana}, \citenamefont {Roushan}, \citenamefont {Vainsencher}, \citenamefont {White}, \citenamefont {Wenner}, \citenamefont {Korotkov},\ and\ \citenamefont {Martinis}}]{google_beyondRWA}%
  \BibitemOpen
  \bibfield  {author} {\bibinfo {author} {\bibfnamefont {D.}~\bibnamefont {Sank}}, \bibinfo {author} {\bibfnamefont {Z.}~\bibnamefont {Chen}}, \bibinfo {author} {\bibfnamefont {M.}~\bibnamefont {Khezri}}, \bibinfo {author} {\bibfnamefont {J.}~\bibnamefont {Kelly}}, \bibinfo {author} {\bibfnamefont {R.}~\bibnamefont {Barends}}, \bibinfo {author} {\bibfnamefont {B.}~\bibnamefont {Campbell}}, \bibinfo {author} {\bibfnamefont {Y.}~\bibnamefont {Chen}}, \bibinfo {author} {\bibfnamefont {B.}~\bibnamefont {Chiaro}}, \bibinfo {author} {\bibfnamefont {A.}~\bibnamefont {Dunsworth}}, \bibinfo {author} {\bibfnamefont {A.}~\bibnamefont {Fowler}}, \bibinfo {author} {\bibfnamefont {E.}~\bibnamefont {Jeffrey}}, \bibinfo {author} {\bibfnamefont {E.}~\bibnamefont {Lucero}}, \bibinfo {author} {\bibfnamefont {A.}~\bibnamefont {Megrant}}, \bibinfo {author} {\bibfnamefont {J.}~\bibnamefont {Mutus}}, \bibinfo {author} {\bibfnamefont {M.}~\bibnamefont {Neeley}}, \bibinfo {author} {\bibfnamefont {C.}~\bibnamefont {Neill}}, \bibinfo
  {author} {\bibfnamefont {P.~J.~J.}\ \bibnamefont {O'Malley}}, \bibinfo {author} {\bibfnamefont {C.}~\bibnamefont {Quintana}}, \bibinfo {author} {\bibfnamefont {P.}~\bibnamefont {Roushan}}, \bibinfo {author} {\bibfnamefont {A.}~\bibnamefont {Vainsencher}}, \bibinfo {author} {\bibfnamefont {T.}~\bibnamefont {White}}, \bibinfo {author} {\bibfnamefont {J.}~\bibnamefont {Wenner}}, \bibinfo {author} {\bibfnamefont {A.~N.}\ \bibnamefont {Korotkov}},\ and\ \bibinfo {author} {\bibfnamefont {J.~M.}\ \bibnamefont {Martinis}},\ }\bibfield  {title} {\bibinfo {title} {Measurement-induced state transitions in a superconducting qubit: Beyond the rotating wave approximation},\ }\href {https://doi.org/10.1103/PhysRevLett.117.190503} {\bibfield  {journal} {\bibinfo  {journal} {Phys. Rev. Lett.}\ }\textbf {\bibinfo {volume} {117}},\ \bibinfo {pages} {190503} (\bibinfo {year} {2016})}\BibitemShut {NoStop}%
\bibitem [{\citenamefont {Dumas}\ \emph {et~al.}(2024)\citenamefont {Dumas}, \citenamefont {Groleau-Par\'e}, \citenamefont {McDonald}, \citenamefont {Mu\~noz Arias}, \citenamefont {Lled\'o}, \citenamefont {D'Anjou},\ and\ \citenamefont {Blais}}]{dumas24}%
  \BibitemOpen
  \bibfield  {author} {\bibinfo {author} {\bibfnamefont {M.~F.}\ \bibnamefont {Dumas}}, \bibinfo {author} {\bibfnamefont {B.}~\bibnamefont {Groleau-Par\'e}}, \bibinfo {author} {\bibfnamefont {A.}~\bibnamefont {McDonald}}, \bibinfo {author} {\bibfnamefont {M.~H.}\ \bibnamefont {Mu\~noz Arias}}, \bibinfo {author} {\bibfnamefont {C.}~\bibnamefont {Lled\'o}}, \bibinfo {author} {\bibfnamefont {B.}~\bibnamefont {D'Anjou}},\ and\ \bibinfo {author} {\bibfnamefont {A.}~\bibnamefont {Blais}},\ }\bibfield  {title} {\bibinfo {title} {Measurement-induced transmon ionization},\ }\href {https://doi.org/10.1103/PhysRevX.14.041023} {\bibfield  {journal} {\bibinfo  {journal} {Phys. Rev. X}\ }\textbf {\bibinfo {volume} {14}},\ \bibinfo {pages} {041023} (\bibinfo {year} {2024})}\BibitemShut {NoStop}%
\bibitem [{\citenamefont {Willsch}\ \emph {et~al.}(2024)\citenamefont {Willsch}, \citenamefont {Rieger}, \citenamefont {Winkel}, \citenamefont {Willsch}, \citenamefont {Dickel}, \citenamefont {Krause}, \citenamefont {Ando}, \citenamefont {Lescanne}, \citenamefont {Leghtas}, \citenamefont {Bronn}, \citenamefont {Deb}, \citenamefont {Lanes}, \citenamefont {Minev}, \citenamefont {Dennig}, \citenamefont {Geisert}, \citenamefont {G{\"u}nzler}, \citenamefont {Ihssen}, \citenamefont {Paluch}, \citenamefont {Reisinger}, \citenamefont {Hanna}, \citenamefont {Bae}, \citenamefont {Sch{\"u}ffelgen}, \citenamefont {Gr{\"u}tzmacher}, \citenamefont {Buimaga-Iarinca}, \citenamefont {Morari}, \citenamefont {Wernsdorfer}, \citenamefont {DiVincenzo}, \citenamefont {Michielsen}, \citenamefont {Catelani},\ and\ \citenamefont {Pop}}]{Willsch2024}%
  \BibitemOpen
  \bibfield  {author} {\bibinfo {author} {\bibfnamefont {D.}~\bibnamefont {Willsch}}, \bibinfo {author} {\bibfnamefont {D.}~\bibnamefont {Rieger}}, \bibinfo {author} {\bibfnamefont {P.}~\bibnamefont {Winkel}}, \bibinfo {author} {\bibfnamefont {M.}~\bibnamefont {Willsch}}, \bibinfo {author} {\bibfnamefont {C.}~\bibnamefont {Dickel}}, \bibinfo {author} {\bibfnamefont {J.}~\bibnamefont {Krause}}, \bibinfo {author} {\bibfnamefont {Y.}~\bibnamefont {Ando}}, \bibinfo {author} {\bibfnamefont {R.}~\bibnamefont {Lescanne}}, \bibinfo {author} {\bibfnamefont {Z.}~\bibnamefont {Leghtas}}, \bibinfo {author} {\bibfnamefont {N.~T.}\ \bibnamefont {Bronn}}, \bibinfo {author} {\bibfnamefont {P.}~\bibnamefont {Deb}}, \bibinfo {author} {\bibfnamefont {O.}~\bibnamefont {Lanes}}, \bibinfo {author} {\bibfnamefont {Z.~K.}\ \bibnamefont {Minev}}, \bibinfo {author} {\bibfnamefont {B.}~\bibnamefont {Dennig}}, \bibinfo {author} {\bibfnamefont {S.}~\bibnamefont {Geisert}}, \bibinfo {author} {\bibfnamefont {S.}~\bibnamefont
  {G{\"u}nzler}}, \bibinfo {author} {\bibfnamefont {S.}~\bibnamefont {Ihssen}}, \bibinfo {author} {\bibfnamefont {P.}~\bibnamefont {Paluch}}, \bibinfo {author} {\bibfnamefont {T.}~\bibnamefont {Reisinger}}, \bibinfo {author} {\bibfnamefont {R.}~\bibnamefont {Hanna}}, \bibinfo {author} {\bibfnamefont {J.~H.}\ \bibnamefont {Bae}}, \bibinfo {author} {\bibfnamefont {P.}~\bibnamefont {Sch{\"u}ffelgen}}, \bibinfo {author} {\bibfnamefont {D.}~\bibnamefont {Gr{\"u}tzmacher}}, \bibinfo {author} {\bibfnamefont {L.}~\bibnamefont {Buimaga-Iarinca}}, \bibinfo {author} {\bibfnamefont {C.}~\bibnamefont {Morari}}, \bibinfo {author} {\bibfnamefont {W.}~\bibnamefont {Wernsdorfer}}, \bibinfo {author} {\bibfnamefont {D.~P.}\ \bibnamefont {DiVincenzo}}, \bibinfo {author} {\bibfnamefont {K.}~\bibnamefont {Michielsen}}, \bibinfo {author} {\bibfnamefont {G.}~\bibnamefont {Catelani}},\ and\ \bibinfo {author} {\bibfnamefont {I.~M.}\ \bibnamefont {Pop}},\ }\bibfield  {title} {\bibinfo {title} {Observation of josephson harmonics in
  tunnel junctions},\ }\href {https://doi.org/10.1038/s41567-024-02400-8} {\bibfield  {journal} {\bibinfo  {journal} {Nature Physics}\ }\textbf {\bibinfo {volume} {20}},\ \bibinfo {pages} {815} (\bibinfo {year} {2024})}\BibitemShut {NoStop}%
\bibitem [{\citenamefont {Wang}\ \emph {et~al.}(2024)\citenamefont {Wang}, \citenamefont {Parker}, \citenamefont {Champion},\ and\ \citenamefont {Blok}}]{wang2024systematicstudyhighejec}%
  \BibitemOpen
  \bibfield  {author} {\bibinfo {author} {\bibfnamefont {Z.}~\bibnamefont {Wang}}, \bibinfo {author} {\bibfnamefont {R.~W.}\ \bibnamefont {Parker}}, \bibinfo {author} {\bibfnamefont {E.}~\bibnamefont {Champion}},\ and\ \bibinfo {author} {\bibfnamefont {M.~S.}\ \bibnamefont {Blok}},\ }\href {https://arxiv.org/abs/2407.17407} {\bibinfo {title} {Systematic study of high $e_j/e_c$ transmon qudits up to $d = 12$}} (\bibinfo {year} {2024}),\ \Eprint {https://arxiv.org/abs/2407.17407} {arXiv:2407.17407 [quant-ph]} \BibitemShut {NoStop}%
\bibitem [{\citenamefont {Grimm}\ \emph {et~al.}(2020)\citenamefont {Grimm}, \citenamefont {Frattini}, \citenamefont {Puri}, \citenamefont {Mundhada}, \citenamefont {Touzard}, \citenamefont {Mirrahimi}, \citenamefont {Girvin}, \citenamefont {Shankar},\ and\ \citenamefont {Devoret}}]{Grimm2020}%
  \BibitemOpen
  \bibfield  {author} {\bibinfo {author} {\bibfnamefont {A.}~\bibnamefont {Grimm}}, \bibinfo {author} {\bibfnamefont {N.~E.}\ \bibnamefont {Frattini}}, \bibinfo {author} {\bibfnamefont {S.}~\bibnamefont {Puri}}, \bibinfo {author} {\bibfnamefont {S.~O.}\ \bibnamefont {Mundhada}}, \bibinfo {author} {\bibfnamefont {S.}~\bibnamefont {Touzard}}, \bibinfo {author} {\bibfnamefont {M.}~\bibnamefont {Mirrahimi}}, \bibinfo {author} {\bibfnamefont {S.~M.}\ \bibnamefont {Girvin}}, \bibinfo {author} {\bibfnamefont {S.}~\bibnamefont {Shankar}},\ and\ \bibinfo {author} {\bibfnamefont {M.~H.}\ \bibnamefont {Devoret}},\ }\bibfield  {title} {\bibinfo {title} {Stabilization and operation of a kerr-cat qubit},\ }\href {https://doi.org/10.1038/s41586-020-2587-z} {\bibfield  {journal} {\bibinfo  {journal} {Nature}\ }\textbf {\bibinfo {volume} {584}},\ \bibinfo {pages} {205} (\bibinfo {year} {2020})}\BibitemShut {NoStop}%
\bibitem [{\citenamefont {Venkatraman}\ \emph {et~al.}(2023)\citenamefont {Venkatraman}, \citenamefont {Cortinas}, \citenamefont {Frattini}, \citenamefont {Xiao},\ and\ \citenamefont {Devoret}}]{jaya2023interference}%
  \BibitemOpen
  \bibfield  {author} {\bibinfo {author} {\bibfnamefont {J.}~\bibnamefont {Venkatraman}}, \bibinfo {author} {\bibfnamefont {R.~G.}\ \bibnamefont {Cortinas}}, \bibinfo {author} {\bibfnamefont {N.~E.}\ \bibnamefont {Frattini}}, \bibinfo {author} {\bibfnamefont {X.}~\bibnamefont {Xiao}},\ and\ \bibinfo {author} {\bibfnamefont {M.~H.}\ \bibnamefont {Devoret}},\ }\href {https://arxiv.org/abs/2211.04605} {\bibinfo {title} {A driven quantum superconducting circuit with multiple tunable degeneracies}} (\bibinfo {year} {2023}),\ \Eprint {https://arxiv.org/abs/2211.04605} {arXiv:2211.04605 [quant-ph]} \BibitemShut {NoStop}%
\bibitem [{\citenamefont {Cohen}\ \emph {et~al.}(2023)\citenamefont {Cohen}, \citenamefont {Petrescu}, \citenamefont {Shillito},\ and\ \citenamefont {Blais}}]{Cohen23}%
  \BibitemOpen
  \bibfield  {author} {\bibinfo {author} {\bibfnamefont {J.}~\bibnamefont {Cohen}}, \bibinfo {author} {\bibfnamefont {A.}~\bibnamefont {Petrescu}}, \bibinfo {author} {\bibfnamefont {R.}~\bibnamefont {Shillito}},\ and\ \bibinfo {author} {\bibfnamefont {A.}~\bibnamefont {Blais}},\ }\bibfield  {title} {\bibinfo {title} {Reminiscence of classical chaos in driven transmons},\ }\href {https://doi.org/10.1103/PRXQuantum.4.020312} {\bibfield  {journal} {\bibinfo  {journal} {PRX Quantum}\ }\textbf {\bibinfo {volume} {4}},\ \bibinfo {pages} {020312} (\bibinfo {year} {2023})}\BibitemShut {NoStop}%
\bibitem [{\citenamefont {Chávez-Carlos}\ \emph {et~al.}(2024)\citenamefont {Chávez-Carlos}, \citenamefont {Reynoso}, \citenamefont {Cortiñas}, \citenamefont {García-Mata}, \citenamefont {Batista}, \citenamefont {Pérez-Bernal}, \citenamefont {Wisniacki},\ and\ \citenamefont {Santos}}]{chávezcarlos2024}%
  \BibitemOpen
  \bibfield  {author} {\bibinfo {author} {\bibfnamefont {J.}~\bibnamefont {Chávez-Carlos}}, \bibinfo {author} {\bibfnamefont {M.~A.~P.}\ \bibnamefont {Reynoso}}, \bibinfo {author} {\bibfnamefont {R.~G.}\ \bibnamefont {Cortiñas}}, \bibinfo {author} {\bibfnamefont {I.}~\bibnamefont {García-Mata}}, \bibinfo {author} {\bibfnamefont {V.~S.}\ \bibnamefont {Batista}}, \bibinfo {author} {\bibfnamefont {F.}~\bibnamefont {Pérez-Bernal}}, \bibinfo {author} {\bibfnamefont {D.~A.}\ \bibnamefont {Wisniacki}},\ and\ \bibinfo {author} {\bibfnamefont {L.~F.}\ \bibnamefont {Santos}},\ }\bibfield  {title} {\bibinfo {title} {Driving superconducting qubits into chaos},\ }\href {https://doi.org/10.1088/2058-9565/ad93fb} {\bibfield  {journal} {\bibinfo  {journal} {Quantum Science and Technology}\ }\textbf {\bibinfo {volume} {10}},\ \bibinfo {pages} {015039} (\bibinfo {year} {2024})}\BibitemShut {NoStop}%
\bibitem [{\citenamefont {Nigg}\ \emph {et~al.}(2012)\citenamefont {Nigg}, \citenamefont {Paik}, \citenamefont {Vlastakis}, \citenamefont {Kirchmair}, \citenamefont {Shankar}, \citenamefont {Frunzio}, \citenamefont {Devoret}, \citenamefont {Schoelkopf},\ and\ \citenamefont {Girvin}}]{BlackboxQuantization}%
  \BibitemOpen
  \bibfield  {author} {\bibinfo {author} {\bibfnamefont {S.~E.}\ \bibnamefont {Nigg}}, \bibinfo {author} {\bibfnamefont {H.}~\bibnamefont {Paik}}, \bibinfo {author} {\bibfnamefont {B.}~\bibnamefont {Vlastakis}}, \bibinfo {author} {\bibfnamefont {G.}~\bibnamefont {Kirchmair}}, \bibinfo {author} {\bibfnamefont {S.}~\bibnamefont {Shankar}}, \bibinfo {author} {\bibfnamefont {L.}~\bibnamefont {Frunzio}}, \bibinfo {author} {\bibfnamefont {M.~H.}\ \bibnamefont {Devoret}}, \bibinfo {author} {\bibfnamefont {R.~J.}\ \bibnamefont {Schoelkopf}},\ and\ \bibinfo {author} {\bibfnamefont {S.~M.}\ \bibnamefont {Girvin}},\ }\bibfield  {title} {\bibinfo {title} {Black-box superconducting circuit quantization},\ }\href {https://doi.org/10.1103/PhysRevLett.108.240502} {\bibfield  {journal} {\bibinfo  {journal} {Phys. Rev. Lett.}\ }\textbf {\bibinfo {volume} {108}},\ \bibinfo {pages} {240502} (\bibinfo {year} {2012})}\BibitemShut {NoStop}%
\bibitem [{\citenamefont {Miano}\ \emph {et~al.}(2023)\citenamefont {Miano}, \citenamefont {Joshi}, \citenamefont {Liu}, \citenamefont {Dai}, \citenamefont {Parakh}, \citenamefont {Frunzio},\ and\ \citenamefont {Devoret}}]{sandro2023}%
  \BibitemOpen
  \bibfield  {author} {\bibinfo {author} {\bibfnamefont {A.}~\bibnamefont {Miano}}, \bibinfo {author} {\bibfnamefont {V.}~\bibnamefont {Joshi}}, \bibinfo {author} {\bibfnamefont {G.}~\bibnamefont {Liu}}, \bibinfo {author} {\bibfnamefont {W.}~\bibnamefont {Dai}}, \bibinfo {author} {\bibfnamefont {P.}~\bibnamefont {Parakh}}, \bibinfo {author} {\bibfnamefont {L.}~\bibnamefont {Frunzio}},\ and\ \bibinfo {author} {\bibfnamefont {M.}~\bibnamefont {Devoret}},\ }\bibfield  {title} {\bibinfo {title} {Hamiltonian extrema of an arbitrary flux-biased josephson circuit},\ }\href {https://doi.org/10.1103/PRXQuantum.4.030324} {\bibfield  {journal} {\bibinfo  {journal} {PRX Quantum}\ }\textbf {\bibinfo {volume} {4}},\ \bibinfo {pages} {030324} (\bibinfo {year} {2023})}\BibitemShut {NoStop}%
\bibitem [{\citenamefont {Minev}\ \emph {et~al.}(2021)\citenamefont {Minev}, \citenamefont {Leghtas}, \citenamefont {Mundhada}, \citenamefont {Christakis}, \citenamefont {Pop},\ and\ \citenamefont {Devoret}}]{Minev2021}%
  \BibitemOpen
  \bibfield  {author} {\bibinfo {author} {\bibfnamefont {Z.~K.}\ \bibnamefont {Minev}}, \bibinfo {author} {\bibfnamefont {Z.}~\bibnamefont {Leghtas}}, \bibinfo {author} {\bibfnamefont {S.~O.}\ \bibnamefont {Mundhada}}, \bibinfo {author} {\bibfnamefont {L.}~\bibnamefont {Christakis}}, \bibinfo {author} {\bibfnamefont {I.~M.}\ \bibnamefont {Pop}},\ and\ \bibinfo {author} {\bibfnamefont {M.~H.}\ \bibnamefont {Devoret}},\ }\bibfield  {title} {\bibinfo {title} {Energy-participation quantization of josephson circuits},\ }\href {https://doi.org/10.1038/s41534-021-00461-8} {\bibfield  {journal} {\bibinfo  {journal} {npj Quantum Information}\ }\textbf {\bibinfo {volume} {7}},\ \bibinfo {pages} {131} (\bibinfo {year} {2021})}\BibitemShut {NoStop}%
\bibitem [{\citenamefont {Venkatraman}\ \emph {et~al.}(2024)\citenamefont {Venkatraman}, \citenamefont {Xiao}, \citenamefont {Corti\~nas},\ and\ \citenamefont {Devoret}}]{jaya2022lind}%
  \BibitemOpen
  \bibfield  {author} {\bibinfo {author} {\bibfnamefont {J.}~\bibnamefont {Venkatraman}}, \bibinfo {author} {\bibfnamefont {X.}~\bibnamefont {Xiao}}, \bibinfo {author} {\bibfnamefont {R.~G.}\ \bibnamefont {Corti\~nas}},\ and\ \bibinfo {author} {\bibfnamefont {M.~H.}\ \bibnamefont {Devoret}},\ }\bibfield  {title} {\bibinfo {title} {Nonlinear dissipation in a driven superconducting circuit},\ }\href {https://doi.org/10.1103/PhysRevA.110.042411} {\bibfield  {journal} {\bibinfo  {journal} {Phys. Rev. A}\ }\textbf {\bibinfo {volume} {110}},\ \bibinfo {pages} {042411} (\bibinfo {year} {2024})}\BibitemShut {NoStop}%
\bibitem [{\citenamefont {Frattini}\ \emph {et~al.}(2024)\citenamefont {Frattini}, \citenamefont {Corti\~nas}, \citenamefont {Venkatraman}, \citenamefont {Xiao}, \citenamefont {Su}, \citenamefont {Lei}, \citenamefont {Chapman}, \citenamefont {Joshi}, \citenamefont {Girvin}, \citenamefont {Schoelkopf}, \citenamefont {Puri},\ and\ \citenamefont {Devoret}}]{frattini2022squeezed}%
  \BibitemOpen
  \bibfield  {author} {\bibinfo {author} {\bibfnamefont {N.~E.}\ \bibnamefont {Frattini}}, \bibinfo {author} {\bibfnamefont {R.~G.}\ \bibnamefont {Corti\~nas}}, \bibinfo {author} {\bibfnamefont {J.}~\bibnamefont {Venkatraman}}, \bibinfo {author} {\bibfnamefont {X.}~\bibnamefont {Xiao}}, \bibinfo {author} {\bibfnamefont {Q.}~\bibnamefont {Su}}, \bibinfo {author} {\bibfnamefont {C.~U.}\ \bibnamefont {Lei}}, \bibinfo {author} {\bibfnamefont {B.~J.}\ \bibnamefont {Chapman}}, \bibinfo {author} {\bibfnamefont {V.~R.}\ \bibnamefont {Joshi}}, \bibinfo {author} {\bibfnamefont {S.~M.}\ \bibnamefont {Girvin}}, \bibinfo {author} {\bibfnamefont {R.~J.}\ \bibnamefont {Schoelkopf}}, \bibinfo {author} {\bibfnamefont {S.}~\bibnamefont {Puri}},\ and\ \bibinfo {author} {\bibfnamefont {M.~H.}\ \bibnamefont {Devoret}},\ }\bibfield  {title} {\bibinfo {title} {Observation of pairwise level degeneracies and the quantum regime of the arrhenius law in a double-well parametric oscillator},\ }\href
  {https://doi.org/10.1103/PhysRevX.14.031040} {\bibfield  {journal} {\bibinfo  {journal} {Phys. Rev. X}\ }\textbf {\bibinfo {volume} {14}},\ \bibinfo {pages} {031040} (\bibinfo {year} {2024})}\BibitemShut {NoStop}%
\bibitem [{\citenamefont {Chapman}\ \emph {et~al.}(2023)\citenamefont {Chapman}, \citenamefont {de~Graaf}, \citenamefont {Xue}, \citenamefont {Zhang}, \citenamefont {Teoh}, \citenamefont {Curtis}, \citenamefont {Tsunoda}, \citenamefont {Eickbusch}, \citenamefont {Read}, \citenamefont {Koottandavida}, \citenamefont {Mundhada}, \citenamefont {Frunzio}, \citenamefont {Devoret}, \citenamefont {Girvin},\ and\ \citenamefont {Schoelkopf}}]{ChapmanStijn2023}%
  \BibitemOpen
  \bibfield  {author} {\bibinfo {author} {\bibfnamefont {B.~J.}\ \bibnamefont {Chapman}}, \bibinfo {author} {\bibfnamefont {S.~J.}\ \bibnamefont {de~Graaf}}, \bibinfo {author} {\bibfnamefont {S.~H.}\ \bibnamefont {Xue}}, \bibinfo {author} {\bibfnamefont {Y.}~\bibnamefont {Zhang}}, \bibinfo {author} {\bibfnamefont {J.}~\bibnamefont {Teoh}}, \bibinfo {author} {\bibfnamefont {J.~C.}\ \bibnamefont {Curtis}}, \bibinfo {author} {\bibfnamefont {T.}~\bibnamefont {Tsunoda}}, \bibinfo {author} {\bibfnamefont {A.}~\bibnamefont {Eickbusch}}, \bibinfo {author} {\bibfnamefont {A.~P.}\ \bibnamefont {Read}}, \bibinfo {author} {\bibfnamefont {A.}~\bibnamefont {Koottandavida}}, \bibinfo {author} {\bibfnamefont {S.~O.}\ \bibnamefont {Mundhada}}, \bibinfo {author} {\bibfnamefont {L.}~\bibnamefont {Frunzio}}, \bibinfo {author} {\bibfnamefont {M.}~\bibnamefont {Devoret}}, \bibinfo {author} {\bibfnamefont {S.}~\bibnamefont {Girvin}},\ and\ \bibinfo {author} {\bibfnamefont {R.}~\bibnamefont {Schoelkopf}},\ }\bibfield  {title}
  {\bibinfo {title} {High-on-off-ratio beam-splitter interaction for gates on bosonically encoded qubits},\ }\href {https://doi.org/10.1103/PRXQuantum.4.020355} {\bibfield  {journal} {\bibinfo  {journal} {PRX Quantum}\ }\textbf {\bibinfo {volume} {4}},\ \bibinfo {pages} {020355} (\bibinfo {year} {2023})}\BibitemShut {NoStop}%
\bibitem [{\citenamefont {Garc{\'{i}}a-Mata}\ \emph {et~al.}(2024)\citenamefont {Garc{\'{i}}a-Mata}, \citenamefont {Corti{\~{n}}as}, \citenamefont {Xiao}, \citenamefont {Ch{\'{a}}vez-Carlos}, \citenamefont {Batista}, \citenamefont {Santos},\ and\ \citenamefont {Wisniacki}}]{GarciaMata2024effectiveversus}%
  \BibitemOpen
  \bibfield  {author} {\bibinfo {author} {\bibfnamefont {I.}~\bibnamefont {Garc{\'{i}}a-Mata}}, \bibinfo {author} {\bibfnamefont {R.~G.}\ \bibnamefont {Corti{\~{n}}as}}, \bibinfo {author} {\bibfnamefont {X.}~\bibnamefont {Xiao}}, \bibinfo {author} {\bibfnamefont {J.}~\bibnamefont {Ch{\'{a}}vez-Carlos}}, \bibinfo {author} {\bibfnamefont {V.~S.}\ \bibnamefont {Batista}}, \bibinfo {author} {\bibfnamefont {L.~F.}\ \bibnamefont {Santos}},\ and\ \bibinfo {author} {\bibfnamefont {D.~A.}\ \bibnamefont {Wisniacki}},\ }\bibfield  {title} {\bibinfo {title} {Effective versus {F}loquet theory for the {K}err parametric oscillator},\ }\href {https://doi.org/10.22331/q-2024-03-25-1298} {\bibfield  {journal} {\bibinfo  {journal} {{Quantum}}\ }\textbf {\bibinfo {volume} {8}},\ \bibinfo {pages} {1298} (\bibinfo {year} {2024})}\BibitemShut {NoStop}%
\bibitem [{\citenamefont {Smith}\ \emph {et~al.}(2016)\citenamefont {Smith}, \citenamefont {Kou}, \citenamefont {Vool}, \citenamefont {Pop}, \citenamefont {Frunzio}, \citenamefont {Schoelkopf},\ and\ \citenamefont {Devoret}}]{smith2016}%
  \BibitemOpen
  \bibfield  {author} {\bibinfo {author} {\bibfnamefont {W.~C.}\ \bibnamefont {Smith}}, \bibinfo {author} {\bibfnamefont {A.}~\bibnamefont {Kou}}, \bibinfo {author} {\bibfnamefont {U.}~\bibnamefont {Vool}}, \bibinfo {author} {\bibfnamefont {I.~M.}\ \bibnamefont {Pop}}, \bibinfo {author} {\bibfnamefont {L.}~\bibnamefont {Frunzio}}, \bibinfo {author} {\bibfnamefont {R.~J.}\ \bibnamefont {Schoelkopf}},\ and\ \bibinfo {author} {\bibfnamefont {M.~H.}\ \bibnamefont {Devoret}},\ }\bibfield  {title} {\bibinfo {title} {Quantization of inductively shunted superconducting circuits},\ }\href {https://doi.org/10.1103/PhysRevB.94.144507} {\bibfield  {journal} {\bibinfo  {journal} {Phys. Rev. B}\ }\textbf {\bibinfo {volume} {94}},\ \bibinfo {pages} {144507} (\bibinfo {year} {2016})}\BibitemShut {NoStop}%
\bibitem [{\citenamefont {Verney}\ \emph {et~al.}(2019)\citenamefont {Verney}, \citenamefont {Lescanne}, \citenamefont {Devoret}, \citenamefont {Leghtas},\ and\ \citenamefont {Mirrahimi}}]{Verney2019}%
  \BibitemOpen
  \bibfield  {author} {\bibinfo {author} {\bibfnamefont {L.}~\bibnamefont {Verney}}, \bibinfo {author} {\bibfnamefont {R.}~\bibnamefont {Lescanne}}, \bibinfo {author} {\bibfnamefont {M.~H.}\ \bibnamefont {Devoret}}, \bibinfo {author} {\bibfnamefont {Z.}~\bibnamefont {Leghtas}},\ and\ \bibinfo {author} {\bibfnamefont {M.}~\bibnamefont {Mirrahimi}},\ }\bibfield  {title} {\bibinfo {title} {Structural instability of driven josephson circuits prevented by an inductive shunt},\ }\href {https://doi.org/10.1103/PhysRevApplied.11.024003} {\bibfield  {journal} {\bibinfo  {journal} {Phys. Rev. Appl.}\ }\textbf {\bibinfo {volume} {11}},\ \bibinfo {pages} {024003} (\bibinfo {year} {2019})}\BibitemShut {NoStop}%
\bibitem [{\citenamefont {Petrescu}\ \emph {et~al.}(2023)\citenamefont {Petrescu}, \citenamefont {Le~Calonnec}, \citenamefont {Leroux}, \citenamefont {Di~Paolo}, \citenamefont {Mundada}, \citenamefont {Sussman}, \citenamefont {Vrajitoarea}, \citenamefont {Houck},\ and\ \citenamefont {Blais}}]{petrescu2023}%
  \BibitemOpen
  \bibfield  {author} {\bibinfo {author} {\bibfnamefont {A.}~\bibnamefont {Petrescu}}, \bibinfo {author} {\bibfnamefont {C.}~\bibnamefont {Le~Calonnec}}, \bibinfo {author} {\bibfnamefont {C.}~\bibnamefont {Leroux}}, \bibinfo {author} {\bibfnamefont {A.}~\bibnamefont {Di~Paolo}}, \bibinfo {author} {\bibfnamefont {P.}~\bibnamefont {Mundada}}, \bibinfo {author} {\bibfnamefont {S.}~\bibnamefont {Sussman}}, \bibinfo {author} {\bibfnamefont {A.}~\bibnamefont {Vrajitoarea}}, \bibinfo {author} {\bibfnamefont {A.~A.}\ \bibnamefont {Houck}},\ and\ \bibinfo {author} {\bibfnamefont {A.}~\bibnamefont {Blais}},\ }\bibfield  {title} {\bibinfo {title} {Accurate methods for the analysis of strong-drive effects in parametric gates},\ }\href {https://doi.org/10.1103/PhysRevApplied.19.044003} {\bibfield  {journal} {\bibinfo  {journal} {Phys. Rev. Appl.}\ }\textbf {\bibinfo {volume} {19}},\ \bibinfo {pages} {044003} (\bibinfo {year} {2023})}\BibitemShut {NoStop}%
\bibitem [{\citenamefont {Chapple}\ \emph {et~al.}(2024)\citenamefont {Chapple}, \citenamefont {McDonald}, \citenamefont {Muñoz-Arias},\ and\ \citenamefont {Blais}}]{chapple2024}%
  \BibitemOpen
  \bibfield  {author} {\bibinfo {author} {\bibfnamefont {A.~A.}\ \bibnamefont {Chapple}}, \bibinfo {author} {\bibfnamefont {A.}~\bibnamefont {McDonald}}, \bibinfo {author} {\bibfnamefont {M.~H.}\ \bibnamefont {Muñoz-Arias}},\ and\ \bibinfo {author} {\bibfnamefont {A.}~\bibnamefont {Blais}},\ }\href {https://arxiv.org/abs/2412.07734} {\bibinfo {title} {Robustness of longitudinal transmon readout to ionization}} (\bibinfo {year} {2024}),\ \Eprint {https://arxiv.org/abs/2412.07734} {arXiv:2412.07734 [quant-ph]} \BibitemShut {NoStop}%
\bibitem [{\citenamefont {Koch}\ \emph {et~al.}(2007)\citenamefont {Koch}, \citenamefont {Yu}, \citenamefont {Gambetta}, \citenamefont {Houck}, \citenamefont {Schuster}, \citenamefont {Majer}, \citenamefont {Blais}, \citenamefont {Devoret}, \citenamefont {Girvin},\ and\ \citenamefont {Schoelkopf}}]{Koch2007}%
  \BibitemOpen
  \bibfield  {author} {\bibinfo {author} {\bibfnamefont {J.}~\bibnamefont {Koch}}, \bibinfo {author} {\bibfnamefont {T.~M.}\ \bibnamefont {Yu}}, \bibinfo {author} {\bibfnamefont {J.}~\bibnamefont {Gambetta}}, \bibinfo {author} {\bibfnamefont {A.~A.}\ \bibnamefont {Houck}}, \bibinfo {author} {\bibfnamefont {D.~I.}\ \bibnamefont {Schuster}}, \bibinfo {author} {\bibfnamefont {J.}~\bibnamefont {Majer}}, \bibinfo {author} {\bibfnamefont {A.}~\bibnamefont {Blais}}, \bibinfo {author} {\bibfnamefont {M.~H.}\ \bibnamefont {Devoret}}, \bibinfo {author} {\bibfnamefont {S.~M.}\ \bibnamefont {Girvin}},\ and\ \bibinfo {author} {\bibfnamefont {R.~J.}\ \bibnamefont {Schoelkopf}},\ }\bibfield  {title} {\bibinfo {title} {Charge-insensitive qubit design derived from the cooper pair box},\ }\href {https://doi.org/10.1103/PhysRevA.76.042319} {\bibfield  {journal} {\bibinfo  {journal} {Phys. Rev. A}\ }\textbf {\bibinfo {volume} {76}},\ \bibinfo {pages} {042319} (\bibinfo {year} {2007})}\BibitemShut {NoStop}%
\bibitem [{\citenamefont {Wilcox}(1967)}]{wilcox1967}%
  \BibitemOpen
  \bibfield  {author} {\bibinfo {author} {\bibfnamefont {R.~M.}\ \bibnamefont {Wilcox}},\ }\bibfield  {title} {\bibinfo {title} {Exponential operators and parameter differentiation in quantum physics},\ }\href {https://doi.org/10.1063/1.1705306} {\bibfield  {journal} {\bibinfo  {journal} {Journal of Mathematical Physics}\ }\textbf {\bibinfo {volume} {8}},\ \bibinfo {pages} {962} (\bibinfo {year} {1967})}\BibitemShut {NoStop}%
\bibitem [{\citenamefont {Frattini}\ \emph {et~al.}(2018)\citenamefont {Frattini}, \citenamefont {Sivak}, \citenamefont {Lingenfelter}, \citenamefont {Shankar},\ and\ \citenamefont {Devoret}}]{fratini18}%
  \BibitemOpen
  \bibfield  {author} {\bibinfo {author} {\bibfnamefont {N.~E.}\ \bibnamefont {Frattini}}, \bibinfo {author} {\bibfnamefont {V.~V.}\ \bibnamefont {Sivak}}, \bibinfo {author} {\bibfnamefont {A.}~\bibnamefont {Lingenfelter}}, \bibinfo {author} {\bibfnamefont {S.}~\bibnamefont {Shankar}},\ and\ \bibinfo {author} {\bibfnamefont {M.~H.}\ \bibnamefont {Devoret}},\ }\bibfield  {title} {\bibinfo {title} {Optimizing the nonlinearity and dissipation of a snail parametric amplifier for dynamic range},\ }\href {https://doi.org/10.1103/PhysRevApplied.10.054020} {\bibfield  {journal} {\bibinfo  {journal} {Phys. Rev. Appl.}\ }\textbf {\bibinfo {volume} {10}},\ \bibinfo {pages} {054020} (\bibinfo {year} {2018})}\BibitemShut {NoStop}%
\bibitem [{\citenamefont {You}\ \emph {et~al.}(2019)\citenamefont {You}, \citenamefont {Sauls},\ and\ \citenamefont {Koch}}]{You2019}%
  \BibitemOpen
  \bibfield  {author} {\bibinfo {author} {\bibfnamefont {X.}~\bibnamefont {You}}, \bibinfo {author} {\bibfnamefont {J.~A.}\ \bibnamefont {Sauls}},\ and\ \bibinfo {author} {\bibfnamefont {J.}~\bibnamefont {Koch}},\ }\bibfield  {title} {\bibinfo {title} {Circuit quantization in the presence of time-dependent external flux},\ }\href {https://doi.org/10.1103/PhysRevB.99.174512} {\bibfield  {journal} {\bibinfo  {journal} {Phys. Rev. B}\ }\textbf {\bibinfo {volume} {99}},\ \bibinfo {pages} {174512} (\bibinfo {year} {2019})}\BibitemShut {NoStop}%
\bibitem [{\citenamefont {Riwar}\ and\ \citenamefont {DiVincenzo}(2022)}]{Riwar2022}%
  \BibitemOpen
  \bibfield  {author} {\bibinfo {author} {\bibfnamefont {R.-P.}\ \bibnamefont {Riwar}}\ and\ \bibinfo {author} {\bibfnamefont {D.~P.}\ \bibnamefont {DiVincenzo}},\ }\bibfield  {title} {\bibinfo {title} {Circuit quantization with time-dependent magnetic fields for realistic geometries},\ }\href {https://doi.org/10.1038/s41534-022-00539-x} {\bibfield  {journal} {\bibinfo  {journal} {npj Quantum Information}\ }\textbf {\bibinfo {volume} {8}},\ \bibinfo {pages} {36} (\bibinfo {year} {2022})}\BibitemShut {NoStop}%
\bibitem [{\citenamefont {Ferguson}\ \emph {et~al.}(2013)\citenamefont {Ferguson}, \citenamefont {Houck},\ and\ \citenamefont {Koch}}]{ferguson2013}%
  \BibitemOpen
  \bibfield  {author} {\bibinfo {author} {\bibfnamefont {D.~G.}\ \bibnamefont {Ferguson}}, \bibinfo {author} {\bibfnamefont {A.~A.}\ \bibnamefont {Houck}},\ and\ \bibinfo {author} {\bibfnamefont {J.}~\bibnamefont {Koch}},\ }\bibfield  {title} {\bibinfo {title} {Symmetries and collective excitations in large superconducting circuits},\ }\href {https://doi.org/10.1103/PhysRevX.3.011003} {\bibfield  {journal} {\bibinfo  {journal} {Phys. Rev. X}\ }\textbf {\bibinfo {volume} {3}},\ \bibinfo {pages} {011003} (\bibinfo {year} {2013})}\BibitemShut {NoStop}%
\bibitem [{\citenamefont {Di~Paolo}\ \emph {et~al.}(2021)\citenamefont {Di~Paolo}, \citenamefont {Baker}, \citenamefont {Foley}, \citenamefont {S{\'e}n{\'e}chal},\ and\ \citenamefont {Blais}}]{DiPaolo2021}%
  \BibitemOpen
  \bibfield  {author} {\bibinfo {author} {\bibfnamefont {A.}~\bibnamefont {Di~Paolo}}, \bibinfo {author} {\bibfnamefont {T.~E.}\ \bibnamefont {Baker}}, \bibinfo {author} {\bibfnamefont {A.}~\bibnamefont {Foley}}, \bibinfo {author} {\bibfnamefont {D.}~\bibnamefont {S{\'e}n{\'e}chal}},\ and\ \bibinfo {author} {\bibfnamefont {A.}~\bibnamefont {Blais}},\ }\bibfield  {title} {\bibinfo {title} {Efficient modeling of superconducting quantum circuits with tensor networks},\ }\href {https://doi.org/10.1038/s41534-020-00352-4} {\bibfield  {journal} {\bibinfo  {journal} {npj Quantum Information}\ }\textbf {\bibinfo {volume} {7}},\ \bibinfo {pages} {11} (\bibinfo {year} {2021})}\BibitemShut {NoStop}%
\bibitem [{\citenamefont {Rymarz}\ and\ \citenamefont {DiVincenzo}(2023)}]{Rymarz23}%
  \BibitemOpen
  \bibfield  {author} {\bibinfo {author} {\bibfnamefont {M.}~\bibnamefont {Rymarz}}\ and\ \bibinfo {author} {\bibfnamefont {D.~P.}\ \bibnamefont {DiVincenzo}},\ }\bibfield  {title} {\bibinfo {title} {Consistent quantization of nearly singular superconducting circuits},\ }\href {https://doi.org/10.1103/PhysRevX.13.021017} {\bibfield  {journal} {\bibinfo  {journal} {Phys. Rev. X}\ }\textbf {\bibinfo {volume} {13}},\ \bibinfo {pages} {021017} (\bibinfo {year} {2023})}\BibitemShut {NoStop}%
\bibitem [{\citenamefont {Féchant}\ \emph {et~al.}(2025)\citenamefont {Féchant}, \citenamefont {Dumas}, \citenamefont {Bénâtre}, \citenamefont {Gosling}, \citenamefont {Lenhard}, \citenamefont {Spiecker}, \citenamefont {Wernsdorfer}, \citenamefont {D'Anjou}, \citenamefont {Blais},\ and\ \citenamefont {Pop}}]{féchant2025offsetchargedependence}%
  \BibitemOpen
  \bibfield  {author} {\bibinfo {author} {\bibfnamefont {M.}~\bibnamefont {Féchant}}, \bibinfo {author} {\bibfnamefont {M.~F.}\ \bibnamefont {Dumas}}, \bibinfo {author} {\bibfnamefont {D.}~\bibnamefont {Bénâtre}}, \bibinfo {author} {\bibfnamefont {N.}~\bibnamefont {Gosling}}, \bibinfo {author} {\bibfnamefont {P.}~\bibnamefont {Lenhard}}, \bibinfo {author} {\bibfnamefont {M.}~\bibnamefont {Spiecker}}, \bibinfo {author} {\bibfnamefont {W.}~\bibnamefont {Wernsdorfer}}, \bibinfo {author} {\bibfnamefont {B.}~\bibnamefont {D'Anjou}}, \bibinfo {author} {\bibfnamefont {A.}~\bibnamefont {Blais}},\ and\ \bibinfo {author} {\bibfnamefont {I.~M.}\ \bibnamefont {Pop}},\ }\href {https://arxiv.org/abs/2505.00674} {\bibinfo {title} {Offset charge dependence of measurement-induced transitions in transmons}} (\bibinfo {year} {2025}),\ \Eprint {https://arxiv.org/abs/2505.00674} {arXiv:2505.00674 [quant-ph]} \BibitemShut {NoStop}%
\bibitem [{\citenamefont {Arfken}\ \emph {et~al.}(2013)\citenamefont {Arfken}, \citenamefont {Weber},\ and\ \citenamefont {Harris}}]{Arfken2013a}%
  \BibitemOpen
  \bibfield  {author} {\bibinfo {author} {\bibfnamefont {G.~B.}\ \bibnamefont {Arfken}}, \bibinfo {author} {\bibfnamefont {H.~J.}\ \bibnamefont {Weber}},\ and\ \bibinfo {author} {\bibfnamefont {F.~E.}\ \bibnamefont {Harris}},\ }\bibinfo {title} {Chapter 14 - bessel functions},\ in\ \href {https://www.sciencedirect.com/science/article/pii/B9780123846549000141} {\emph {\bibinfo {booktitle} {Mathematical Methods for Physicists (Seventh Edition)}}}\ (\bibinfo  {publisher} {Academic Press},\ \bibinfo {address} {Boston},\ \bibinfo {year} {2013})\ pp.\ \bibinfo {pages} {643--713}\BibitemShut {NoStop}%
\bibitem [{\citenamefont {Hillmann}\ \emph {et~al.}(2020)\citenamefont {Hillmann}, \citenamefont {Quijandr\'{\i}a}, \citenamefont {Johansson}, \citenamefont {Ferraro}, \citenamefont {Gasparinetti},\ and\ \citenamefont {Ferrini}}]{Hillmann2020}%
  \BibitemOpen
  \bibfield  {author} {\bibinfo {author} {\bibfnamefont {T.}~\bibnamefont {Hillmann}}, \bibinfo {author} {\bibfnamefont {F.}~\bibnamefont {Quijandr\'{\i}a}}, \bibinfo {author} {\bibfnamefont {G.}~\bibnamefont {Johansson}}, \bibinfo {author} {\bibfnamefont {A.}~\bibnamefont {Ferraro}}, \bibinfo {author} {\bibfnamefont {S.}~\bibnamefont {Gasparinetti}},\ and\ \bibinfo {author} {\bibfnamefont {G.}~\bibnamefont {Ferrini}},\ }\bibfield  {title} {\bibinfo {title} {Universal gate set for continuous-variable quantum computation with microwave circuits},\ }\href {https://doi.org/10.1103/PhysRevLett.125.160501} {\bibfield  {journal} {\bibinfo  {journal} {Phys. Rev. Lett.}\ }\textbf {\bibinfo {volume} {125}},\ \bibinfo {pages} {160501} (\bibinfo {year} {2020})}\BibitemShut {NoStop}%
\bibitem [{\citenamefont {Zhu}\ \emph {et~al.}(2013)\citenamefont {Zhu}, \citenamefont {Ferguson}, \citenamefont {Manucharyan},\ and\ \citenamefont {Koch}}]{Zhu2013_fluxonium}%
  \BibitemOpen
  \bibfield  {author} {\bibinfo {author} {\bibfnamefont {G.}~\bibnamefont {Zhu}}, \bibinfo {author} {\bibfnamefont {D.~G.}\ \bibnamefont {Ferguson}}, \bibinfo {author} {\bibfnamefont {V.~E.}\ \bibnamefont {Manucharyan}},\ and\ \bibinfo {author} {\bibfnamefont {J.}~\bibnamefont {Koch}},\ }\bibfield  {title} {\bibinfo {title} {Circuit qed with fluxonium qubits: Theory of the dispersive regime},\ }\href {https://doi.org/10.1103/PhysRevB.87.024510} {\bibfield  {journal} {\bibinfo  {journal} {Phys. Rev. B}\ }\textbf {\bibinfo {volume} {87}},\ \bibinfo {pages} {024510} (\bibinfo {year} {2013})}\BibitemShut {NoStop}%
\bibitem [{\citenamefont {Nguyen}\ \emph {et~al.}(2022)\citenamefont {Nguyen}, \citenamefont {Koolstra}, \citenamefont {Kim}, \citenamefont {Morvan}, \citenamefont {Chistolini}, \citenamefont {Singh}, \citenamefont {Nesterov}, \citenamefont {J\"unger}, \citenamefont {Chen}, \citenamefont {Pedramrazi}, \citenamefont {Mitchell}, \citenamefont {Kreikebaum}, \citenamefont {Puri}, \citenamefont {Santiago},\ and\ \citenamefont {Siddiqi}}]{Nguyen2022_fluxonium}%
  \BibitemOpen
  \bibfield  {author} {\bibinfo {author} {\bibfnamefont {L.~B.}\ \bibnamefont {Nguyen}}, \bibinfo {author} {\bibfnamefont {G.}~\bibnamefont {Koolstra}}, \bibinfo {author} {\bibfnamefont {Y.}~\bibnamefont {Kim}}, \bibinfo {author} {\bibfnamefont {A.}~\bibnamefont {Morvan}}, \bibinfo {author} {\bibfnamefont {T.}~\bibnamefont {Chistolini}}, \bibinfo {author} {\bibfnamefont {S.}~\bibnamefont {Singh}}, \bibinfo {author} {\bibfnamefont {K.~N.}\ \bibnamefont {Nesterov}}, \bibinfo {author} {\bibfnamefont {C.}~\bibnamefont {J\"unger}}, \bibinfo {author} {\bibfnamefont {L.}~\bibnamefont {Chen}}, \bibinfo {author} {\bibfnamefont {Z.}~\bibnamefont {Pedramrazi}}, \bibinfo {author} {\bibfnamefont {B.~K.}\ \bibnamefont {Mitchell}}, \bibinfo {author} {\bibfnamefont {J.~M.}\ \bibnamefont {Kreikebaum}}, \bibinfo {author} {\bibfnamefont {S.}~\bibnamefont {Puri}}, \bibinfo {author} {\bibfnamefont {D.~I.}\ \bibnamefont {Santiago}},\ and\ \bibinfo {author} {\bibfnamefont {I.}~\bibnamefont {Siddiqi}},\ }\bibfield  {title} {\bibinfo
  {title} {Blueprint for a high-performance fluxonium quantum processor},\ }\href {https://doi.org/10.1103/PRXQuantum.3.037001} {\bibfield  {journal} {\bibinfo  {journal} {PRX Quantum}\ }\textbf {\bibinfo {volume} {3}},\ \bibinfo {pages} {037001} (\bibinfo {year} {2022})}\BibitemShut {NoStop}%
\bibitem [{\citenamefont {Bao}\ \emph {et~al.}(2022)\citenamefont {Bao}, \citenamefont {Deng}, \citenamefont {Ding}, \citenamefont {Gao}, \citenamefont {Gao}, \citenamefont {Huang}, \citenamefont {Jiang}, \citenamefont {Ku}, \citenamefont {Li}, \citenamefont {Ma}, \citenamefont {Ni}, \citenamefont {Qin}, \citenamefont {Song}, \citenamefont {Sun}, \citenamefont {Tang}, \citenamefont {Wang}, \citenamefont {Wu}, \citenamefont {Xia}, \citenamefont {Yu}, \citenamefont {Zhang}, \citenamefont {Zhang}, \citenamefont {Zhang}, \citenamefont {Zhou}, \citenamefont {Zhu}, \citenamefont {Shi}, \citenamefont {Chen}, \citenamefont {Zhao},\ and\ \citenamefont {Deng}}]{Bao2022_fluxoniumm}%
  \BibitemOpen
  \bibfield  {author} {\bibinfo {author} {\bibfnamefont {F.}~\bibnamefont {Bao}}, \bibinfo {author} {\bibfnamefont {H.}~\bibnamefont {Deng}}, \bibinfo {author} {\bibfnamefont {D.}~\bibnamefont {Ding}}, \bibinfo {author} {\bibfnamefont {R.}~\bibnamefont {Gao}}, \bibinfo {author} {\bibfnamefont {X.}~\bibnamefont {Gao}}, \bibinfo {author} {\bibfnamefont {C.}~\bibnamefont {Huang}}, \bibinfo {author} {\bibfnamefont {X.}~\bibnamefont {Jiang}}, \bibinfo {author} {\bibfnamefont {H.-S.}\ \bibnamefont {Ku}}, \bibinfo {author} {\bibfnamefont {Z.}~\bibnamefont {Li}}, \bibinfo {author} {\bibfnamefont {X.}~\bibnamefont {Ma}}, \bibinfo {author} {\bibfnamefont {X.}~\bibnamefont {Ni}}, \bibinfo {author} {\bibfnamefont {J.}~\bibnamefont {Qin}}, \bibinfo {author} {\bibfnamefont {Z.}~\bibnamefont {Song}}, \bibinfo {author} {\bibfnamefont {H.}~\bibnamefont {Sun}}, \bibinfo {author} {\bibfnamefont {C.}~\bibnamefont {Tang}}, \bibinfo {author} {\bibfnamefont {T.}~\bibnamefont {Wang}}, \bibinfo {author} {\bibfnamefont
  {F.}~\bibnamefont {Wu}}, \bibinfo {author} {\bibfnamefont {T.}~\bibnamefont {Xia}}, \bibinfo {author} {\bibfnamefont {W.}~\bibnamefont {Yu}}, \bibinfo {author} {\bibfnamefont {F.}~\bibnamefont {Zhang}}, \bibinfo {author} {\bibfnamefont {G.}~\bibnamefont {Zhang}}, \bibinfo {author} {\bibfnamefont {X.}~\bibnamefont {Zhang}}, \bibinfo {author} {\bibfnamefont {J.}~\bibnamefont {Zhou}}, \bibinfo {author} {\bibfnamefont {X.}~\bibnamefont {Zhu}}, \bibinfo {author} {\bibfnamefont {Y.}~\bibnamefont {Shi}}, \bibinfo {author} {\bibfnamefont {J.}~\bibnamefont {Chen}}, \bibinfo {author} {\bibfnamefont {H.-H.}\ \bibnamefont {Zhao}},\ and\ \bibinfo {author} {\bibfnamefont {C.}~\bibnamefont {Deng}},\ }\bibfield  {title} {\bibinfo {title} {Fluxonium: An alternative qubit platform for high-fidelity operations},\ }\href {https://doi.org/10.1103/PhysRevLett.129.010502} {\bibfield  {journal} {\bibinfo  {journal} {Phys. Rev. Lett.}\ }\textbf {\bibinfo {volume} {129}},\ \bibinfo {pages} {010502} (\bibinfo {year}
  {2022})}\BibitemShut {NoStop}%
\bibitem [{\citenamefont {Earnest}\ \emph {et~al.}(2018)\citenamefont {Earnest}, \citenamefont {Chakram}, \citenamefont {Lu}, \citenamefont {Irons}, \citenamefont {Naik}, \citenamefont {Leung}, \citenamefont {Ocola}, \citenamefont {Czaplewski}, \citenamefont {Baker}, \citenamefont {Lawrence}, \citenamefont {Koch},\ and\ \citenamefont {Schuster}}]{Earnest2018_fluxonium}%
  \BibitemOpen
  \bibfield  {author} {\bibinfo {author} {\bibfnamefont {N.}~\bibnamefont {Earnest}}, \bibinfo {author} {\bibfnamefont {S.}~\bibnamefont {Chakram}}, \bibinfo {author} {\bibfnamefont {Y.}~\bibnamefont {Lu}}, \bibinfo {author} {\bibfnamefont {N.}~\bibnamefont {Irons}}, \bibinfo {author} {\bibfnamefont {R.~K.}\ \bibnamefont {Naik}}, \bibinfo {author} {\bibfnamefont {N.}~\bibnamefont {Leung}}, \bibinfo {author} {\bibfnamefont {L.}~\bibnamefont {Ocola}}, \bibinfo {author} {\bibfnamefont {D.~A.}\ \bibnamefont {Czaplewski}}, \bibinfo {author} {\bibfnamefont {B.}~\bibnamefont {Baker}}, \bibinfo {author} {\bibfnamefont {J.}~\bibnamefont {Lawrence}}, \bibinfo {author} {\bibfnamefont {J.}~\bibnamefont {Koch}},\ and\ \bibinfo {author} {\bibfnamefont {D.~I.}\ \bibnamefont {Schuster}},\ }\bibfield  {title} {\bibinfo {title} {Realization of a $\mathrm{\ensuremath{\Lambda}}$ system with metastable states of a capacitively shunted fluxonium},\ }\href {https://doi.org/10.1103/PhysRevLett.120.150504} {\bibfield  {journal}
  {\bibinfo  {journal} {Phys. Rev. Lett.}\ }\textbf {\bibinfo {volume} {120}},\ \bibinfo {pages} {150504} (\bibinfo {year} {2018})}\BibitemShut {NoStop}%
\bibitem [{\citenamefont {Iyama}\ \emph {et~al.}(2024)\citenamefont {Iyama}, \citenamefont {Kamiya}, \citenamefont {Fujii}, \citenamefont {Mukai}, \citenamefont {Zhou}, \citenamefont {Nagase}, \citenamefont {Tomonaga}, \citenamefont {Wang}, \citenamefont {Xue}, \citenamefont {Watabe}, \citenamefont {Kwon},\ and\ \citenamefont {Tsai}}]{Iyama2024}%
  \BibitemOpen
  \bibfield  {author} {\bibinfo {author} {\bibfnamefont {D.}~\bibnamefont {Iyama}}, \bibinfo {author} {\bibfnamefont {T.}~\bibnamefont {Kamiya}}, \bibinfo {author} {\bibfnamefont {S.}~\bibnamefont {Fujii}}, \bibinfo {author} {\bibfnamefont {H.}~\bibnamefont {Mukai}}, \bibinfo {author} {\bibfnamefont {Y.}~\bibnamefont {Zhou}}, \bibinfo {author} {\bibfnamefont {T.}~\bibnamefont {Nagase}}, \bibinfo {author} {\bibfnamefont {A.}~\bibnamefont {Tomonaga}}, \bibinfo {author} {\bibfnamefont {R.}~\bibnamefont {Wang}}, \bibinfo {author} {\bibfnamefont {J.-J.}\ \bibnamefont {Xue}}, \bibinfo {author} {\bibfnamefont {S.}~\bibnamefont {Watabe}}, \bibinfo {author} {\bibfnamefont {S.}~\bibnamefont {Kwon}},\ and\ \bibinfo {author} {\bibfnamefont {J.-S.}\ \bibnamefont {Tsai}},\ }\bibfield  {title} {\bibinfo {title} {Observation and manipulation of quantum interference in a superconducting kerr parametric oscillator},\ }\href {https://doi.org/10.1038/s41467-023-44496-1} {\bibfield  {journal} {\bibinfo  {journal} {Nature
  Communications}\ }\textbf {\bibinfo {volume} {15}},\ \bibinfo {pages} {86} (\bibinfo {year} {2024})}\BibitemShut {NoStop}%
\bibitem [{\citenamefont {Kang}\ \emph {et~al.}(2022)\citenamefont {Kang}, \citenamefont {Chen}, \citenamefont {Wang}, \citenamefont {Song}, \citenamefont {Xia}, \citenamefont {Miranowicz}, \citenamefont {Zheng},\ and\ \citenamefont {Nori}}]{Kang2022}%
  \BibitemOpen
  \bibfield  {author} {\bibinfo {author} {\bibfnamefont {Y.-H.}\ \bibnamefont {Kang}}, \bibinfo {author} {\bibfnamefont {Y.-H.}\ \bibnamefont {Chen}}, \bibinfo {author} {\bibfnamefont {X.}~\bibnamefont {Wang}}, \bibinfo {author} {\bibfnamefont {J.}~\bibnamefont {Song}}, \bibinfo {author} {\bibfnamefont {Y.}~\bibnamefont {Xia}}, \bibinfo {author} {\bibfnamefont {A.}~\bibnamefont {Miranowicz}}, \bibinfo {author} {\bibfnamefont {S.-B.}\ \bibnamefont {Zheng}},\ and\ \bibinfo {author} {\bibfnamefont {F.}~\bibnamefont {Nori}},\ }\bibfield  {title} {\bibinfo {title} {Nonadiabatic geometric quantum computation with cat-state qubits via invariant-based reverse engineering},\ }\href {https://doi.org/10.1103/PhysRevResearch.4.013233} {\bibfield  {journal} {\bibinfo  {journal} {Phys. Rev. Res.}\ }\textbf {\bibinfo {volume} {4}},\ \bibinfo {pages} {013233} (\bibinfo {year} {2022})}\BibitemShut {NoStop}%
\bibitem [{\citenamefont {de~Graaf}\ \emph {et~al.}(2024)\citenamefont {de~Graaf}, \citenamefont {Xue}, \citenamefont {Chapman}, \citenamefont {Teoh}, \citenamefont {Tsunoda}, \citenamefont {Winkel}, \citenamefont {Garmon}, \citenamefont {Chang}, \citenamefont {Frunzio}, \citenamefont {Puri},\ and\ \citenamefont {Schoelkopf}}]{degraaf2024}%
  \BibitemOpen
  \bibfield  {author} {\bibinfo {author} {\bibfnamefont {S.~J.}\ \bibnamefont {de~Graaf}}, \bibinfo {author} {\bibfnamefont {S.~H.}\ \bibnamefont {Xue}}, \bibinfo {author} {\bibfnamefont {B.~J.}\ \bibnamefont {Chapman}}, \bibinfo {author} {\bibfnamefont {J.~D.}\ \bibnamefont {Teoh}}, \bibinfo {author} {\bibfnamefont {T.}~\bibnamefont {Tsunoda}}, \bibinfo {author} {\bibfnamefont {P.}~\bibnamefont {Winkel}}, \bibinfo {author} {\bibfnamefont {J.~W.~O.}\ \bibnamefont {Garmon}}, \bibinfo {author} {\bibfnamefont {K.~M.}\ \bibnamefont {Chang}}, \bibinfo {author} {\bibfnamefont {L.}~\bibnamefont {Frunzio}}, \bibinfo {author} {\bibfnamefont {S.}~\bibnamefont {Puri}},\ and\ \bibinfo {author} {\bibfnamefont {R.~J.}\ \bibnamefont {Schoelkopf}},\ }\href {https://arxiv.org/abs/2406.14621} {\bibinfo {title} {A mid-circuit erasure check on a dual-rail cavity qubit using the joint-photon number-splitting regime of circuit qed}} (\bibinfo {year} {2024}),\ \Eprint {https://arxiv.org/abs/2406.14621} {arXiv:2406.14621 [quant-ph]}
  \BibitemShut {NoStop}%
\bibitem [{\citenamefont {Pietikäinen}\ \emph {et~al.}(2024)\citenamefont {Pietikäinen}, \citenamefont {Černotík}, \citenamefont {Eickbusch}, \citenamefont {Maiti}, \citenamefont {Garmon}, \citenamefont {Filip},\ and\ \citenamefont {Girvin}}]{pietikäinen2024}%
  \BibitemOpen
  \bibfield  {author} {\bibinfo {author} {\bibfnamefont {I.}~\bibnamefont {Pietikäinen}}, \bibinfo {author} {\bibfnamefont {O.}~\bibnamefont {Černotík}}, \bibinfo {author} {\bibfnamefont {A.}~\bibnamefont {Eickbusch}}, \bibinfo {author} {\bibfnamefont {A.}~\bibnamefont {Maiti}}, \bibinfo {author} {\bibfnamefont {J.~W.~O.}\ \bibnamefont {Garmon}}, \bibinfo {author} {\bibfnamefont {R.}~\bibnamefont {Filip}},\ and\ \bibinfo {author} {\bibfnamefont {S.~M.}\ \bibnamefont {Girvin}},\ }\href {https://arxiv.org/abs/2403.02278} {\bibinfo {title} {Strategies and trade-offs for controllability and memory time of ultra-high-quality microwave cavities in circuit qed}} (\bibinfo {year} {2024}),\ \Eprint {https://arxiv.org/abs/2403.02278} {arXiv:2403.02278 [quant-ph]} \BibitemShut {NoStop}%
\bibitem [{\citenamefont {Ranadive}\ \emph {et~al.}(2024)\citenamefont {Ranadive}, \citenamefont {Fazliji}, \citenamefont {Gal}, \citenamefont {Cappelli}, \citenamefont {Butseraen}, \citenamefont {Bonet}, \citenamefont {Eyraud}, \citenamefont {Böhling}, \citenamefont {Planat}, \citenamefont {Metelmann},\ and\ \citenamefont {Roch}}]{ranadive2024}%
  \BibitemOpen
  \bibfield  {author} {\bibinfo {author} {\bibfnamefont {A.}~\bibnamefont {Ranadive}}, \bibinfo {author} {\bibfnamefont {B.}~\bibnamefont {Fazliji}}, \bibinfo {author} {\bibfnamefont {G.~L.}\ \bibnamefont {Gal}}, \bibinfo {author} {\bibfnamefont {G.}~\bibnamefont {Cappelli}}, \bibinfo {author} {\bibfnamefont {G.}~\bibnamefont {Butseraen}}, \bibinfo {author} {\bibfnamefont {E.}~\bibnamefont {Bonet}}, \bibinfo {author} {\bibfnamefont {E.}~\bibnamefont {Eyraud}}, \bibinfo {author} {\bibfnamefont {S.}~\bibnamefont {Böhling}}, \bibinfo {author} {\bibfnamefont {L.}~\bibnamefont {Planat}}, \bibinfo {author} {\bibfnamefont {A.}~\bibnamefont {Metelmann}},\ and\ \bibinfo {author} {\bibfnamefont {N.}~\bibnamefont {Roch}},\ }\href {https://arxiv.org/abs/2406.19752} {\bibinfo {title} {A traveling wave parametric amplifier isolator}} (\bibinfo {year} {2024}),\ \Eprint {https://arxiv.org/abs/2406.19752} {arXiv:2406.19752 [quant-ph]} \BibitemShut {NoStop}%
\bibitem [{\citenamefont {Sivak}\ \emph {et~al.}(2020)\citenamefont {Sivak}, \citenamefont {Shankar}, \citenamefont {Liu}, \citenamefont {Aumentado},\ and\ \citenamefont {Devoret}}]{Sivak2020}%
  \BibitemOpen
  \bibfield  {author} {\bibinfo {author} {\bibfnamefont {V.~V.}\ \bibnamefont {Sivak}}, \bibinfo {author} {\bibfnamefont {S.}~\bibnamefont {Shankar}}, \bibinfo {author} {\bibfnamefont {G.}~\bibnamefont {Liu}}, \bibinfo {author} {\bibfnamefont {J.}~\bibnamefont {Aumentado}},\ and\ \bibinfo {author} {\bibfnamefont {M.~H.}\ \bibnamefont {Devoret}},\ }\bibfield  {title} {\bibinfo {title} {Josephson array-mode parametric amplifier},\ }\href {https://doi.org/10.1103/PhysRevApplied.13.024014} {\bibfield  {journal} {\bibinfo  {journal} {Phys. Rev. Appl.}\ }\textbf {\bibinfo {volume} {13}},\ \bibinfo {pages} {024014} (\bibinfo {year} {2020})}\BibitemShut {NoStop}%
\bibitem [{\citenamefont {Kaufman}\ \emph {et~al.}(2024)\citenamefont {Kaufman}, \citenamefont {Liu}, \citenamefont {Cicak}, \citenamefont {Mesits}, \citenamefont {Xia}, \citenamefont {Zhou}, \citenamefont {Nowicki}, \citenamefont {Aumentado}, \citenamefont {Pekker},\ and\ \citenamefont {Hatridge}}]{kaufman2024}%
  \BibitemOpen
  \bibfield  {author} {\bibinfo {author} {\bibfnamefont {R.}~\bibnamefont {Kaufman}}, \bibinfo {author} {\bibfnamefont {C.}~\bibnamefont {Liu}}, \bibinfo {author} {\bibfnamefont {K.}~\bibnamefont {Cicak}}, \bibinfo {author} {\bibfnamefont {B.}~\bibnamefont {Mesits}}, \bibinfo {author} {\bibfnamefont {M.}~\bibnamefont {Xia}}, \bibinfo {author} {\bibfnamefont {C.}~\bibnamefont {Zhou}}, \bibinfo {author} {\bibfnamefont {M.}~\bibnamefont {Nowicki}}, \bibinfo {author} {\bibfnamefont {J.}~\bibnamefont {Aumentado}}, \bibinfo {author} {\bibfnamefont {D.}~\bibnamefont {Pekker}},\ and\ \bibinfo {author} {\bibfnamefont {M.}~\bibnamefont {Hatridge}},\ }\href {https://arxiv.org/abs/2402.19435} {\bibinfo {title} {Simple, high saturation power, quantum-limited, rf squid array-based josephson parametric amplifiers}} (\bibinfo {year} {2024}),\ \Eprint {https://arxiv.org/abs/2402.19435} {arXiv:2402.19435 [quant-ph]} \BibitemShut {NoStop}%
\bibitem [{\citenamefont {Eriksson}\ \emph {et~al.}(2024)\citenamefont {Eriksson}, \citenamefont {S{\'e}pulcre}, \citenamefont {Kervinen}, \citenamefont {Hillmann}, \citenamefont {Kudra}, \citenamefont {Dupouy}, \citenamefont {Lu}, \citenamefont {Khanahmadi}, \citenamefont {Yang}, \citenamefont {Castillo-Moreno}, \citenamefont {Delsing},\ and\ \citenamefont {Gasparinetti}}]{Eriksson2024}%
  \BibitemOpen
  \bibfield  {author} {\bibinfo {author} {\bibfnamefont {A.~M.}\ \bibnamefont {Eriksson}}, \bibinfo {author} {\bibfnamefont {T.}~\bibnamefont {S{\'e}pulcre}}, \bibinfo {author} {\bibfnamefont {M.}~\bibnamefont {Kervinen}}, \bibinfo {author} {\bibfnamefont {T.}~\bibnamefont {Hillmann}}, \bibinfo {author} {\bibfnamefont {M.}~\bibnamefont {Kudra}}, \bibinfo {author} {\bibfnamefont {S.}~\bibnamefont {Dupouy}}, \bibinfo {author} {\bibfnamefont {Y.}~\bibnamefont {Lu}}, \bibinfo {author} {\bibfnamefont {M.}~\bibnamefont {Khanahmadi}}, \bibinfo {author} {\bibfnamefont {J.}~\bibnamefont {Yang}}, \bibinfo {author} {\bibfnamefont {C.}~\bibnamefont {Castillo-Moreno}}, \bibinfo {author} {\bibfnamefont {P.}~\bibnamefont {Delsing}},\ and\ \bibinfo {author} {\bibfnamefont {S.}~\bibnamefont {Gasparinetti}},\ }\bibfield  {title} {\bibinfo {title} {Universal control of a bosonic mode via drive-activated native cubic interactions},\ }\href {https://doi.org/10.1038/s41467-024-46507-1} {\bibfield  {journal} {\bibinfo  {journal}
  {Nature Communications}\ }\textbf {\bibinfo {volume} {15}},\ \bibinfo {pages} {2512} (\bibinfo {year} {2024})}\BibitemShut {NoStop}%
\bibitem [{\citenamefont {hui Yu}\ \emph {et~al.}(2024)\citenamefont {hui Yu}, \citenamefont {Zhu}, \citenamefont {jiao Xue},\ and\ \citenamefont {rong Li}}]{yu2024}%
  \BibitemOpen
  \bibfield  {author} {\bibinfo {author} {\bibfnamefont {K.}~\bibnamefont {hui Yu}}, \bibinfo {author} {\bibfnamefont {F.}~\bibnamefont {Zhu}}, \bibinfo {author} {\bibfnamefont {J.}~\bibnamefont {jiao Xue}},\ and\ \bibinfo {author} {\bibfnamefont {H.}~\bibnamefont {rong Li}},\ }\href {https://arxiv.org/abs/2406.13390} {\bibinfo {title} {Stabilizing the kerr arbitrary cat states and holonomic universal control}} (\bibinfo {year} {2024}),\ \Eprint {https://arxiv.org/abs/2406.13390} {arXiv:2406.13390 [quant-ph]} \BibitemShut {NoStop}%
\bibitem [{\citenamefont {Pietik\"ainen}\ \emph {et~al.}(2022)\citenamefont {Pietik\"ainen}, \citenamefont {\v{C}ernot\'\i{}k}, \citenamefont {Puri}, \citenamefont {Filip},\ and\ \citenamefont {Girvin}}]{Pietikainen:2022iqj}%
  \BibitemOpen
  \bibfield  {author} {\bibinfo {author} {\bibfnamefont {I.}~\bibnamefont {Pietik\"ainen}}, \bibinfo {author} {\bibfnamefont {O.}~\bibnamefont {\v{C}ernot\'\i{}k}}, \bibinfo {author} {\bibfnamefont {S.}~\bibnamefont {Puri}}, \bibinfo {author} {\bibfnamefont {R.}~\bibnamefont {Filip}},\ and\ \bibinfo {author} {\bibfnamefont {S.~M.}\ \bibnamefont {Girvin}},\ }\bibfield  {title} {\bibinfo {title} {{Controlled beam splitter gate transparent to dominant ancilla errors}},\ }\href {https://doi.org/10.1088/2058-9565/ac760a} {\bibfield  {journal} {\bibinfo  {journal} {Quantum Sci. Technol.}\ }\textbf {\bibinfo {volume} {7}},\ \bibinfo {pages} {035025} (\bibinfo {year} {2022})}\BibitemShut {NoStop}%
\bibitem [{\citenamefont {Liu}\ \emph {et~al.}(2024)\citenamefont {Liu}, \citenamefont {Singh}, \citenamefont {Smith}, \citenamefont {Crane}, \citenamefont {Martyn}, \citenamefont {Eickbusch}, \citenamefont {Schuckert}, \citenamefont {Li}, \citenamefont {Sinanan-Singh}, \citenamefont {Soley}, \citenamefont {Tsunoda}, \citenamefont {Chuang}, \citenamefont {Wiebe},\ and\ \citenamefont {Girvin}}]{Liu2024}%
  \BibitemOpen
  \bibfield  {author} {\bibinfo {author} {\bibfnamefont {Y.}~\bibnamefont {Liu}}, \bibinfo {author} {\bibfnamefont {S.}~\bibnamefont {Singh}}, \bibinfo {author} {\bibfnamefont {K.~C.}\ \bibnamefont {Smith}}, \bibinfo {author} {\bibfnamefont {E.}~\bibnamefont {Crane}}, \bibinfo {author} {\bibfnamefont {J.~M.}\ \bibnamefont {Martyn}}, \bibinfo {author} {\bibfnamefont {A.}~\bibnamefont {Eickbusch}}, \bibinfo {author} {\bibfnamefont {A.}~\bibnamefont {Schuckert}}, \bibinfo {author} {\bibfnamefont {R.~D.}\ \bibnamefont {Li}}, \bibinfo {author} {\bibfnamefont {J.}~\bibnamefont {Sinanan-Singh}}, \bibinfo {author} {\bibfnamefont {M.~B.}\ \bibnamefont {Soley}}, \bibinfo {author} {\bibfnamefont {T.}~\bibnamefont {Tsunoda}}, \bibinfo {author} {\bibfnamefont {I.~L.}\ \bibnamefont {Chuang}}, \bibinfo {author} {\bibfnamefont {N.}~\bibnamefont {Wiebe}},\ and\ \bibinfo {author} {\bibfnamefont {S.~M.}\ \bibnamefont {Girvin}},\ }\href@noop {} {\bibinfo {title} {Hybrid oscillator-qubit quantum processors: Instruction set
  architectures, abstract machine models, and applications}} (\bibinfo {year} {2024}),\ \Eprint {https://arxiv.org/abs/2407.10381} {arXiv:2407.10381 [quant-ph]} \BibitemShut {NoStop}%
\bibitem [{\citenamefont {Tsunoda}\ \emph {et~al.}(2023)\citenamefont {Tsunoda}, \citenamefont {Teoh}, \citenamefont {Kalfus}, \citenamefont {de~Graaf}, \citenamefont {Chapman}, \citenamefont {Curtis}, \citenamefont {Thakur}, \citenamefont {Girvin},\ and\ \citenamefont {Schoelkopf}}]{tsunoda2023error}%
  \BibitemOpen
  \bibfield  {author} {\bibinfo {author} {\bibfnamefont {T.}~\bibnamefont {Tsunoda}}, \bibinfo {author} {\bibfnamefont {J.~D.}\ \bibnamefont {Teoh}}, \bibinfo {author} {\bibfnamefont {W.~D.}\ \bibnamefont {Kalfus}}, \bibinfo {author} {\bibfnamefont {S.~J.}\ \bibnamefont {de~Graaf}}, \bibinfo {author} {\bibfnamefont {B.~J.}\ \bibnamefont {Chapman}}, \bibinfo {author} {\bibfnamefont {J.~C.}\ \bibnamefont {Curtis}}, \bibinfo {author} {\bibfnamefont {N.}~\bibnamefont {Thakur}}, \bibinfo {author} {\bibfnamefont {S.~M.}\ \bibnamefont {Girvin}},\ and\ \bibinfo {author} {\bibfnamefont {R.~J.}\ \bibnamefont {Schoelkopf}},\ }\bibfield  {title} {\bibinfo {title} {Error-detectable bosonic entangling gates with a noisy ancilla},\ }\href {https://journals.aps.org/prxquantum/abstract/10.1103/PRXQuantum.4.020354} {\bibfield  {journal} {\bibinfo  {journal} {PRX Quantum}\ }\textbf {\bibinfo {volume} {4}},\ \bibinfo {pages} {020354} (\bibinfo {year} {2023})}\BibitemShut {NoStop}%
\bibitem [{\citenamefont {Teoh}\ \emph {et~al.}(2023)\citenamefont {Teoh}, \citenamefont {Winkel}, \citenamefont {Babla}, \citenamefont {Chapman}, \citenamefont {Claes}, \citenamefont {de~Graaf}, \citenamefont {Garmon}, \citenamefont {Kalfus}, \citenamefont {Lu}, \citenamefont {Maiti}, \citenamefont {Sahay}, \citenamefont {Thakur}, \citenamefont {Tsunoda}, \citenamefont {Xue}, \citenamefont {Frunzio}, \citenamefont {Girvin}, \citenamefont {Puri},\ and\ \citenamefont {Schoelkopf}}]{Teoh2023}%
  \BibitemOpen
  \bibfield  {author} {\bibinfo {author} {\bibfnamefont {J.~D.}\ \bibnamefont {Teoh}}, \bibinfo {author} {\bibfnamefont {P.}~\bibnamefont {Winkel}}, \bibinfo {author} {\bibfnamefont {H.~K.}\ \bibnamefont {Babla}}, \bibinfo {author} {\bibfnamefont {B.~J.}\ \bibnamefont {Chapman}}, \bibinfo {author} {\bibfnamefont {J.}~\bibnamefont {Claes}}, \bibinfo {author} {\bibfnamefont {S.~J.}\ \bibnamefont {de~Graaf}}, \bibinfo {author} {\bibfnamefont {J.~W.~O.}\ \bibnamefont {Garmon}}, \bibinfo {author} {\bibfnamefont {W.~D.}\ \bibnamefont {Kalfus}}, \bibinfo {author} {\bibfnamefont {Y.}~\bibnamefont {Lu}}, \bibinfo {author} {\bibfnamefont {A.}~\bibnamefont {Maiti}}, \bibinfo {author} {\bibfnamefont {K.}~\bibnamefont {Sahay}}, \bibinfo {author} {\bibfnamefont {N.}~\bibnamefont {Thakur}}, \bibinfo {author} {\bibfnamefont {T.}~\bibnamefont {Tsunoda}}, \bibinfo {author} {\bibfnamefont {S.~H.}\ \bibnamefont {Xue}}, \bibinfo {author} {\bibfnamefont {L.}~\bibnamefont {Frunzio}}, \bibinfo {author} {\bibfnamefont {S.~M.}\
  \bibnamefont {Girvin}}, \bibinfo {author} {\bibfnamefont {S.}~\bibnamefont {Puri}},\ and\ \bibinfo {author} {\bibfnamefont {R.~J.}\ \bibnamefont {Schoelkopf}},\ }\bibfield  {title} {\bibinfo {title} {Dual-rail encoding with superconducting cavities},\ }\href {https://doi.org/10.1073/pnas.2221736120} {\bibfield  {journal} {\bibinfo  {journal} {Proc. Natl. Acad. Sci.}\ }\textbf {\bibinfo {volume} {120}},\ \bibinfo {pages} {e2221736120} (\bibinfo {year} {2023})}\BibitemShut {NoStop}%
\bibitem [{\citenamefont {Chou}\ \emph {et~al.}(2024)\citenamefont {Chou}, \citenamefont {Shemma}, \citenamefont {McCarrick}, \citenamefont {Chien}, \citenamefont {Teoh}, \citenamefont {Winkel}, \citenamefont {Anderson}, \citenamefont {Chen}, \citenamefont {Curtis}, \citenamefont {de~Graaf}, \citenamefont {Garmon}, \citenamefont {Gudlewski}, \citenamefont {Kalfus}, \citenamefont {Keen}, \citenamefont {Khedkar}, \citenamefont {Lei}, \citenamefont {Liu}, \citenamefont {Lu}, \citenamefont {Lu}, \citenamefont {Maiti}, \citenamefont {Mastalli-Kelly}, \citenamefont {Mehta}, \citenamefont {Mundhada}, \citenamefont {Narla}, \citenamefont {Noh}, \citenamefont {Tsunoda}, \citenamefont {Xue}, \citenamefont {Yuan}, \citenamefont {Frunzio}, \citenamefont {Aumentado}, \citenamefont {Puri}, \citenamefont {Girvin}, \citenamefont {Moseley},\ and\ \citenamefont {Schoelkopf}}]{Chou2024}%
  \BibitemOpen
  \bibfield  {author} {\bibinfo {author} {\bibfnamefont {K.~S.}\ \bibnamefont {Chou}}, \bibinfo {author} {\bibfnamefont {T.}~\bibnamefont {Shemma}}, \bibinfo {author} {\bibfnamefont {H.}~\bibnamefont {McCarrick}}, \bibinfo {author} {\bibfnamefont {T.-C.}\ \bibnamefont {Chien}}, \bibinfo {author} {\bibfnamefont {J.~D.}\ \bibnamefont {Teoh}}, \bibinfo {author} {\bibfnamefont {P.}~\bibnamefont {Winkel}}, \bibinfo {author} {\bibfnamefont {A.}~\bibnamefont {Anderson}}, \bibinfo {author} {\bibfnamefont {J.}~\bibnamefont {Chen}}, \bibinfo {author} {\bibfnamefont {J.~C.}\ \bibnamefont {Curtis}}, \bibinfo {author} {\bibfnamefont {S.~J.}\ \bibnamefont {de~Graaf}}, \bibinfo {author} {\bibfnamefont {J.~W.~O.}\ \bibnamefont {Garmon}}, \bibinfo {author} {\bibfnamefont {B.}~\bibnamefont {Gudlewski}}, \bibinfo {author} {\bibfnamefont {W.~D.}\ \bibnamefont {Kalfus}}, \bibinfo {author} {\bibfnamefont {T.}~\bibnamefont {Keen}}, \bibinfo {author} {\bibfnamefont {N.}~\bibnamefont {Khedkar}}, \bibinfo {author} {\bibfnamefont
  {C.~U.}\ \bibnamefont {Lei}}, \bibinfo {author} {\bibfnamefont {G.}~\bibnamefont {Liu}}, \bibinfo {author} {\bibfnamefont {P.}~\bibnamefont {Lu}}, \bibinfo {author} {\bibfnamefont {Y.}~\bibnamefont {Lu}}, \bibinfo {author} {\bibfnamefont {A.}~\bibnamefont {Maiti}}, \bibinfo {author} {\bibfnamefont {L.}~\bibnamefont {Mastalli-Kelly}}, \bibinfo {author} {\bibfnamefont {N.}~\bibnamefont {Mehta}}, \bibinfo {author} {\bibfnamefont {S.~O.}\ \bibnamefont {Mundhada}}, \bibinfo {author} {\bibfnamefont {A.}~\bibnamefont {Narla}}, \bibinfo {author} {\bibfnamefont {T.}~\bibnamefont {Noh}}, \bibinfo {author} {\bibfnamefont {T.}~\bibnamefont {Tsunoda}}, \bibinfo {author} {\bibfnamefont {S.~H.}\ \bibnamefont {Xue}}, \bibinfo {author} {\bibfnamefont {J.~O.}\ \bibnamefont {Yuan}}, \bibinfo {author} {\bibfnamefont {L.}~\bibnamefont {Frunzio}}, \bibinfo {author} {\bibfnamefont {J.}~\bibnamefont {Aumentado}}, \bibinfo {author} {\bibfnamefont {S.}~\bibnamefont {Puri}}, \bibinfo {author} {\bibfnamefont {S.~M.}\ \bibnamefont
  {Girvin}}, \bibinfo {author} {\bibfnamefont {S.~H.}\ \bibnamefont {Moseley}},\ and\ \bibinfo {author} {\bibfnamefont {R.~J.}\ \bibnamefont {Schoelkopf}},\ }\bibfield  {title} {\bibinfo {title} {A superconducting dual-rail cavity qubit with erasure-detected logical measurements},\ }\href {https://doi.org/10.1038/s41567-024-02539-4} {\bibfield  {journal} {\bibinfo  {journal} {Nat. Phys.}\ }\textbf {\bibinfo {volume} {20}},\ \bibinfo {pages} {1454} (\bibinfo {year} {2024})}\BibitemShut {NoStop}%
\bibitem [{\citenamefont {Puri}\ \emph {et~al.}(2017)\citenamefont {Puri}, \citenamefont {Boutin},\ and\ \citenamefont {Blais}}]{Puri2017}%
  \BibitemOpen
  \bibfield  {author} {\bibinfo {author} {\bibfnamefont {S.}~\bibnamefont {Puri}}, \bibinfo {author} {\bibfnamefont {S.}~\bibnamefont {Boutin}},\ and\ \bibinfo {author} {\bibfnamefont {A.}~\bibnamefont {Blais}},\ }\bibfield  {title} {\bibinfo {title} {Engineering the quantum states of light in a kerr-nonlinear resonator by two-photon driving},\ }\href {https://doi.org/10.1038/s41534-017-0019-1} {\bibfield  {journal} {\bibinfo  {journal} {npj Quantum Information}\ }\textbf {\bibinfo {volume} {3}},\ \bibinfo {pages} {18} (\bibinfo {year} {2017})}\BibitemShut {NoStop}%
\bibitem [{\citenamefont {de~Albornoz}\ \emph {et~al.}(2024)\citenamefont {de~Albornoz}, \citenamefont {Cortiñas}, \citenamefont {Schäfer}, \citenamefont {Frattini}, \citenamefont {Allen}, \citenamefont {Cabral}, \citenamefont {Videla}, \citenamefont {Khazaei}, \citenamefont {Geva}, \citenamefont {Batista},\ and\ \citenamefont {Devoret}}]{alex2024}%
  \BibitemOpen
  \bibfield  {author} {\bibinfo {author} {\bibfnamefont {A.~C.~C.}\ \bibnamefont {de~Albornoz}}, \bibinfo {author} {\bibfnamefont {R.~G.}\ \bibnamefont {Cortiñas}}, \bibinfo {author} {\bibfnamefont {M.}~\bibnamefont {Schäfer}}, \bibinfo {author} {\bibfnamefont {N.~E.}\ \bibnamefont {Frattini}}, \bibinfo {author} {\bibfnamefont {B.}~\bibnamefont {Allen}}, \bibinfo {author} {\bibfnamefont {D.~G.~A.}\ \bibnamefont {Cabral}}, \bibinfo {author} {\bibfnamefont {P.~E.}\ \bibnamefont {Videla}}, \bibinfo {author} {\bibfnamefont {P.}~\bibnamefont {Khazaei}}, \bibinfo {author} {\bibfnamefont {E.}~\bibnamefont {Geva}}, \bibinfo {author} {\bibfnamefont {V.~S.}\ \bibnamefont {Batista}},\ and\ \bibinfo {author} {\bibfnamefont {M.~H.}\ \bibnamefont {Devoret}},\ }\href {https://arxiv.org/abs/2409.13113} {\bibinfo {title} {Oscillatory dissipative tunneling in an asymmetric double-well potential}} (\bibinfo {year} {2024}),\ \Eprint {https://arxiv.org/abs/2409.13113} {arXiv:2409.13113 [quant-ph]} \BibitemShut {NoStop}%
\bibitem [{\citenamefont {Chirikov}(1979)}]{CHIRIKOV1979}%
  \BibitemOpen
  \bibfield  {author} {\bibinfo {author} {\bibfnamefont {B.~V.}\ \bibnamefont {Chirikov}},\ }\bibfield  {title} {\bibinfo {title} {A universal instability of many-dimensional oscillator systems},\ }\href {https://doi.org/https://doi.org/10.1016/0370-1573(79)90023-1} {\bibfield  {journal} {\bibinfo  {journal} {Physics Reports}\ }\textbf {\bibinfo {volume} {52}},\ \bibinfo {pages} {263} (\bibinfo {year} {1979})}\BibitemShut {NoStop}%
\bibitem [{\citenamefont {Chepelianskii}\ and\ \citenamefont {Shepelyansky}(2002)}]{Chepelianskii2002}%
  \BibitemOpen
  \bibfield  {author} {\bibinfo {author} {\bibfnamefont {A.~D.}\ \bibnamefont {Chepelianskii}}\ and\ \bibinfo {author} {\bibfnamefont {D.~L.}\ \bibnamefont {Shepelyansky}},\ }\bibfield  {title} {\bibinfo {title} {Simulation of chaos-assisted tunneling in a semiclassical regime on existing quantum computers},\ }\href {https://doi.org/10.1103/PhysRevA.66.054301} {\bibfield  {journal} {\bibinfo  {journal} {Phys. Rev. A}\ }\textbf {\bibinfo {volume} {66}},\ \bibinfo {pages} {054301} (\bibinfo {year} {2002})}\BibitemShut {NoStop}%
\bibitem [{\citenamefont {Anand}\ \emph {et~al.}(2021)\citenamefont {Anand}, \citenamefont {Styliaris}, \citenamefont {Kumari},\ and\ \citenamefont {Zanardi}}]{Anand21}%
  \BibitemOpen
  \bibfield  {author} {\bibinfo {author} {\bibfnamefont {N.}~\bibnamefont {Anand}}, \bibinfo {author} {\bibfnamefont {G.}~\bibnamefont {Styliaris}}, \bibinfo {author} {\bibfnamefont {M.}~\bibnamefont {Kumari}},\ and\ \bibinfo {author} {\bibfnamefont {P.}~\bibnamefont {Zanardi}},\ }\bibfield  {title} {\bibinfo {title} {Quantum coherence as a signature of chaos},\ }\href {https://doi.org/10.1103/PhysRevResearch.3.023214} {\bibfield  {journal} {\bibinfo  {journal} {Phys. Rev. Res.}\ }\textbf {\bibinfo {volume} {3}},\ \bibinfo {pages} {023214} (\bibinfo {year} {2021})}\BibitemShut {NoStop}%
\bibitem [{\citenamefont {Soskin}\ and\ \citenamefont {Mannella}(2009)}]{soskin2009}%
  \BibitemOpen
  \bibfield  {author} {\bibinfo {author} {\bibfnamefont {S.~M.}\ \bibnamefont {Soskin}}\ and\ \bibinfo {author} {\bibfnamefont {R.}~\bibnamefont {Mannella}},\ }\bibfield  {title} {\bibinfo {title} {Maximal width of the separatrix chaotic layer},\ }\href {https://doi.org/10.1103/PhysRevE.80.066212} {\bibfield  {journal} {\bibinfo  {journal} {Phys. Rev. E}\ }\textbf {\bibinfo {volume} {80}},\ \bibinfo {pages} {066212} (\bibinfo {year} {2009})}\BibitemShut {NoStop}%
\bibitem [{\citenamefont {Zaslavsky}(2007)}]{zaslavsky2007physics}%
  \BibitemOpen
  \bibfield  {author} {\bibinfo {author} {\bibfnamefont {G.}~\bibnamefont {Zaslavsky}},\ }\href {https://doi.org/10.1142/p507} {\emph {\bibinfo {title} {The Physics of Chaos in Hamiltonian Systems}}},\ G - Reference,Information and Interdisciplinary Subjects Series\ (\bibinfo  {publisher} {Imperial College Press},\ \bibinfo {year} {2007})\BibitemShut {NoStop}%
\bibitem [{\citenamefont {Weiss}\ \emph {et~al.}(2021)\citenamefont {Weiss}, \citenamefont {DeGottardi}, \citenamefont {Koch},\ and\ \citenamefont {Ferguson}}]{weiss2021}%
  \BibitemOpen
  \bibfield  {author} {\bibinfo {author} {\bibfnamefont {D.~K.}\ \bibnamefont {Weiss}}, \bibinfo {author} {\bibfnamefont {W.}~\bibnamefont {DeGottardi}}, \bibinfo {author} {\bibfnamefont {J.}~\bibnamefont {Koch}},\ and\ \bibinfo {author} {\bibfnamefont {D.~G.}\ \bibnamefont {Ferguson}},\ }\bibfield  {title} {\bibinfo {title} {Variational tight-binding method for simulating large superconducting circuits},\ }\href {https://doi.org/10.1103/PhysRevResearch.3.033244} {\bibfield  {journal} {\bibinfo  {journal} {Phys. Rev. Res.}\ }\textbf {\bibinfo {volume} {3}},\ \bibinfo {pages} {033244} (\bibinfo {year} {2021})}\BibitemShut {NoStop}%
\bibitem [{\citenamefont {Marcos}\ \emph {et~al.}(2013)\citenamefont {Marcos}, \citenamefont {Rabl}, \citenamefont {Rico},\ and\ \citenamefont {Zoller}}]{marcos2013}%
  \BibitemOpen
  \bibfield  {author} {\bibinfo {author} {\bibfnamefont {D.}~\bibnamefont {Marcos}}, \bibinfo {author} {\bibfnamefont {P.}~\bibnamefont {Rabl}}, \bibinfo {author} {\bibfnamefont {E.}~\bibnamefont {Rico}},\ and\ \bibinfo {author} {\bibfnamefont {P.}~\bibnamefont {Zoller}},\ }\bibfield  {title} {\bibinfo {title} {Superconducting circuits for quantum simulation of dynamical gauge fields},\ }\href {https://doi.org/10.1103/PhysRevLett.111.110504} {\bibfield  {journal} {\bibinfo  {journal} {Phys. Rev. Lett.}\ }\textbf {\bibinfo {volume} {111}},\ \bibinfo {pages} {110504} (\bibinfo {year} {2013})}\BibitemShut {NoStop}%
\bibitem [{\citenamefont {Frattini}\ \emph {et~al.}(2017)\citenamefont {Frattini}, \citenamefont {Vool}, \citenamefont {Shankar}, \citenamefont {Narla}, \citenamefont {Sliwa},\ and\ \citenamefont {Devoret}}]{Frattini3wm}%
  \BibitemOpen
  \bibfield  {author} {\bibinfo {author} {\bibfnamefont {N.~E.}\ \bibnamefont {Frattini}}, \bibinfo {author} {\bibfnamefont {U.}~\bibnamefont {Vool}}, \bibinfo {author} {\bibfnamefont {S.}~\bibnamefont {Shankar}}, \bibinfo {author} {\bibfnamefont {A.}~\bibnamefont {Narla}}, \bibinfo {author} {\bibfnamefont {K.~M.}\ \bibnamefont {Sliwa}},\ and\ \bibinfo {author} {\bibfnamefont {M.~H.}\ \bibnamefont {Devoret}},\ }\bibfield  {title} {\bibinfo {title} {{3-wave mixing Josephson dipole element}},\ }\href {https://doi.org/10.1063/1.4984142} {\bibfield  {journal} {\bibinfo  {journal} {Applied Physics Letters}\ }\textbf {\bibinfo {volume} {110}},\ \bibinfo {pages} {222603} (\bibinfo {year} {2017})}\BibitemShut {NoStop}%
\bibitem [{\citenamefont {Frattini}(2021)}]{Frattini2021}%
  \BibitemOpen
  \bibfield  {author} {\bibinfo {author} {\bibfnamefont {N.}~\bibnamefont {Frattini}},\ }\href@noop {} {\emph {\bibinfo {title} {Three-wave Mixing in Superconducting Circuits: Stabilizing Cats with SNAILs}}}\ (\bibinfo  {publisher} {Yale University, thesis},\ \bibinfo {year} {2021})\BibitemShut {NoStop}%
\bibitem [{\citenamefont {Venkatraman}\ \emph {et~al.}(2022)\citenamefont {Venkatraman}, \citenamefont {Xiao}, \citenamefont {Corti\~nas}, \citenamefont {Eickbusch},\ and\ \citenamefont {Devoret}}]{jaya2022}%
  \BibitemOpen
  \bibfield  {author} {\bibinfo {author} {\bibfnamefont {J.}~\bibnamefont {Venkatraman}}, \bibinfo {author} {\bibfnamefont {X.}~\bibnamefont {Xiao}}, \bibinfo {author} {\bibfnamefont {R.~G.}\ \bibnamefont {Corti\~nas}}, \bibinfo {author} {\bibfnamefont {A.}~\bibnamefont {Eickbusch}},\ and\ \bibinfo {author} {\bibfnamefont {M.~H.}\ \bibnamefont {Devoret}},\ }\bibfield  {title} {\bibinfo {title} {Static effective hamiltonian of a rapidly driven nonlinear system},\ }\href {https://doi.org/10.1103/PhysRevLett.129.100601} {\bibfield  {journal} {\bibinfo  {journal} {Phys. Rev. Lett.}\ }\textbf {\bibinfo {volume} {129}},\ \bibinfo {pages} {100601} (\bibinfo {year} {2022})}\BibitemShut {NoStop}%
\end{thebibliography}%
\end{document}